\documentclass[a4paper,11pt]{article}
\pdfoutput=1 

\usepackage{jheppub} 

\usepackage[T1]{fontenc} 

\usepackage{amsmath,amsfonts,amsbsy,amssymb,array,accents,dsfont}
\usepackage{enumerate,latexsym,graphicx}
\usepackage{xcolor,tikz}
\usepackage{tcolorbox}

\usepackage{subfigure}
\usepackage{stmaryrd}

 \def\cN{\mathcal {N}}

\def\cV{\mathcal {V}}

\newcommand{\eq}[1]{\eqref{#1}}

\newcommand{\beq}{\begin{equation}}
\newcommand{\eeq}{\end{equation}}
\newcommand{\bea}{\begin{eqnarray}}
\newcommand{\eea}{\end{eqnarray}}

\newcommand{\vep}{\varepsilon}

\newcommand{\der}{\partial}

\newcommand{\nn}{\nonumber}

\def \sl{\text{sl}}
\def \su{\text{su}}

\newcommand{\n}{{\mathsf n}}
\newcommand{\m}{{\mathsf m}}
\renewcommand{\k}{{\mathsf k}}
\newcommand{\pp}{{\mathsf p}}

\newcommand{\msl}{m_{\sl}}
\newcommand{\msu}{m_{\su}}

\newcommand{\NS}{\mathrm{NS}}

\newcommand{\jsl}{j_{\sl}}
\newcommand{\jsu}{j_{\su}}

\newcommand{\rhoo}{\ensuremath{\! \rho \:\! }}

\usepackage{braket}
\usepackage{tikz}
\usetikzlibrary{matrix,calc,positioning,decorations.markings,decorations.pathmorphing,decorations.pathreplacing}
\usetikzlibrary{arrows,cd}
\usepackage{bbding}
\usetikzlibrary{positioning}
\tikzset{>=stealth}
\newcommand{\Vsl}{V}
\newcommand{\Vsu}{V'}

\newcommand{\sucur}{k} 
\newcommand{\psisl}{\psi}
\newcommand{\psisu}{\chi}
\renewcommand{\jsl}{j}
\renewcommand{\jsu}{j'}
\renewcommand{\msl}{m}
\renewcommand{\msu}{m'}

\newcommand{\LLL}{\scriptscriptstyle{\text{L}} }
\newcommand{\RRR}{\scriptscriptstyle{\text{R}} }

\newcommand{\inv}{^{-1}}
\newcommand{\del}{\partial}             
\newcommand{\bdel}{\bar{\partial}}

\newcommand{\qqquad}{\;, \quad\qquad}  

\newcommand{\half}{\frac{1}{2}}

\newcommand{\SSS}{\mathbb{S}}

\newcommand{\RR}{\mathbb{R}}
\newcommand{\ZZ}{\mathbb{Z}}

\newcommand{\TT}{\mathbb{T}}
\newcommand{\Aa}{\mathcal{A}}
\newcommand{\Bb}{\mathcal{B}}

\newcommand{\Gg}{\mathcal{G}}
\newcommand{\Hh}{\mathcal{H}}

\newcommand{\Jj}{\mathcal{J}}
\newcommand{\Ll}{\mathcal{L}}

\newcommand{\Oo}{\mathcal{O}}

\newcommand{\Qq}{\mathcal{Q}}

\newcommand{\Tt}{\mathcal{T}}

\newcommand{\Vv}{\mathcal{V}}
\newcommand{\Xx}{\mathcal{X}}
\newcommand{\Yy}{\mathcal{Y}}
\newcommand{\Ww}{\mathcal{W}}

\newcommand{\R}{\ensuremath{\mathbb{R}}}

\newcommand{\mb}{\bar{m}}

\newcommand{\xb}{\bar{x}}

\newcommand{\w}{\omega}
\newcommand{\cc}{\mathrm{c.c.}}
\newcommand{\tvphi}{\tilde{\varphi}}
\newcommand{\Z}{\mathbb{Z}}

\usepackage{color}

\usepackage[normalem]{ulem}  

\definecolor{darkred}{rgb}{0.6,0,0}
\definecolor{darkblue}{rgb}{0,0,0.6}

\newcommand{\be}{\begin{equation}}
\newcommand{\ee}{\end{equation}}

\usepackage[bbgreekl]{mathbbol}
\usepackage{amsfonts}
\DeclareSymbolFontAlphabet{\mathbb}{AMSb} 
\DeclareSymbolFontAlphabet{\mathbbl}{bbold} 

\newcommand{\Jd}{\mathbb{J}} 
\newcommand{\jd}{\mathbbl{j}} 
\newcommand{\Gd}{\mathbb{G}} 
\newcommand{\psidr}{\psi} 

\title{Worldsheet computation of heavy-light correlators
}

\author[a]{Davide Bufalini,}
\author[b,c
]{Sergio Iguri,}
\author[d]{Nicolas Kovensky,}
\author[a]{David Turton}

\affiliation[a]{Mathematical Sciences and STAG Research Centre, University of Southampton, Southampton
SO17 1BJ, United Kingdom.}

\affiliation[b]{Instituto de Astronomía y Física del Espacio (IAFE), CONICET - Universidad de Buenos Aires, C. C. 67, Suc. 28, 1428 Buenos Aires, Argentina.}

\affiliation[c]{Mathematics with Computer Science Program, Guangdong Technion - Israel Institute of Technology, 515063 Shantou, Guangdong, People's Republic of China.
}
\affiliation[d]{Institut de Physique Th\'eorique, Universit\'e Paris Saclay, CEA, CNRS, Orme des Merisiers, 91191 Gif-sur-Yvette CEDEX, France.}

\emailAdd{d.bufalini@soton.ac.uk}
\emailAdd{siguri@iafe.uba.ar}
\emailAdd{nicolas.kovensky@ipht.fr}
\emailAdd{d.j.turton@soton.ac.uk}

\abstract{
We compute a large collection of string worldsheet correlators describing light probes interacting with heavy black hole microstates. 
The heavy states consist of NS5 branes carrying momentum and/or fundamental string charge.
In the fivebrane decoupling limit, worldsheet string theory on a family of such backgrounds is given by exactly solvable null-gauged WZW models.
We construct physical vertex operators in these cosets, including all massless fluctuations. 
We first compute a large class of novel heavy-light-light-heavy correlators in the AdS$_3$ limit, where the light operators include those dual to chiral primaries of the holographically dual CFT.
We compare a subset of these correlators to the holographic CFT at the symmetric product orbifold point, and find precise agreement in all cases, including for light operators in twisted sectors of the orbifold CFT.
The agreement is highly non-trivial, and includes amplitudes that describe the analogue of Hawking radiation for these microstates.
We further derive a formula for worldsheet correlators consisting of $n$ light insertions on these backgrounds, and discuss which subset of these correlators are likely to be protected. 
As a test, we compute a heavy-light five-point function, obtaining precisely the same result both from the worldsheet and the symmetric orbifold CFT. 
This paper is a companion to and extension of~\cite{Bufalini:2022wyp}.
}

\begin{document} 

\setcounter{tocdepth}{2}

\maketitle
\flushbottom

\section{Introduction}

String Theory provides a microscopic description of black holes as being bound states of strings and branes with an exponentially large number of internal microstates~\cite{Strominger:1996sh}. Amongst these microstates, there are coherent pure states, large families of which have been shown to be well-described by smooth and horizonless supergravity solutions, see e.g.~\cite{Lunin:2001jy,Lunin:2002iz,Bena:2016ypk,Bena:2017xbt,Ceplak:2018pws,Heidmann:2019zws,Ganchev:2022exf}. Upon taking an appropriate AdS decoupling limit, these solutions are proposed to correspond to specific families of pure states in the holographically dual CFT (HCFT); precision holography has provided sharp evidence supporting this correspondence~\cite{Kanitscheider:2006zf,Kanitscheider:2007wq,Giusto:2015dfa,Giusto:2019qig,Rawash:2021pik}.

While supergravity constructions provide valuable insight into the structure of black hole microstates, it is natural to expect that string-theoretic physics beyond supergravity will be necessary to obtain a complete description of black hole microstructure. 
A fruitful arena in which to investigate such stringy physics is provided by bound states of NS5 branes carrying fundamental string (F1) and/or momentum charge (P). More specifically, we work in Type IIB compactified on $\SSS^1\times\TT^4$, with $n_5$ NS5 branes wrapped on $\SSS^1\times\TT^4$, $n_1$ units of F1 winding on $\SSS^1$, and $n_P$ units of momentum charge along $\SSS^1$.

Upon taking the fivebrane decoupling limit, one obtains asymptotically linear dilaton configurations, which are holographically dual to (doubly scaled) Little String Theory~\cite{Giveon:1999px,Giveon:2001up}.
In an appropriate region of the parameter space, there is an AdS$_3$ regime in the IR, and one can take a further AdS$_3$ decoupling limit~\cite{Maldacena:1997re}. Upon doing so, one obtains the well-studied NS5-F1 instance of AdS$_3$/CFT$_2$ holography~\cite{Giveon:1998ns,Kutasov:1999xu}.

The NSNS vacuum of the holographic CFT corresponds to the global AdS$_3\times \SSS^3 \times \TT^4$ background, whose worldsheet theory involves an SL(2,$\R$)$\times$SU(2) Wess-Zumino-Witten (WZW)  model~\cite{Zamolodchikov:1986bd,Teschner:1999ug,Maldacena:2000hw,Maldacena:2000kv,Maldacena:2001km}.
In recent work, a family of gauged WZW models has been constructed and studied  involving the same Lie groups, providing an exact worldsheet description of a set of NS5-F1-P black hole microstates  \cite{Martinec:2017ztd,Martinec:2018nco,Martinec:2019wzw,Martinec:2020gkv,Bufalini:2021ndn}.

Processes in which light probes interact with a heavy background such as a black hole or a black hole microstate give rise to interesting and computable dynamical observables. Mixed heavy-light (HL) correlators have been previously studied in  holography, see e.g.~\cite{Galliani:2016cai,Fitzpatrick:2016mjq,Balasubramanian:2017fan,Galliani:2017jlg,Bombini:2019vnc}. 
In the NS5-F1 system, there is a locus in moduli space at which the holographic CFT is conjectured to be the ${\cal{N}}=(4,4)$ symmetric product orbifold CFT with target space $\left(\TT^4\right)^{N}/S_{N}$, where $N=n_1 n_5$. There is now a substantial body of evidence for this conjecture, see e.g.~\cite{Kanitscheider:2006zf,Kanitscheider:2007wq,Giusto:2015dfa,Giusto:2019qig,Rawash:2021pik,Argurio:2000tb,Gaberdiel:2007vu,Dabholkar:2007ey,Giribet:2007wp}. 
For recent discussions of holography in related systems, see~\cite{Eberhardt:2021vsx,Eberhardt:2019qcl,Balthazar:2021xeh}.

For instance, heavy-light-light-heavy (HLLH) four-point functions have been computed in the supergravity approximation and/or in the symmetric product orbifold CFT, for particular sets of heavy and light operators~\cite{Galliani:2016cai,Galliani:2017jlg,Bombini:2019vnc}.
Having solvable worldsheet models associated to black hole microstates means we can go much further by taking into account $\alpha'$ corrections \cite{Martinec:2017ztd}.
Given a worldsheet model describing string dynamics on a heavy background, the relevant quantities correspond to (a particular limit of) integrated correlators of light operators in the worldsheet vacuum.

Worldsheet correlators in global AdS$_3$ were first studied in~\cite{Maldacena:2001km}, building in part on~\cite{Fateev,Teschner:1999ug}, and 
the role of the vertex operators associated with  spectrally flowed representations was highlighted. Further studies  include~\cite{Giribet:2001ft,Cardona:2010qf,Cagnacci:2013ufa,Dei:2021xgh,Dei:2021yom}. The spectrum of chiral primaries and their three-point functions in global AdS$_3\times \SSS^3 \times \TT^4$ were computed in \cite{Argurio:2000tb,Gaberdiel:2007vu,Dabholkar:2007ey,Giribet:2007wp}, and shown to match those of the symmetric product orbifold CFT, as studied in~\cite{Lunin:2000yv,Lunin:2001pw}.  

The supergravity backgrounds we consider are known as NS5-F1 circular supertubes and spectral flows thereof~\cite{Lunin:2004uu,Giusto:2004id,Giusto:2004ip,Jejjala:2005yu,Giusto:2012yz,Chakrabarty:2015foa}. This includes non-BPS spectrally flowed supertubes, known as the JMaRT solutions, after the authors of~\cite{Jejjala:2005yu}. 
The associated worldsheet models are null-gauged WZW models, where before gauging one considers a (10+2)-dimensional target space AdS$_3\times \SSS^3 \times \R_t \times \SSS^1_y\times \TT^4$. Roughly speaking, in the IR AdS$_3$ regime, the gauging is concentrated mostly along the $t$ and $y$ directions, while in the linear dilaton regime the gauging is concentrated mostly in the time and angular directions of SL(2,$\R$).

These coset models can also be thought of as marginal current-current deformations of the worldsheet theory for strings in AdS$_3$. These are instances of a larger class of deformations that undo the decoupling limit with respect to the F1 harmonic function, i.e.~they~\textit{``add back the 1+''} in that function, leading to linear dilaton asymptotics; see e.g.~\cite{Martinec:2020gkv}. At the level of the dual field theory, a closely related procedure has been argued to correspond to the so-called single-trace $T\bar{T}$ irrelevant deformation of the original holographic CFT~\cite{Giveon:2017myj,Giveon:2017nie}, flowing towards a non-local Little String Theory.

In this paper we study string correlators in these highly excited backgrounds. To do so, we first compute a large set of physical vertex operators, in both NSNS and RR sectors, building on~\cite{Martinec:2018nco,Martinec:2020gkv}. These describe linearized perturbations of the background configurations. 
We focus primarily on coset states in discrete series representations, including worldsheet spectral flow, that are dual to chiral primary operator excitations in the HCFT.
When the background is BPS, a subset of these are BPS fluctuations.

The currents being gauged in these cosets are linear combinations of the Cartan generators of the symmetry algebra. Therefore the ``$m$-basis'' for vertex operators, in which the actions of these currents are diagonalized, is the natural framework to use. In the IR AdS$_3$ limit, we describe how these operators are related to their global AdS$_3\times\SSS^3$ counterparts.

We then compute a large set of correlators in the AdS$_3$ limit. It is well known that in worldsheet models of global AdS$_3$, one can define an ``$x$'' variable that corresponds to the local coordinate of the holographic CFT~\cite{Kutasov:1999xu}. 
One of the main novelties of our approach is the identification of the analogous $x$ variable in the coset models we study. This identification requires some care due to the gauging. Indeed, the construction of~\cite{Kutasov:1999xu} breaks down, because the SL(2,$\R$) raising and lowering operators do not commute with the BRST charge. A considerable amount of interesting physics follows from this step. It leads, for instance, to the combination of seemingly simple $m$-basis two-point functions into spacetime-local $x$-basis correlators with highly non-trivial $x$-dependence.

Our first main result is of a family of HLLH correlators, for which we obtain fully explicit expressions. In doing so, we show that these correlators assume a remarkably simple structure when written in terms of a covering space related only to the heavy states.
From this observation, we obtain our second main result: a closed-form expression for a set of HL worldsheet correlators with an arbitrary number $n$ of massless insertions, in terms of a correlator consisting of $n$ light insertions in global AdS$_3 \times \SSS^3$.
For $n=3$ this result can be made completely explicit, and we present a particular example in full detail. This constitutes the first correlator in the literature involving three light worldsheet vertices on a black hole microstate background, dual to a heavy-light five-point function of the holographic CFT.

\textit{A priori}, our worldsheet correlators give predictions for correlators of the dual holographic CFT at strong coupling. Generically, four-point correlators are not protected across moduli space, however, a specific set of HLLH correlators have been shown to precisely agree between supergravity and the symmetric product orbifold CFT~\cite{Galliani:2016cai}. Similarly, the
emission spectrum and rate for the unitary
analog of Hawking radiation from the JMaRT solutions agrees between supergravity and symmetric product orbifold CFT~\cite{Chowdhury:2007jx,Avery:2009tu,Avery:2009xr,Chakrabarty:2015foa}. Thus it is natural to investigate more generally which HL correlators are protected (at large $N$) between worldsheet and symmetric product orbifold CFT, and which are not.

We carry out this comparison for three sub-families of our worldsheet correlators. Firstly, we compare various sets of HLLH correlators to the symmetric product orbifold CFT, finding exact agreement in all cases for which the orbifold CFT correlator is available in the literature. Importantly, this matching holds at leading order in large $N$, but exactly in $\alpha'$. 
This comparison includes a substantial generalization of the supergravity and holographic CFT correlators computed in~\cite{Galliani:2016cai}. Our comparisons notably include an example in which the light operators in the symmetric orbifold CFT are twist-two. In this case, and as shown recently in~\cite{Lima:2021wrz,AlvesLima:2022elo}, the Lunin-Mathur covering map used in the symmetric orbifold computation is different to the one appearing in the worldsheet computation, making the comparison highly non-trivial. Remarkably, both results agree exactly in the large $N$ limit.

Secondly, we compute the five-point HLLLH symmetric orbifold CFT correlator corresponding to the three-point worldsheet correlator mentioned above, and also find exact agreement. While most of our main results were announced in the short paper~\cite{Bufalini:2022wyp}, this five-point correlator is completely new.

Finally, we compute the analogue of the Hawking radiation rate for the JMaRT solutions. Once again, we find perfect agreement with the dual symmetric product orbifold CFT, extending the supergravity and holographic CFT results of~\cite{Chowdhury:2007jx,Avery:2009tu,Avery:2009xr,Chakrabarty:2015foa}.

A likely explanation for this remarkable agreement is that the heavy states we consider are quite special. Specifically, the heavy backgrounds are related to the global AdS$_3\times \SSS^3$ vacuum via orbifolding and fractional spectral flow~\cite{Giusto:2012yz,Chakrabarty:2015foa}. This fact also underlies our general formula for the HL correlators with $n$ light insertions. When $n>3$, we do not expect these HL correlators to be generically protected across moduli space; we shall discuss this in detail in due course.

The structure of the paper is as follows.
In Section \ref{Section2} we review the null-gauged WZW models we study. We present the supergravity fields in the fivebrane decoupling limit, take the AdS$_3$ limit, and describe the dual heavy states of the holographic CFT. 
In Section \ref{sec:light states} we present the light operators we are interested in. We review the chiral primaries of the symmetric product orbifold CFT. We then describe how the corresponding operators are constructed in the worldsheet theory for strings in AdS$_3 \times \SSS^3$ in the RNS formalism, including spectrally flowed states.
In Section \ref{sec:nullgaugedmodel} we construct a large set of vertex operators of the worldsheet cosets we study, both in the NS and in the R sectors.
We then examine their AdS$_3$ limit and relate these  vertices to those constructed in Sec.~\ref{sec:light states}.

In Sections \ref{Section 5} and \ref{sec:Sec6}, we present our main results.
We identify the ``$x$'' variable dual to the local coordinate of the holographic CFT, and obtain an extensive set of novel HLLH correlators, including massless insertions with arbitrary spacetime weights and charges. The final results are presented in Eqs.~\eqref{finalHLLH} and \eqref{finalHLLHbeta}. We then compare a subset of these results to the symmetric product orbifold CFT, finding exact agreement for all correlators available in the literature. 
We present a closed formula for a large class of worldsheet correlators with an arbitrary number of massless insertions, Eq.~\eqref{finalHLLLLLH}. We compute a five-point correlator in the symmetric orbifold CFT and find agreement with our general worldsheet formula.
Finally, we compute the amplitude describing the unitary analogue of Hawking radiation for the JMaRT microstates.  We discuss our results in Section~\ref{sec:discussion}.

\section{The heavy background states}
\label{Section2}

In this section we introduce the heavy backgrounds known as NS5-F1 circular supertubes and spectral flows thereof~\cite{Lunin:2004uu,Giusto:2004id,Giusto:2004ip,Jejjala:2005yu,Giusto:2012yz,Chakrabarty:2015foa}, including the general NS5-F1-P JMaRT solutions and their BPS limits. We work in the fivebrane decoupling limit, and review the worldsheet description of tree-level string theory on these backgrounds in terms of null-gauged WZW models~\cite{Martinec:2017ztd,Martinec:2018nco,Bufalini:2021ndn}. In the IR, the backgrounds become asymptotically AdS$_3\times \SSS^3$, and we review the corresponding heavy states of the holographically dual CFT at the symmetric orbifold point~\cite{Chakrabarty:2015foa}.

\subsection{JMaRT backgrounds from the worldsheet}

We begin by reviewing the family of coset CFT models that describe strings probing the JMaRT backgrounds (and their BPS limits), introduced in~\cite{Martinec:2017ztd} and analyzed in~\cite{Martinec:2018nco,Bufalini:2021ndn}. We will mostly use the notation and conventions from~\cite{Bufalini:2021ndn}, to which we refer the reader who is interested in more details.
We work in units in which $\alpha'=1$.

The null-gauged WZW model relevant for the present work has the following coset as a target space:
\begin{equation}
\label{eq: coset}
\Gg/\Hh \; \times \TT^4 = \frac{\text{SL}(2,\R) \times \text{SU}(2)  \times \R_t \times \text{U}(1)_y}{\text{U}(1)_L \times \text{U}(1)_R} \; \times \TT^4 \,.
\end{equation}
To be precise, globally we work with the universal cover of SL(2,$\R$), and we gauge $\mathbb{R} \times U(1)$. 
 The line element and NSNS three-form flux of the 10+2-dimensional ``upstairs'' model before gauging are given in local coordinates by
\begin{align}\label{eq: base ds2}
    ds^2 & = n_5 \big( - \cosh^2 \rhoo \, d\tau^2 + d\rho^2 + \sinh^2 \rhoo\,  d\sigma^2       + d\theta^2 + \cos^2 \theta d\psi^2 + \sin^2 \theta d\phi^2   \big) - dt^2 + dy^2 , \nonumber\\[1ex]
    H & = n_5 \big( \sinh2\rho \; d\rho \wedge d\tau \wedge d\sigma
    +
    \sin2\theta \;  d\theta \wedge d\psi \wedge d\phi \big).
\end{align}
The Killing vectors associated to the group action being gauged are\footnote{Note that some conventions differ from our letter~\cite{Bufalini:2022wyp}. In the latter, the following changes must be performed to make contact with our current notation: $s_+ \mapsto l_2, s_- \mapsto -r_2, \mu \mapsto l_3, k_+ \mapsto l_4, k_- \mapsto r_4$.}
\begin{align}
\begin{aligned}
\label{eq: killing vectors}
	\xi_{\LLL} & =  (\del_\tau - \del_\sigma) - l_2(\del_\psi - \del_\phi) + l_3 \del_t - l_4 \del_y \,,\cr
	\xi_{\RRR} & =  (\del_\tau + \del_\sigma) + r_2(\del_\psi + \del_\phi) + r_3 \del_t - r_4 \del_y \,.
\end{aligned}
\end{align}
We could be slightly more general and include similar parameters $l_1$, $r_1$ in~\eqref{eq: killing vectors}, however we have assumed these to be non-vanishing and have set $l_1=r_1=1$ by a choice of normalization. The corresponding currents are
\begin{equation}
\label{eq:jjbar-def}
    \Jj =  \mathsf{j}^3_\sl +l_2 \, \mathsf{j}^3_\su + l_3 \mathsf{P}^t_{\LLL} +l_4 \mathsf{P}^y_{\LLL} \qqquad 
    \bar{\Jj} = \bar{\, \mathsf{j}}^3_\sl +r_2 \bar{\, \mathsf{j}}^3_\su + r_3 \mathsf{P}^t_{\RRR} +r_4 \mathsf{P}^y_{\RRR} \, ,
\end{equation}
where\footnote{Compared to \cite{Bufalini:2021ndn} we have implemented the change $\theta \mapsto \frac{\pi}{2} - \theta, \phi \leftrightarrow - \psi$. This effectively exchanges the sign of $r_2$. }
\begin{align}
\begin{aligned}
    & \, \mathsf{j}^3_\sl  = n_5 \big( \cosh^2\rhoo \, \del \tau + \sinh^2 \rhoo \,  \del \sigma  \big) \qqquad \bar{\, \mathsf{j}}^3_\sl = n_5 \big( \cosh^2\rhoo \,  \bdel \tau - \sinh^2 \rhoo \, \bdel \sigma  \big)\, , \\
    & \mathsf{j}^3_\su = n_5 \big(  \cos^2 \theta \, \del \psi - \sin^2 \theta\,  \del \phi \big) \qqquad \hspace{-1.4mm} \quad \bar{\, \mathsf{j}}^3_\su = - n_5 \big(   \cos^2 \theta \, \bdel \psi + \sin^2 \theta \, \bdel \phi \big) \, ,
\end{aligned}
\end{align}
and
\begin{align}
	\mathsf{P}^t_{L}= \del t \ , \qquad \mathsf{P}^t_{R} = \bdel t \ , \qquad
	 \mathsf{P}^y_{L}= \del y \ , \qquad \mathsf{P}^y_{R}= \bdel y \,.
\end{align}
For the currents in Eq.~\eqref{eq:jjbar-def} to be null, we impose the constraints
\begin{equation}
\label{eq: null gauge constraints}
 n_5 (1 - l_2^2) + l_3^2 - l_4^2 = 0 \qqquad 
 n_5 (1 - r_2^2) + r_3^2 - r_4^2 = 0 \,.
\end{equation}
Upon integrating out the gauge fields, the gauging procedure effectively adds a term quadratic in the currents, resulting in an action of the schematic form
\begin{align}
\label{eq: gWZW action}
& S_{\text{\tiny{WZW}}} +\frac{2}{\pi} \int \frac{\Jj \:\! \bar{\Jj}}{\Sigma} \, d^2 z   \, ,
\end{align}
with 
\begin{equation}
    \Sigma \;\equiv\; -\frac12\xi_1^i G_{ij} \xi_2^j   \;,
\label{eq:sigma-def}
\end{equation}
where $G_{ij}$ is the metric in Eq.~\eqref{eq: base ds2}.
One can then read off the resulting line element and  $B$-field of the gauged model. The change in the measure also generates a non-trivial dilaton, which can be obtained by solving for the vanishing of the appropriate worldsheet one-loop beta function, see \cite{Bufalini:2021ndn}.

The geometry obtained from the WZW model is free of horizons and closed timelike curves (CTCs) if and only if~\cite{Bufalini:2021ndn}
\begin{equation}
    l_3 \;=\; r_3 \,.
\end{equation}
To obtain smooth geometries up to orbifold singularities or NS5 sources, we further impose\footnote{There are also consistent models with $l_2=r_2=0$, which we do not consider in this work.}
\begin{equation}
\label{integersCFT1odd}
   l_2 = \m + \n \in 2\mathbb{Z}+1 
    \ , \qquad
    r_2 = \m - \n \in 2\mathbb{Z}+1 \,, \qquad \m,\n \in \mathbb{Z} \, ,
\end{equation}
and
\begin{equation}
\label{integersCFT2-b}
    l_4\,=\,-\Big(\k R_y - \frac{\pp}{R_y} \Big) \,,\qquad  r_4 \,=\, \k R_y +\frac{\pp}{R_y} \qqquad \k,\pp \in \ZZ \, .
\end{equation}
Combining these expression with the null constraints Eq.~\eqref{eq: null gauge constraints} leads to 
\begin{equation}
\label{L32def}
l_3 \;=\; r_3 \;=\; -\sqrt{ 
    \k^2 R_y^2 + \frac{\pp^2}{R_y^2} +   n_5\left(\m^2 + \n^2 - 1\right)}\, ,
\end{equation}
and 
\begin{equation}
\label{pkequalsn5mn}
    \k \, \pp = n_5\, \m \, \n \,,
\end{equation}
hence that only three of the four integers $\k,\m,\n,\pp$ are independent. 
The very same conditions are necessary and sufficient for the consistency of the spectrum of the worldsheet CFT \cite{Bufalini:2021ndn}. 
For the AdS$_3$ limit of these backgrounds to be dual to pure states of the holographic CFT, there is an additional requirement from momentum quantization in the $\k$-twisted sectors that we shall review in Section \ref{sec:heavyD1D5} (see e.g.~\cite{Chakrabarty:2015foa}),
\begin{equation}
\label{mnk-quant}
    \frac{\m \:\! \n}{\k} \,\in\, \mathbb{Z} \,.
\end{equation}
Without loss of generality, we work in the range of parameters $ \k\geq 0 , \, \m > \n \geq 0$.

\subsection{Supergravity configurations and AdS$_3$ limit}

The family of coset CFTs we have just defined corresponds precisely to the 
NS5-decoupled JMaRT configurations and their limits~\cite{Bufalini:2021ndn}. 
Using the integer parametrization introduced above, the supergravity fields are given by
\begin{align}
\label{FinalMetricandBfield}
     ds^2  \,  =&   \,\;  n_5(d\theta^2 + d\rho^2) + \frac{1}{\Sigma_0}\Bigg[  - \left(\sinh^2\rhoo + (\m^2 - \n^2) \cos^2 \theta + 1 - \m^2 - \frac{\pp^2}{n_5 R_y^2}  \right)   dt^2 \nn  \\
    & + \left(\sinh^2\rhoo + (\m^2 - \n^2) \cos^2 \theta + \n^2  + \frac{\pp^2}{n_5 R_y^2}  \right)   dy^2 - 2 \:\! \frac{\pp }{n_5 R_y} \;\! \Delta\;\! dt dy  \nonumber\\[1.5mm]
    & + \left( n_5 \sinh^2 \rhoo + n_5 \m^2 + \k^2 R_y^2 \right)
    \:\! \sin^2 \theta \;\! d\phi^2 + \left( n_5 \sinh^2 \rhoo + n_5 \n^2 + \k^2 R_y^2 \right) \;\! \cos^2 \theta \:\! d\psi^2  \nonumber \\[1.5mm]
    & + 2  \left( \m \:\! \Delta \:\! dt  - \Big(  \m \frac{\pp}{R_y} + \n \:\! \k R_y \Big)  dy \right)  \sin^2 \theta \:\! d\phi - 2 \left( \n  \:\! \Delta \:\! dt  - \Big( \n  \frac{\pp}{R_y} + \m \:\! \k R_y \Big)  dy \right)  \cos^2 \theta \;\! d\psi \Bigg] , \nn \\[3mm]
 B  \:  =&   \,\; \frac{1}{\Sigma_0} \Bigg[  - \frac{\k R_y}{n_5} \;\! \Delta \:\! dt \wedge dy +  \left( n_5 \sinh^2\rhoo + n_5 \,  \m^2 + \k^2 R_y^2  \right) \, \cos^2 \!\theta \,  d\phi \wedge d\psi  
 \nonumber\\
 & + \left( \m \:\!\Delta \:\! dt  - \Big(  \m \frac{\pp}{R_y} + \n \:\! \k R_y \Big) dy \right) \wedge  \cos^2 \theta \:\! d\psi
 -\left( \n \:\! \Delta \:\! dt - \Big(\n  \frac{\pp}{R_y} + \m \:\! \k R_y\Big)  dy \right) \wedge  \sin^2\theta  \:\! d\phi  \Bigg] \:\!  , 
\end{align}
where $R_y$ is the asymptotic proper radius of the $\SSS^1_y$ circle, and where
\begin{align}
    \Sigma_0& = 
    \sinh^2\rho + (\m^2-\n^2) \cos^2\theta +  \n^2 + \frac{\k^2 R_y^2}{n_5}  \, ,  
    \label{Sigma0def} \\[1ex]
    \Delta & = \sqrt{n_5(\m^2+\n^2-1)+ \k^2 R_y^2 + \frac{\pp^2}{R_y^2}} \, .
    \label{Deltadef} 
\end{align}
We also note the relations between the supergravity charges and integer charge quanta,
\begin{equation}
\label{eq:q1-qp}
     Q_1  = n_1 \frac{g_s^2 }{V_4} \qqquad  Q_p = \frac{n_p}{R_y^2} \frac{g_s^2 }{V_4} \;,
\end{equation}
where $(2\pi)^4 V_4$ is the coordinate volume of the $\TT^4$.
Note that the three-charge NS5-decoupled JMaRT solutions are specified by the integers $n_5,\k,\m,\n$, the modulus $R_y$, and the charge $Q_1$ appearing in the dilaton.

One can take a further IR AdS$_3$ decoupling limit by taking $R_y$ to be large, keeping fixed the charge $Q_1$ and the rescaled energy $E R_y$ and momentum  $P_y R_y$. We define rescaled coordinates
\begin{equation}
    \tilde{t} = \frac{t}{R_y} \qqquad
    \tilde{y} = \frac{y}{R_y}\, .
    \label{FixedAdSlimit}
\end{equation} 
and perform the large-$R_y$ expansion at the level of the coefficients in Eqs.~\eqref{integersCFT1odd}--\eqref{pkequalsn5mn}, such that the leading terms in $l_3,r_3,l_4$ and $r_4$ become independent of $\pp$. However, the product $\k \pp $ is kept fixed and the relation Eq.~\eqref{pkequalsn5mn} still holds, defining the momentum per strand for the holographic CFT \cite{Chakrabarty:2015foa}. 
Order-by-order in $1/R_y$, the coefficients still satisfy the null conditions Eq.~\eqref{eq: null gauge constraints}. 
The six-dimensional fields in \eqref{FinalMetricandBfield} then become
\begin{align}
 ds^2 & = 
	n_5 \bigg[ -\frac{1}{\k^2}\cosh^2 \rho \,  d\tilde{t}\;\!^2 +\frac{1}{\k^2}  \sinh^2 \rho \, d\tilde{y}^2  +  d\rho^2 
	 \label{eq:ads-JMaRT-met} \\
	&\qquad~ + d\theta^2+ \sin^2 \theta \left( d\phi -\frac{\n}{\k} d\tilde{t}
	+\frac{\m}{\k} d\tilde{y}
	 \right)^2 
	+ \cos^2 \theta \left( d\psi
	+\frac{\m}{\k} d\tilde{t}
	-\frac{\n}{\k} d\tilde{y}
	 \right)^2 \bigg] , \nn  \\
	 B & =  n_5 \Bigg[ \frac{\sinh^2\rhoo + (\m^2 - \n^2)\cos^2\theta}{\k^2} \, d\tilde{t}  \wedge  d\tilde{y} +  \, \cos^2\theta \, d\phi \wedge d\psi   \label{eq:ads-JMaRT-B}  \\
	  & \hspace{1.5cm} +  \sin^2 \theta \left(  - \frac{\n}{\k} \, d \tilde{t}  + \frac{\m}{\k} \, d \tilde{y} \right) \wedge d\phi +  \cos^2 \theta \left( \frac{\m}{\k} \, d \tilde{t}  - \frac{\n}{\k} \, d \tilde{y} \right) \wedge d\psi \Bigg]  \, , \nn \\
	 e^{2\Phi} & =  \frac{n_5}{Q_1} = \frac{Q_5}{Q_1} \, , 
	 \label{eq:ads-JMaRT-dil}
\end{align}
where a trivial gauge transformation has been performed on the $B$-field. 

These solutions are related by a large coordinate transformation to $\ZZ_\k$ orbifolds of global AdS$_3\times \SSS^3$, which are the decoupling limits of the supertube solutions of~\cite{Balasubramanian:2000rt,Maldacena:2000dr}. This large coordinate transformation is known as spacetime spectral flow, and takes the form 
\be
\label{LGT}
\tilde\psi \;=\;
\psi
	+\frac{\m}{\k} \,\tilde{t}
	-\frac{\n}{\k} \, \tilde{y}  \qqquad
\tilde\phi \;=\;
\phi 
	-\frac{\n}{\k} \, \tilde{t}
	+\frac{\m}{\k} \, \tilde{y}\,.
\ee
For the special case $\m=\n=0$, spectral flow is not relevant and the solutions are already $\ZZ_\k$ orbifolds of global AdS$_3\times \SSS^3$. 
When $\m,\n$ are not both zero, one typically works in the range $\m>\n\ge 0$ without loss of generality.
For $\m=1,\n=0$, the solutions  \eqref{eq:ads-JMaRT-met}--\eqref{eq:ads-JMaRT-dil}, are the AdS decoupling limits of the two-charge circular supertube solutions of~\cite{Balasubramanian:2000rt,Maldacena:2000dr}.
For $\m=\n+1$ with $\n>0$, the solutions the AdS limit of the supersymmetric spectral flowed solutions of~\cite{Lunin:2004uu,Giusto:2004id,Giusto:2004ip,Giusto:2012yz}, and for other values of $\m,\n$ one obtains the AdS limit of the non-supersymmetric JMaRT solutions~\cite{Jejjala:2005yu}.
For $\k=1$ the solutions are smooth; for $\k>1$ the solutions have orbifold singularities near $\tilde{y}=0$, the details of which depend on the common divisors of $\m,\n,\k$~\cite{Jejjala:2005yu,Chakrabarty:2015foa,Martinec:2018nco}.

\subsection{Holographic description and boundary spectral flow}
\label{sec:heavyD1D5}

As mentioned in the Introduction, the holographic CFT that corresponds to the AdS$_3$ limit of the system in which we work is an $\cN=(4,4)$ symmetric product orbifold CFT with target space $\left(\TT^4\right)^{N}/S_{N}$, where $N=n_1 n_5$. To make the presentation self-contained, we now review some aspects of this theory.

Recall that we work in Type IIB compactified on $\SSS^1\times\TT^4$, with $n_5$ NS5 branes wrapped on $\SSS^1\times\TT^4$, $n_1$ units of F1 winding on $\SSS^1$, and $n_P$ units of momentum charge along $\SSS^1$.  
The moduli space is $ 20 $-dimensional and the symmetric product orbifold CFT lies at a particular locus of this moduli space~\cite{Larsen:1999uk}, see also~\cite{Bena:2017xbt}.
The configuration
breaks the $ \text{SO}(1,9) $ Lorentz group to $ \text{SO}(1,1) \times \text{SO}(4)_{E} \times \text{U}(1)^4 $, where the external 
R-symmetry
$\text{SO}(4)_{E} \simeq \text{SU}(2)_L \times \text{SU}(2)_R$ corresponds to rotations in the spatial $ \RR^{4} $ transverse to the branes (in the IR limit, rotations of the $\SSS^3$).
It is customary to introduce an approximate internal $ \text{SO}(4)_I \simeq \text{SU}(2)_1 \times \text{SU}(2)_2$, which is broken to $ \text{U}(1)^4 $ by the compactification, but which is useful for classifying states and organizing fields~\cite{David:2002wn,Avery:2010qw}. 

In the symmetric product orbifold theory, for each copy of $ \TT^4 $ there are four free bosons, together with their left and right-moving fermionic superpartners. 
Indices $\alpha,\dot\alpha,A,\dot{A}$ correspond respectively to $\text{SU}(2)_L \:\! , \text{SU}(2)_R \:\! , \text{SU}(2)_1 \:\! , \text{SU}(2)_2  $.
The free fields are denoted as (we use the conventions of \cite{Avery:2009tu})
\begin{equation}
	X_{A\dot{A}\, (r)}(z,\bar{z}) \qqquad \psidr^{\alpha \dot{A}}_{(r)}(z) \qqquad \bar{\psidr}^{\dot{\alpha} \dot{A}}_{(r)}(\bar{z})  \, ,
\end{equation}
where the subscript $(r)$ denotes the $r$-th copy of the seed  $ \TT^4 $ theory. Omitting this copy subscript and focusing on the holomorphic sector, the energy-momentum tensor $\TT(z)$, the supercurrents $\Gd^{\alpha A}(z)$ and the SU$(2)_L$ currents $\Jd^a$ generate the small (4,4) supersymmetric algebra. We denote holographic CFT conformal weights by $h$ and R-symmetry quantum numbers by $ (\jd,m') $ and $ (\bar{\jd}, \bar{m}') $, respectively. 

The heavy states we are interested in are obtained by fractional spectral flow~\cite{Chakrabarty:2015foa}, see also~\cite{Martinec:2001cf,Giusto:2012yz}. We start from the  NSNS vacuum in the $\k$-twisted sector, $\ket{0_{\k}}_{\text{NS}}$. 
In order to have a gauge invariant state we consider $n_1n_5/\k$ identical strands of length $\k$. Its dimension is 
\begin{equation}
    h  = \bar{h}   = \frac{c}{24} \left[  1 - \frac{1}{\k^2} \right],
\end{equation}
where the central charge $c=6N$. The R-charges of this state are zero. Because all strands are of length $\k$ there is an enhancement of the usual spectral flow, such that one can perform spectral flow with fractional parameters,
\begin{equation}
\label{SFchargesHCFT}
    \alpha = \frac{\m+\n}{\k} = \frac{2s+1}{\k} \qqquad
    \bar{\alpha} = \frac{\m-\n}{\k} = \frac{2 \bar{s}+1}{\k} \, , 
\end{equation}
where $s,\bar{s} \in \ZZ$ and the range $\m>\n\ge 0$ is the range $s\ge \bar{s} \ge 0$. This generates a new state with quantum numbers
\begin{align}
    h &  = \frac{c}{24} \left[  1 - \frac{1}{\k^2} + \alpha^2 \right] \qqquad m' = \frac{\alpha  c}{12}  \, , \nn\\
     \bar{h} & = \frac{c}{24} \left[  1 - \frac{1}{\k^2} + \bar{\alpha}^2 \right] \qqquad \bar{m}' = \frac{\bar{\alpha}  c}{12}  \, .
\end{align}
These states are ``heavy'' in the sense that their conformal dimensions and charges scale linearly with the large central charge $c=6N$. In the dual theory they correspond to the classical configurations presented in Eqs.\;\eqref{eq:ads-JMaRT-met}--\eqref{eq:ads-JMaRT-dil}.
By constrast, the ``light'' perturbative string states probing these backgrounds will correspond to holographic CFT states with conformal dimensions that are independent of $c$.

We conclude this section by summarizing how the bulk, the worldsheet, and the dual CFT encode in different ways the same information about the heavy state. 

\begin{itemize}
    \item In the worldsheet model, the heavy state defines the theory itself by means of the gauging parameters $l_i, r_i$ appearing in Eq.~\eqref{eq:jjbar-def} and the radius $R_y$.
    
    \item In supergravity, the information about the heavy state is contained in the integers $\m,\n,\k$ parameterizing the fields in Eq.~\eqref{FinalMetricandBfield}, together with $R_y$, which gets scaled out in the AdS limit.
    
    \item In the symmetric orbifold CFT, the information about the heavy states is contained in the spectral flow parameters $\alpha$ and $\bar{\alpha}$, and the twist index $\k$.
\end{itemize} 
The map between the three descriptions, in the AdS limit, is then 
\begin{equation}
    - R_y \, \frac{l_2}{l_4} ~ \xrightarrow{R_y \to \infty} ~ \frac{\m + \n}{\k} \, = \, \alpha \qqquad
    R_y \, \frac{r_2}{r_4} ~ \xrightarrow{R_y \to \infty} ~ \frac{\m - \n}{\k} \, = \, \bar{\alpha} \; .
\end{equation}

\section{The light probe states}
\label{sec:light states}

We now introduce the light states that we will study. These correspond to chiral primary operators of the boundary theory. 
We focus first on fluctuations around the global AdS$_3 \times \SSS^3$ vacuum. We review the dictionary between holographic CFT operators and their counterparts in the worldsheet theory, following~\cite{Dabholkar:2007ey,Giribet:2007wp}.

\subsection{Chiral primaries in the D1D5 CFT}
\label{secCPD1D5CFT}
We first briefly review the construction of chiral primary operators in the symmetric orbifold CFT \cite{Lunin:2001pw}. We focus primarily on the holomorphic sector in the following; the antiholomorphic sector is entirely analogous. 
In the untwisted sector, on each copy of the seed $\TT^4$ theory, the chiral primary operators correspond to the states (suppressing the copy $(r)$ label) 
\begin{equation}
    |0_{\NS}\rangle \qqquad
    \psi_{-\half}^{+\dot{A}}|0_{\NS} \rangle \qqquad
    \Jd_{-1}^{+}|0_{\NS}\rangle \,=\, \psi_{-\half}^{+\dot{1}}\psi_{-\half}^{+\dot{2}}|0_{\NS}\rangle \, ,
\label{CPO1}
\end{equation}
where $|0_{\NS}\rangle$ is the NS vacuum. The corresponding weights and R-charges are  $h=m'=0,\half,1$, respectively. Physical configurations in the orbifold theory are obtained by symmetrizing the states in \eqref{CPO1} over the different copies of the seed theory. 

By including the antiholomorphic sector we can obtain, for instance, the dimension $(\half,\half)$ operator 
(see e.g.~\cite{Moscato:2017usq})
\begin{equation}
\label{O++ operator}
    O^{++} = \sum_{r=1}^{N} O^{++}_{(r)} = \frac{-i}{\sqrt{2}} \sum_{r=1}^{N} \psidr^{+\dot{A}}_{(r)} \vep_{\dot{A}\dot{B}} \bar{\psidr}^{+\dot{B}}_{(r)} \qqquad (O^{++})^\dagger = O^{--}\, .
\end{equation}
We will use this operator in an explicit example later in the paper.

In order to construct more general chiral primaries one needs to consider the twisted sectors of the theory. Consider the `bare' twist operators $\sigma_n$, defined on the cylinder, that impose the following boundary conditions corresponding to a single-cycle permutation,
\begin{align}
\begin{aligned}
\label{eq:n-bcs}
     X_{(1)} \to  X_{(2)} \to \cdots \to X_{(n)} \to  X_{(1)} \, , \\  
 \psi_{(1)} \to  \psi_{(2)} \to \cdots \to \psi_{(n)} \to  -\psi_{(1)} \, ,  
\end{aligned}   
\end{align}
and likewise for the antiholomorphic fermions.
The bare twist operators are defined to be the lowest-dimension twist operators that impose the above boundary conditions; they have dimension $h = \bar{h} = \frac{1}{4}\left(n-\frac{1}{n}\right)$ and zero R-charge. Chiral operators are obtained by exciting the bare twist operators operators to add R-charge. The lowest-dimension chiral operators have $h=m'=\frac{n-1}{2}$.  For $n$ odd, these operators are obtained by acting with modes of the SU(2) currents, which are bilinears in the free fermions. Due to the twist operator, the SU(2) currents are fractional-moded in units of $1/n$. The relation between these modes and those of free fermions on the $n$ copies of the seed theory can be found in~\cite{Lunin:2001pw}.
To construct the chiral operators, one acts with the currents  $\Jd^+_{-l/n}$ for which $l$ is odd and $l<n$, 
\begin{equation}
n~\mathrm{odd}: \qquad\quad    \sigma_n^{-} \,=\, 
\prod_{p=1}^{(n-1)/2}
    \Jd^+_{-\frac{2p-1}{n}}  {\sigma}_n
    ~=~
    \Jd^+_{-\frac{n-2}{n}}  \cdots \Jd^+_{-\frac{3}{n}} \Jd^+_{-\frac{1}{n}}  {\sigma}_n \,.
    \label{chiral0-n-odd}
\end{equation}
For $n$ even, one first acts with a spin field $S^+_n$, which has weight $\frac{1}{4n}$ and charge $\half$, putting the fermions into the Ramond sector (i.e.~their boundary conditions are similar to Eq.\;\eqref{eq:n-bcs} but with the final sign being $+\psi_{(1)}$). One then acts with the currents $\Jd^+_{-l/n}$ for which $l$ is even and $l<n$,
\begin{equation}
n~\mathrm{even}: \qquad\quad    \sigma_n^{-} \,=\, 
\prod_{p=1}^{(n-2)/2}
    \Jd^+_{-\frac{2p}{n}} S^+_n {\sigma}_n
    ~=~
    \Jd^+_{-\frac{n-2}{n}}  \cdots \Jd^+_{-\frac{4}{n}} \Jd^+_{-\frac{2}{n}}  S^+_n{\sigma}_n \,.
    \label{chiral0-n-even}
\end{equation}
As in the untwisted case, for both odd and even $n$ we can act with $\psi_{-\half}^{+\dot{A}}\equiv \sum\limits_{r=1}^{n}\psi_{-\half(r)}^{+\dot{A}}$ to obtain a chiral operator 
$\psi_{-\half}^{+\dot{A}}\sigma_n^-$ 
which has $h=m'=\frac{n}{2}$. Similarly we can act with $\Jd_{-1}^+$ to obtain a chiral operator 
$\Jd^+_{-1}\sigma_n^-$ 
which has $h=m'=\frac{n+1}{2}$.
Together with the analogous antiholomorphic operators, this exhausts the single-cycle chiral operators. Indeed, by making use of anti-commutators of the supercurrent modes $\Gd_{-m/n}^{\pm A}$ in the corresponding twisted sectors, one can show that chiral weights are bounded by  \cite{Avery:2010qw}
\begin{equation}\label{eq:cp-bound}
 \frac{n-1}{2} \,\leq\, h \,\leq\, \frac{n+1}{2} \, .
\end{equation}

In the twisted sectors, it is often convenient to work in a basis that diagonalizes the twisted boundary conditions. We shall make use of this basis in Section \ref{sec:HLLLH}. One defines
\begin{equation}
    \tilde{\psi}_{\rho}^{\alpha\dot{A}} \,=\,\frac{1}{\sqrt{n}} 
    \sum_{r=1}^n e^{ \alpha \frac{2\pi i  r \rho}{n}}
    \psi_{(r)}^{\alpha\dot{A}} \qqquad
    \rho = 0 , \dots, n-1 \, ,
    \label{rhofermions}
\end{equation}
where $\alpha=\pm$ should not be confused with the spectral flow parameter in Eq.\;\eqref{SFchargesHCFT}. These are mutually orthogonal, and diagonalize the twisted boundary conditions as 
\beq
\tilde{\psi}_{\rho}^{\alpha\dot{A}} (e^{2\pi i} z) \,=\, 
     e^{-\alpha \frac{2\pi i \rho}{n}} \tilde{\psi}_{\rho}^{\alpha\dot{A}} ( z) \, .
\end{equation}
These fermions can be bosonized to construct an explicit expression for the spin fields mentioned above. Note that the fields $\tilde{\psi}_{\rho=0}^{\alpha \dot{A}}$ are invariant under the twisting. For further discussion, see~\cite{Dabholkar:2007ey}.   

We now combine the above holomorphic construction with its antiholomorphic counterpart and define the complete list of scalar left-right chiral primaries we will be interested in:
\begin{equation}
    O_n^{--} = \sigma_n^{--}  \qqquad
    O_n^{\dot{A}\dot{B}} = 
    \tilde{\psi}_{\!\!\;\rho=0}^{+\dot{A}}\;\! \bar{\tilde{\psi}}_{\!\!\;\rho=0}^{+\dot{B}}\;\!
    \sigma_n^{--} \qqquad
    O_n^{++} = 
    \tilde{\psi}_{\!\!\;\rho=0}^{+\dot{1}}\;\!\tilde{\psi}_{\rho=0}^{+\dot{2}}\;\!
    \bar{\tilde{\psi}}_{\!\!\;\rho=0}^{+\dot{1}}\;\!
    \bar{\tilde{\psi}}_{\!\!\;\rho=0}^{+\dot{2}}\;\!
    \sigma_n^{--} \, ,
    \label{ccops}
\end{equation}
where $\sigma_n^{--}$ is defined similarly to Eqs.~\eqref{chiral0-n-odd}, \eqref{chiral0-n-even} but now also with the same construction in the antiholomorphic sector. The operators in \eqref{ccops} are normalized such that they have unit two-point functions. 

For later reference, we note that in each case the respective weights and twist numbers can be written in terms of $j=\frac{n+1}{2}$ as
\begin{equation}
    h \left[O_n^{--} \right] = j-1  \qqquad
    h \left[O_n^{\dot{A}\dot{B}} \right] = j - \frac{1}{2} \qqquad
    h \left[O_n^{++} \right] = j  \, .
        \label{D1D5CFTweights3}
\end{equation}
An analogous list of anti-chiral primaries (which have $h=-m'$) is obtained by acting on the bare twist fields with current and fermion modes with opposite charge, i.e. $\Jd_{-l/n}^-$ and $\psi^{-\dot{A}}$. As we will shortly review, and up to a shift related to spectral flow, this $j$ will be identified with the principal quantum number of the bosonic (global) SL(2,$\R$) algebra of the worldsheet theory, to which we now turn.

\subsection{Superstring theory on AdS$_3\times \SSS^3\times \TT^4$}
\label{sec:AdS3S3T4}

We now review the basics of superstring theory on AdS$_3 \times \SSS^3 \times \TT^4$ using the RNS formalism with BRST quantization. We first discuss the bosonic SL($2,\RR$) and SU$(2)$ WZW models and then present their supersymmetric counterparts.
We present the current algebra and review the spectrum, including states arising from worldsheet spectral flow.

\subsubsection{Bosonic WZW model for SL(2,$\R$)}
\label{sec:sl2}

The SL(2,$\R$) WZW model was studied in detail in \cite{Maldacena:2000hw,Maldacena:2000kv,Maldacena:2001km}. In what follows we will mostly follow the notation of~\cite{Dabholkar:2007ey,Giribet:2007wp,Dei:2021xgh}, and normal ordering will be implicitly assumed. The holomorphic SL(2,$\R$) currents will be denoted $j^a(z)$. They satisfy the OPEs
\begin{equation}
    j^a(z)j^b(w) \,\sim\,\frac{k}{2} \frac{\eta^{ab}}{(z-w)^2} + \frac{ f^{ab}_{\phantom{ab}c}\;\! j^c(w)}{z-w}  \, ,
    \label{OPEjSL2}
\end{equation}
where $k$ is the level of the affine algebra, and where 
\begin{equation}
-2\eta^{33} = \eta^{+-} = 2 \qqquad
f^{+-}_{\phantom{+-}3}=-2 \, , \quad
    f^{3+}_{\phantom{3+}+}=-
    f^{3-}_{\phantom{3-}-}=1 \, .
\end{equation} 
The holomorphic stress tensor and the  central charge follow from the Sugawara construction, and are given by (likewise for the antiholomorphic sector)
\begin{equation}
   T_{\mathrm{sl}}(z) = \frac{1}{k-2} \left[-j^3 (z) j^3(z) + \frac{1}{2} \, j^+(z) j^-(z) + \half \, j^-(z) j^+(z) \right] \, , 
    \qquad c_{\mathrm{sl}} = \frac{3k}{k-2} \,.
\end{equation}

We denote bosonic SL$(2,\R)$ primary vertex operators by $V_{j,m, \bar{m}}(z,\bar{z})$. Their zero-mode wavefunctions do not factorize between holomorphic and antiholomorphic sectors, however as is often done we shall work primarily with the holomorphic sector, and suppress the $\mb$ and $\bar{z}$ dependence.
The relevant representations of the holomorphic zero-mode algebra are as follows.
The principal series discrete representations of lowest (highest) weight are spanned by
\begin{equation}
     {\cal{D}}_j^{\pm} = 
        \{
        |j,m\rangle \ , \ m= \pm j
        ,\pm j \pm 1, \pm j \pm 2,\cdots
        \} \, ,
\end{equation}
respectively, where $j_0^3|j,m\rangle = m|j,m\rangle$.
These are unitary representations for any positive real $j$, and one is the charge conjugate of the other (we will restrict the range of $j$ momentarily). There are also the principal continuous series representations, spanned by 
\begin{equation}
    {\cal{C}}_j^{\hat{\alpha}} = 
        \{
        |j, \hat{\alpha}, m\rangle \ , \ 0 \leq \hat{\alpha} < 1 \ , \ j=\frac{1}{2} + i s \ , \ s \in \R \ , \ m=\hat{\alpha}
        ,\hat{\alpha}\pm 1,\hat{\alpha} \pm 2,\cdots
        \} \, .
\end{equation}
The particular case $\hat{\alpha} = 1/2 = j$ is actually reducible. It was shown in \cite{Maldacena:2000hw} that the spectrum of the model is built out of  continuous and lowest weight representations with 
\begin{equation}
        \frac{1}{2} < j < \frac{k-1}{2} \, ,
        \label{Djrange}
    \end{equation}
together with their spectrally flowed images, to be introduced below. The allowed range \eqref{Djrange} follows from $L^2$ normalization conditions, no-ghost theorems and spectral flow considerations.

Before considering worldsheet spectral flow (we refer to this as the ``unflowed'' sector), the action of the currents on the primary states is given by
\begin{subequations}
\label{zeromodesSL2}
\begin{eqnarray}
    j^3_0 |j,m \rangle &=& m |j,m \rangle\;, \\ 
    j^\pm_0 |j,m \rangle &=& 
    \begin{cases}
			\,(m \mp (j-1)) |j,m \pm 1 \rangle & \text{if}~~ m\neq\mp j\\
            ~0 & \text{if}~~ m=\mp j \;,
		 \end{cases}\\
    j^{a}_{n} |j,m \rangle &=&  0 \ \qquad \forall n>0\,.
\end{eqnarray}
\end{subequations}
These vertex operators 
can be obtained from those of the Euclidean counterpart of the model, namely the $H_3^+$ WZW model~\cite{Teschner:1997ft,Teschner:1997fv,Teschner:1999ug} (see also~\cite{McElgin:2015eho}), as follows.
One introduces a set of operators depending on a complex label $x$, written as $V_{j}(x|z)$, and having conformal weight
\begin{equation}
    \Delta \,=\, -\frac{j(j-1)}{k-2} \, . 
\end{equation}
The action of the currents on $V_{j}(x,z)$ is given by
\begin{equation}
    j^{a}(z) V_{j}(x,w) \,\sim \,
     \frac{D_j^a V_{j}(x,w)}{(z-w)} \, ,
\end{equation}
where 
\begin{equation}
    D_j^+ = \der_x \qqquad  D_j^3 = x \der_x + j \qqquad
    D_j^- = x^2 \der_x + 2 j x \, .
    \label{diffSL2}
\end{equation}
The two-point function is given by~\cite{Teschner:1999ug}
\begin{equation}
    \langle V_{j_1}(x_1,z_1) V_{j_2}(x_2,z_2)\rangle \,=\, \frac{1}{|z_{12}|^{4\Delta_1}} \left[
    \delta^2(x_1 - x_2) \delta(j_1+j_2-1) +
    \frac{B(j_1)}{|x_{12}|^{4j_1} }\delta (j_1 - j_2)
    \right],
    \label{2ptXbasisSL2}
\end{equation}
with 
\begin{equation}
    B(j)=\frac{2j-1}{\pi}
    \frac{\Gamma[1-b^2(2j-1)]}{\Gamma[1+b^2(2j-1)]} \, \nu^{1-2j} \qqquad 
    \nu = \frac{\Gamma[1-b^2]}{\Gamma[1+b^2]}
     \ , \  b^2 = (k-2)^{-1}
     \label{defBj} \, .
\end{equation} 
The operators $V_{j,m}(z)$ are related to $V_{j}(x|z)$ by means of the following Mellin-like transform: 
\begin{equation}
 \label{MellinSL2}
    V_{j,m}(z) = \int_{\mathbb{C}} d^2x \, x^{j-m-1}\xb^{j-\mb-1} V_{j}(x,z) \, .
\end{equation}
In the Euclidean $H_3^+$ model, $j$ takes values $j=1/2+is$. To obtain the unflowed $V_{j,m}$ for Lorentzian AdS$_3$, one assumes a well-defined analytically continuation to real values of $j$. This procedure was discussed in \cite{Maldacena:2001km}, which identified the physical origin of the different divergences arising in correlation functions. For related work, see~\cite{Harlow:2011ny}. 
The two-point functions in the $m$-basis then follow from~\eqref{2ptXbasisSL2}, \eqref{MellinSL2}.
Using the shorthand $V_i\equiv V_{j_i , m_i}$,  one finds
\begin{equation}
    \langle V_1 V_2\rangle \,=\, 
    \frac{\delta^2(m_1+m_2)}{|z_{12}|^{4\Delta_1}}\left[\delta(j_1+j_2-1)+
     \delta(j_1-j_2) 
    \frac{\pi B(j_1)  }{\gamma(2j_1)}
    \frac{\gamma(j_1+m_1)}{\gamma(1-j_1+m_1)}\right]
    \label{2ptMbasisSL2},
\end{equation}
where $\gamma(x) = \Gamma(x)/\Gamma(1-\xb)$, and where $\delta^2(m)$ is a Dirac delta in $m+\mb$ times a Kroenecker delta in $m-\mb$.

At first sight, the complex variable $x$ may appear simply as an SL(2,$\R$) version of the isospin variables  defined for SU(2) in \cite{Zamolodchikov:1986bd}. However, given that the integrated zero modes of the currents realize the spacetime Virasoro modes $L_0$ and $L_{\pm 1}$, and by examining the expressions of the associated differential operators \eqref{diffSL2}, one is led to interpret $x$ as the local coordinate on the boundary theory~\cite{Giveon:1998ns}. According to \eqref{2ptXbasisSL2}, in the bosonic theory a $z$-integrated vertex operator $V_{j}(x)$ is identified with a local  operator on the boundary theory with weight $j$. Conversely, the corresponding boundary modes are given by the $m$-basis operators.  Indeed, for states in the discrete sector, the transform in Eq.~\eqref{MellinSL2} can be inverted, giving
\begin{equation}
    V_{j}(x,z) \;= \sum_{m=j+n, \, n\in \mathbb{N}_0}  x^{m-j} \:\!
    \bar{x}^{\bar{m}-j} \, V_{j,m \bar{m}}(z) \, .
    \label{ExpSL2}
\end{equation}
The vertex $V_j(x,z)$ is realized via Eq.~\eqref{ExpSL2} as $V_{j,j}(z)$ translated from the origin to $x$. Poles in the integrand of \eqref{MellinSL2} coming from the expansion around  $x=0$ ($x=\infty$) are associated to states in the ${\cal D}_j^+$ (${\cal D}_j^-$) representations~\cite{Eberhardt:2019ywk,Dei:2021xgh}.

Spectral flow automorphisms of the current algebra \eqref{OPEjSL2} are defined as
\begin{equation}
       j^\pm(z) \to \tilde{j}^\pm(z) = z^{\pm w} j^\pm(z) \qqquad 
    j^3(z) \to \tilde{j}^3(z) = j^3(z) - \frac{k\,\w}{2} \frac{1}{z} \, ,
    \label{sl2bosflow2}
\end{equation}
where the so-called spectral flow charge $\w$ is an integer. 
Analogous formulas hold for the antiholomorphic sector. We work with the universal cover of SL$(2,\R)$, which imposes that the holomorphic and antiholomorphic spectral flow parameters must be equal, $\bar{\w}= \w$.
The action of \eqref{sl2bosflow2} on the above representations defines in general inequivalent representations that must be considered in order to generate a consistent spectrum. This holds up to the so-called series identifications due to the fact that the affine modules $\hat{{\cal{D}}}_j^{+,w}$ and $\hat{{\cal{D}}}_{k/2-j}^{-,w+1}$ are isomorphic. Thus, as mentioned above, the discrete series spectrum is constructed solely upon lowest weight representations with $j$ restricted to the range \eqref{Djrange}.

At the level of vertex operators and for $w>0$, the spectral flow operation introduced in \eqref{sl2bosflow2} defines the so-called flowed primaries, whose OPEs with the currents take the form  
\begin{subequations}
\label{flowedOPESL2}
\begin{eqnarray}
    j^{+}(z)V_{j,m }^{\w}(w) &=& \frac{(m+1-j)V_{j,m+1 }^{\w}(w)}{(z-w)^{\w+1}} + 
    \sum_{n=1}^\w 
    \frac{(j^+_{n-1} V_{j,m }^\w)(w)}{(z-w)^n} + \dots \, , \\ [1ex]
    j^{3}(z)V_{j,m }^{\w}(w) &=& 
    \frac{\left(m+ \frac{k}{2}\w \right)V_{j,m }^{\w}(w)}{(z-w)}
    + \dots \, , \\ [1ex]
    j^{-}(z)V_{j,m }^{\w}(w) &=& (z-w)^{\w-1} (m-1+j) V_{j,m-1 }^{\w}(w) + \dots \, , 
\end{eqnarray}
\end{subequations}
where the ellipses indicate higher-order terms. Similar expressions hold for $\w<0$, with the roles of $j^+$ and $j^-$ inverted.
The operators  $V_{j,m}^{\w}(z)$ are not affine primaries. They are, however, Virasoro primaries, with worldsheet conformal weight
\begin{equation}
    \hat{\Delta} \,=\, -\frac{j(j-1)}{k-2} - m \w - \frac{k}{4}\w^2 \, . 
    \label{defDeltaw}
\end{equation}
Note that for $\w>0$ ($\w< 0$), independently of  the characteristics of the original state, these correspond to lowest (highest) weight states, with SL(2,$\R$) spin 
\begin{equation}
\label{h-sf-2}
h= m +  \frac{k}{2} \w  \, ,
\end{equation}
($h=- m - k \w/2$, respectively).
The notation $h$ anticipates that the SL(2,$\R$) spin is identified with the holographic CFT conformal weight~\cite{Kutasov:1999xu} (see also e.g.~\cite{Dabholkar:2007ey}), as we shall see in Eq.~\eqref{2ptwSL2}.

The flowed affine modules alluded above are built by acting freely with the currents on flowed primary states. In particular, the remaining states in the zero-mode algebra, which are obtained by acting with $j_0^-$, are \textit{not} flowed primaries.
Nevertheless, one can proceed as done for the unflowed states, and combine them into a local operator, defined initially for $\w>0$ as 
\begin{equation}
    V^w_{j,h} (x,z)  
    \,=\, \sum_{ n\in \mathbb{N}_0}  x^{n}
    \bar{x}^{\bar{n}} \, V_{j,h+n,h+\bar{n}}^w(z) \,.
    \label{flowedxbasis}
\end{equation}
Moreover, by inverting $x \to 1/x$ in the expansion, one also obtains the states in the highest-weight representation with the same spin and opposite $\w$ and $m$. This shows that the resulting $x$-basis states are actually defined in terms of the absolute value of $\w$, its sign being irrelevant. A direct $x$-basis definition for spectrally flowed vertex operators was recently derived in \cite{Iguri:2022eat}, extending the original proposal of \cite{Maldacena:2001km} valid only for the singly flowed case.

The classical analog of the spectral flow operation \eqref{sl2bosflow2} maps space-like geodesics of point-like strings into solutions in which a long string wound around the AdS$_3$ angular direction at large radius comes in to the centre of global AdS$_3$, collapses to a point, and then re-expands to large radial distance~\cite{Maldacena:2000hw}. The spectral flow parameter $\w$ is thus sometimes referred to in the literature as a ``winding'' number. Note that since the AdS$_3$ angular direction is contractible in the interior of global AdS$_3$, the parameter $\w$ is not a conserved quantity. However, the $m$-basis two-point functions are diagonal in $\w$: 
it was shown in~\cite{Maldacena:2001km} that the $m$-basis two-point function of flowed primaries is as in \eqref{2ptMbasisSL2} with an extra factor of $\delta_{\w_1,-\w_2}$ and the worldsheet conformal weight $\Delta_1$ replaced by $\hat{\Delta}_1$ given in Eq.~\eqref{defDeltaw}. On the other hand, in the $x$-basis one finds 
\begin{equation}
    \langle V_{j_1,h_1}^{\w_1} (x_1,z_1) V_{j_2,h_2}^{\w_2} (x_2,z_2) \rangle \,=\, \frac{1}{
    |x_{12}|^{4h_1}} 
    \frac{\langle V^{\w_1}_{j_1, m_1}  V^{\w_2}_{j_2, m_2}\rangle}{V_{\mathrm{conf}}} \, .
    \label{2ptwSL2}
\end{equation}
Thus, as mentioned above, the SL(2,$\R$) spin $h$ is identified with the holographic CFT conformal weight~\cite{Kutasov:1999xu}, even though in the flowed sectors the spin is independent of the value of $j$ of the corresponding unflowed operator. The factor $V_{\mathrm{conf}}$ stands for the divergent volume of the conformal group; it reflects the fact we are picking up the contribution from a pole, and it will cancel in the relevant computations that follow.

\subsubsection{Bosonic WZW model for SU(2)}
\label{sec:su2}

The bosonic WZW model based on the SU$(2)$ group manifold was studied in \cite{Fateev:1985mm,Zamolodchikov:1986bd}. We denote the generators of the current algebra by $k^a$, and for most quantities we use primes to distinguish them from their SL$(2,\R)$ counterparts. The currents satisfy the OPEs
\begin{equation}
    k^a(z)k^b(w) \,\sim\, \frac{k'}{2}\frac{\delta^{ab} }{(z-w)^2} + \frac{ f'^{ab}_{\phantom{ab}c} \;\! k^c(w)}{z-w} \; , 
    \label{OPEkSU2}
\end{equation}
where $k'$ is the level of the affine Lie algebra, $\delta^{ab}$ is the Killing form, and $f'^{abc}$ are the corresponding structure constants,
\begin{equation}
2\delta^{33} = \delta^{+-} = 2 \qqquad f'^{+-}_{\phantom{+-}3}=2 \ , \quad
    f'^{3+}_{\phantom{3+}+}=-
    f'^{3-}_{\phantom{3-}-}=1 \, .
\end{equation} 
The energy-momentum tensor and central charge are  
\begin{equation}
    T_{\mathrm{su}}(z) = \frac{1}{k'+2} \left[ k^3 (z) k^3(z) + \frac{1}{2} \, k^+(z) k^-(z) + \frac{1}{2} \, k^-(z) k^+(z)\right]  \;,  \qquad 
    c_{\su} = \frac{3k'}{k'+2}.
\end{equation}
We denote SU$(2)$ vertex operators by $V'_{j',m',\bar{m}'}(z,\bar{z})$. Again, their zero-mode wavefunctions do not factorize into holomorphic and antiholomorphic parts, however we shall mostly work holomorphically and suppress antiholomorphic quantities ($\bar{m}',\bar{z}$).

For SU$(2)$, the unitary representations of the zero-mode algebra are labeled by 
\begin{equation}
     \ 0\leq j' \leq \frac{k'}{2} \qqquad j'  \in \Z/2 \ ,
    \label{rangejp}
\end{equation}
and their states are $
    |j',m'\rangle$ with $m'= -j',-j'+1,\dots, j'-1,j'$. Using conventions that mimic those used above for SL$(2,\R)$, we have
\begin{subequations}
\begin{eqnarray}
    k_0^3  |j',m'\rangle &=& m'  |j',m'\rangle \, ,
     \label{kV0a}\\
   k^\pm_0 |j',m' \rangle &=& 
   \begin{cases}
			\,(j'+1\pm m') |j',m' \pm 1 \rangle & \text{if}~~ m\neq\pm j\\
            ~0 & \text{if}~~ m=\pm j \;,
		 \end{cases}\\
   k_{n}^a  |j',m'\rangle &=& 0 \ \qquad \forall  n>0 \, ,
    \label{kV0b}
\end{eqnarray}
\end{subequations}
and 
\begin{equation}
    \Delta' \,=\, \frac{j'(j'+1)}{k'+2} \, .
\end{equation}

Unlike SL$(2,\R)$, in the SU$(2)$ WZW model spectral flow is not necessary for constructing a consistent spectrum due to the compactness of the group manifold. Indeed, the spectral flow automorphisms merely reshuffle primary and descendant fields, and they do not introduce new inequivalent representations. Nevertheless, for superstring theory applications it is of practical use to include it in the discussion \cite{Giribet:2007wp,Martinec:2018nco,Bufalini:2021ndn}. We will discuss this in more detail shortly.

For SU$(2)$, spectral flow is defined as 
\begin{equation}
    k^\pm(z) \to \tilde{k}^\pm(z) = z^{\mp w'} k^\pm(z) \qqquad
    k^3(z) \to \tilde{k}^3(z) = k^3(z) - \frac{k'\w'}{2} \frac{1}{z} \, .
    \label{su2bosflow2}
\end{equation}
In this case, however,
it is possible to have $\bar{\w}'\neq \w'$.
As before, spectrally flowed primaries  $V_{j',m'}^{\w'}(z)$ are Virasoro primaries, with weight 
\begin{equation}
    \hat{\Delta}' \,=\, \frac{j'(j'+1)}{k'-2} + m' \w' + \frac{k'}{4}\w'^2 \,, 
    \label{defDeltapw}
\end{equation}
but they are not affine primaries, and for $\w'>0$ they are defined in terms of the OPEs 
\begin{subequations}
\label{flowedOPEsSU2}
\begin{eqnarray}
     k^{+}(z)V_{j',m' }^{\w'}(w) &=& (z-w)^{\w'-1} (j'-m') V_{j',m'+1 }^{\w'}(w) + \dots  \, , \\ [1ex]
    k^{3}(z)V_{j',m' }^{\w'}(w) &=& 
    \frac{\left(m'+ \frac{k}{2}\w' \right)V_{j',m' }^{\w'}(w)}{(z-w)}
    + \dots \, , \label{flowedOPEsSU2-b}\\ [1ex]
    k^{-}(z)V_{j',m'}^{\w'}(w) &=& \frac{(j'+m')V_{j',m'-1 }^{\w'}(w)}{(z-w)^{\w'+1}} + 
    \sum_{n=0}^{\w'} 
    \frac{(k^-_{n-1} V_{j',m' }^{\w'})(w)}{(z-w)^n} + \dots \, . 
\end{eqnarray}  
\end{subequations}
The corresponding  two-point functions are, again, the unflowed ones times $\delta_{\w_1,-\w_2}$, with the appropriate powers of $z_{12}$.

\subsubsection{Superstrings in AdS$_3\times \SSS^3\times \TT^4$}

We now review supersymmetric generalizations of the bosonic WZW models discussed above. 
We introduce fermions $\psisl^a$ and $\psisu^a$ which are superpartners of the SL$(2,\R)$ and SU$(2)$ currents $J^a$ and $K^a$ respectively.
The appropriate ${\cal{N}}=1$ supersymmetric extensions of the affine sl(2,$\RR$)$_k$ and su(2)$_{k'}$ algebras are generated by the supercurrents $ \psisl^a + \theta \:\! J^a$ and $ \psisu^a + \theta K^a$, where $\theta$ is a Grassmann variable. The currents $J^a$ and $K^a$ satisfy the OPEs~\eqref{OPEjSL2} and \eqref{OPEkSU2} respectively, with level $n_5$ in both cases, and the OPEs involving the fermions $\psisl^a$ and $\psisu^a$ are 
\begin{subequations}
\label{fermionsSL2SU2def}
\begin{align}
J^a(z) \psisl^b(w)  \,\sim\,  \frac{ f^{ab}_{\phantom{ab}c} \;\! \psisl^c(w)}{(z-w)} \qqquad & 
	K^a(z) \psisu^b(w)  \,\sim\,  \frac{f'^{ab}_{\phantom{lab}c} \;\! \psisu^c(w)}{(z-w)} , \\
	\psisl^a(z) \psisl^b(w)  \,\sim\, \frac{n_5}{2}\frac{\eta^{ab}}{(z-w)} \qqquad  &
	\psisu^a(z) \psisu^b(w)  \,\sim\, \frac{n_5}{2}\frac{\delta^{ab}}{(z-w)} \,.
\end{align}
\end{subequations}
One can split the currents into two independent contributions via 
\begin{equation}
\label{eq: Jtot and Ktot current}
	J^a = j^a - \frac{1}{n_5}f^{a}_{\phantom{a}bc}\:\! \psisl^b \psisl^c \qqquad 
		K^a  =\sucur^a - \frac{1}{n_5}f'^{a}_{\phantom{la}bc}\;\! \psisu^b \psisu^c \, .
\end{equation}
The ``bosonic'' currents $j^a$ and $k^a$ commute with the free fermions, and are currents of bosonic WZW models as described in Section~\ref{sec:AdS3S3T4}, with levels $k=n_5+2$ and $k'=n_5-2$ respectively. In the fermionic sector, the spectral flow automorphisms act as
\begin{equation}
	\tilde{\psisl}^\pm(z)  = z^{\mp}\psisl^\pm(z) \, , \quad 
	\tilde{\psisl}^3(z)  = \psisl^3(z)
	\qqquad 
    \tilde{\psisu}^\pm(z)  = z^{\mp}\psisu^\pm(z) \, , \quad 
	\tilde{\psisu}^3(z)  = \psisu^3(z) \, .
	\label{flowfermionmodes}
\end{equation}

The remaining flat compact directions are treated as usual. For the $\TT^4$, we simply have four (canonically normalized) free bosons $Y^i$ and their fermionic partners $\lambda^i$ $(i=6,\dots,9)$, with OPEs 
\begin{equation}
    Y^i(z)Y^j(w) \,\sim\, - \delta^{ij} \log(z-w) \qqquad \lambda^i(z) \lambda^j(w) \,\sim\, \frac{\delta^{ij}}{(z-w)} \, .
\end{equation}

We can now write down the energy-momentum tensor $T$ and the supercurrent $G$ of the worldsheet theory for type II superstrings in AdS$_3\times \SSS^3\times \TT^4$. The matter contributions read
\begin{eqnarray}
    T &\,=\,& \frac{1}{n_5} \Big(j^a j_a - \psisl^a \der \psisl_a + 
    k^a k_a - \psisu^a \der \psisu_a 
    \Big) + \frac{1}{2}
    \left(\der Y^i \der Y_i - \lambda^i \der \lambda_i\right),
    \label{TAdS3S3T4def}
    \\
    G &\,=\,& \frac{2}{n_5} \left(
    \psisl^a j_a - \frac{1}{3n_5} f_{abc} \psisl^a \psisl^b \psisl^c + 
    \psisu^a k_a - \frac{1}{3n_5} f'_{abc} \psisu^a \psisu^b \psisu^c
    \right) + i \:\lambda^j \der Y_j \, ,
    \label{GAdS3S3T4def}
\end{eqnarray}
and the resulting central charge is compensated by the usual $bc$ and $\beta \gamma$ ghost systems, leading to the BRST charge 
\begin{equation}
\label{eq:BRSToperator}
    {\cal{Q}} = \oint dz \left( c \left(T + T_{\beta\gamma}\right) - \gamma \, G + c(\der c) b - \frac{1}{4} b \gamma^2 \right) \, .
\end{equation}
Here $T_{\beta\gamma}$ is the energy-momentum tensor of the $\beta\gamma$ system, which is bosonized as 
\begin{align}
	\beta = e^{-\varphi} \del \xi  \qqquad \gamma = \eta \:\! e^{\varphi} \,,
\end{align}
where $\varphi(z) \varphi(w) \simeq - \ln(z-w)$ has background charge $2$, and $\xi(z)\eta(w) \sim (z-w)^{-1}$.
For computational purposes it is useful to also bosonize the rest of the fermions~\cite{Giveon:1998ns,Dabholkar:2007ey}. We thus define (canonically normalized) bosonic fields $H_I$ with $I=1,\dots 5$, and write 
\begin{equation}
        \hat{H}_I = H_I + \pi \sum_{J<I} N_J \qqquad N_J \equiv \oint i\del H_J \, ,
\end{equation}
where the number operators $N_I$ are introduced in order to keep track of the cocycle factors, namely
\begin{equation}
        e^{i a \hat{H}_I} e^{i b \hat{H}_J} = e^{i b \hat{H}_J} e^{i a \hat{H}_I} \, e^{i \pi a b} \qqquad \text{if} \quad I > J \, . \label{HIAdS3}
\end{equation}
We bosonize as 
\begin{subequations}
\begin{gather}
	 \psisl^\pm = \sqrt{n_5} \, e^{\pm i\hat{H}_1} 
	 \, , \quad 
	\psisu^\pm = \sqrt{n_5} \, e^{\pm i \hat{H}_2} 
	 \, , \quad
	 \lambda^{6} \pm i \lambda^7 = e^{\pm i \hat{H}_4} 
	 \, , \quad 
	 \lambda^{8} \pm i \lambda^9 = e^{\pm i \hat{H}_5} \, ,  \label{psiH1xiH2} \\
	 \qquad \psisl^3 = \frac{\sqrt{n_5}}{2} \, \left(e^{i \hat{H}_3} - e^{-i\hat{H}_3} \right)  \qqquad \psisu^3 =  \frac{\sqrt{n_5}}{2}  \, \left(e^{i \hat{H}_3} + e^{-i\hat{H}_3} \right) \,,
\end{gather}
\end{subequations}
where $ \hat{H}_I^\dagger = \hat{H}_I$ for $I \ne 3 $ and $ \hat{H}_3^\dagger = - \hat{H}_3 $. Then we have
\begin{subequations}
  \begin{gather}
        i\del \hat{H}_1 = \frac{1}{n_5} \, \psisl^+\psisl^- \qqquad   i \del \hat{H}_2 =  \frac{1}{n_5} \, \psisu^+\psisu^-  \qqquad
	 i \del \hat{H}_3 = \frac{2}{n_5} \, \psisl^3\psisu^3  \, ,  
    \\  i\del \hat{H}_4 = i\lambda^6 \lambda^7 \qqquad  i\del \hat{H}_5 = i\lambda^8 \lambda^9 \, .
	\end{gather}
\end{subequations}  
The phases in \eqref{HIAdS3} ensure that bosonized fermions anticommute, and will be important when working with states in the Ramond sector. From now on we will simply omit the hats, and explicitly include the phase factors when they are needed.

The spacetime supercharges can be written as: 
\begin{equation}
    Q_\vep = \oint dz \, e^{-\varphi/2} S_\vep \qqquad   S_\vep = \exp \left(\frac{i}{2} 
    \sum_{I=1}^{5}\vep_I H_I\right),
    \label{supercharges}
\end{equation}
where $S_\vep$ are spin fields and $\vep_I=\pm 1$.
Imposing BRST invariance -- where the relevant contributions come from the $f_{abc}\psi^a\psi^b\psi^c$ and $f_{abc}'\chi^a\chi^b\chi^c$ pieces of $G$ in \eqref{GAdS3S3T4def} -- and mutual locality (chiral GSO) leads to the conditions
\begin{equation}
\label{sign constraint}
    \prod_{I=1}^{3} \vep_I = \prod_{I=1}^{5} \vep_I = 1 \, .
\end{equation}
In the holomorphic sector this gives the expected four `ordinary' supercharges and four `superconformal' supercharges. The same applies in the antiholomorphic sector, giving the total 16 real supercharges of global AdS$_3\times \SSS^3$~\cite{Giveon:1998ns}.

For later use, let us also recall that the R-symmetry of the boundary theory is generated on the worldsheet by the SU(2) currents. More precisely, the zero modes of the spacetime R-currents are given by the integrated worldsheet currents \cite{Giveon:1998ns}, i.e. 
\begin{equation}
    \Jd^a_0 = \oint dz \, K^a (z) \, .
\end{equation} 
Consequently, the holomorphic R-charge in the holographic CFT is identified with $m'$, the eigenvalue of $K^3$. 
It is for this reason that we used the notation $m'$ in Sections \ref{sec:heavyD1D5} and \ref{secCPD1D5CFT}.

\subsection{Vertex operators and two-point functions}
\label{sec:vertexops}

We now discuss physical vertex operators and their two-point functions, both in NS and R sectors. This section is largely review, though we also give explicit expressions for some R sector operators that to our knowledge have not appeared before in the literature.

Our main interest is in worldsheet operators that correspond to chiral primaries of the holographic CFT.  We thus focus on states belonging to the discrete representations of SL(2,$\R$). We also discuss the role of SL(2,$\R$) and SU$(2)$ spectral flows in the string theoretical construction. 
These vertex operators and their two-point functions will be used as building blocks for constructing the vertex operators and two-point functions of the null-gauged models in Section~\ref{sec:nullgaugedmodel}.

\subsubsection{NS sector}

The unflowed NS-NS sector was considered  in \cite{Kutasov:1998zh}, see also \cite{Dabholkar:2007ey}.
We continue to suppress antiholomorphic parts of the SL$(2,\R)$ and SU$(2)$ vertex operators $V_{j,m,\bar{m}}$ and $V'_{j',m',\bar{m}'}$. The complete NSNS vertex is obtained by including the antiholomorphic fermions and ghosts.

We work in the canonical ``$-1$'' ghost picture, and consider only states with vanishing momentum in the $\TT^4$ directions. 
Then the (holomorphic part of the) BRST invariant states with up to a single fermionic excitation include the
tachyon (which is projected out by GSO),
\begin{equation}
    \Tt_{\jsl,\msl,\jsu,\msu} = 
    e^{-\varphi} \Vsl_{\jsl,\msl} \Vsu_{\jsu,\msu} \, , 
    \label{groundstate}
\end{equation}
and the spacetime vectors ($\vep = \pm 1$, and recall $i=6,\dots,9$)
\begin{subequations}
\label{states BRST KLL}
\begin{eqnarray}
	\Vv^{i}_{\jsl,\msl,\jsu,\msu} &=&  e^{-\varphi} \lambda^i \Vsl_{\jsl,\msl} \Vsu_{\jsu,\msu} \,,
	\label{T4statesAdS3}\\
	\Ww^{\vep}_{\jsl,\msl,\jsu,\msu} &=&   e^{-\varphi} \left( \psisl \Vsl_{\jsl}\right)_{\jsl +\vep ,\msl} \Vsu_{\jsu,\msu} \,,
	\label{WstatesAdS3} \\
	\mathcal{X}^{\vep}_{\jsl,\msl,\jsu,\msu} 
	&=& e^{-\varphi} \Vsl_{\jsl,\msl} (\psisu \Vsu_{\jsu})_{\jsu +\vep,\msu} \, ,
	\label{XstatesAdS3}
\end{eqnarray}
\end{subequations}
where we have introduced the linear combinations  
\begin{equation}
\label{linearcombsKLL}
		(\psisl \Vsl_{\jsl})_{\jsl +\vep ,\msl} = c_{\vep}^r \psisl^r \Vsl_{\jsl,\msl - r} \qqquad   
		(\psisu \Vsu_{\jsu})_{\jsu +\vep ,\msu}  = d_{\vep}^r \psisu^r \Vsu_{\jsu,\msu - r} \, ,
\end{equation}
where a summation over $r=+1,-1,0$ is implicit, ``$0$'' corresponding to the ``$3$'' direction of the respective algebras. These combine the products of bosonic primaries and free fermions into fields of total spins $ J = j + \vep$ and $J' = j' + \vep$ under the action of the supersymmetric currents $J^a$ and $K^a$~\cite{Kutasov:1998zh}. The Clebsh-Gordan coefficients are given in our conventions by
\begin{align}
	c^r_-  &= \left( \half,  \half,  -1\right), \quad 
	d^r_+  = \left( -\half ,   \half, 1 \right), \nn \\
	c_+^r &= \left(\half (\jsl+\msl)(\jsl+\msl-1) ,\; \half (\jsl-\msl) (\jsl-\msl-1) ,\; (\jsl+\msl)(\jsl-\msl) \right), \label{cdcoefs} \\
	d^r_-  &= \left( \half(\jsu-\msu)(\jsu-\msu+1) , \; -\half (\jsu+\msu)(\jsu+\msu+1), \; (\jsu-\msu) (\jsu+\msu)\right) \, . \nn
\end{align}
The Virasoro condition associated to all vertex operators in Eq.~\eqref{states BRST KLL} reads
\begin{equation}
 \frac{1}{2} + \frac{1}{2} - \frac{j(j-1)}{n_5} + 
 \frac{j'(j'+1)}{n_5} = 1 \, , 
 \label{Virasoro}
\end{equation}
and is solved by $j=j'+1$ (or its reflection under $j\to 1-j$), thus implying that we are dealing with bosonic primaries in the discrete representations of SL(2,$\R$)$_{k}$. 

Let us briefly discuss the worldsheet two-point functions involving these operators. The different bosonic sectors factorize and the fermions are free, so we can express the results directly in terms of the non-trivial contributions coming from the bosonic SL(2,$\R$) WZW model, namely Eq.~\eqref{2ptMbasisSL2}. By construction, the only non-vanishing two-point functions are the diagonal ones. For the 6D scalars coming from the NS-NS sector polarizations on the $\TT^4$, \eqref{T4statesAdS3}, we have
\begin{equation}
    \langle \Vv^{i\bar{i}}_1\Vv^{j\bar{j}}_2 \rangle \,=\, 
    \langle V_1 V_2\rangle
    \langle V_1' V_2'\rangle
    \left[
    \langle e^{-\varphi_1}e^{-\varphi_2}\rangle
    \langle \lambda^i_1 \lambda^j_2 \rangle 
     \times \cc \right] \,=\, \langle V_1 V_2\rangle
    \langle V_1' V_2'\rangle
    \times \frac{\delta^{ij}
    \delta^{\bar{i}\bar{j}}}{|z_{12}|^4} \, .
\end{equation}
Since we are dealing with discrete representations, the contact term in \eqref{2ptMbasisSL2} vanishes, thus imposing $j_1=j_2\equiv j$. 
As discussed above Eq.\;\eqref{ExpSL2}, the conformal weight in the holographic CFT is to be identified with the SL(2,$\R$) spin, i.e.~$h=j$. On the other hand, the R-charge is given by $m'$, with $|m'|\leq j' = j-1$. Thus $h\neq |m'|$, so $\Vv^{i}$ cannot correspond to a chiral primary of the HCFT. 

We now turn to the operators introduced in the second and third line of \eqref{states BRST KLL}. When computing correlators of two $\Ww$ states, we must deal with expressions of the form
\begin{eqnarray}
	\langle \left( \psisl_1 \Vsl_{j_1}\right)_{j_1+\vep_1,m_1}\left( \psisl_2 \Vsl_{j_2}\right)_{j_2+\vep_2,m_2} \rangle  =
    \sum_{r_{1,2}}
    c^{r_1}_{\vep_1}c^{r_2}_{\vep_2} 
    \braket{\psisl^{r_1}\psisl^{r_2} } \braket{\Vsl_{j_1,m_1-r_1}\Vsl_{j_2,m_2-r_2}}. 
	\label{WW1}
\end{eqnarray}
We use the action of the bosonic currents~\eqref{zeromodesSL2} to express $\braket{\Vsl_{j_1,m_1-r_1}\Vsl_{j_2,m_2-r_2}}$ in terms of $\braket{\Vsl_{j_1,m_1}\Vsl_{j_2,m_2}}$, insert the coefficients \eqref{cdcoefs},  and perform the sum. Recalling the shorthand $V_i\equiv V_{j_i , m_i}$, $V'_i\equiv V'_{j'_i , m'_i}$, we obtain
\begin{equation}
\label{eq: W2pt}
   \hspace{-1.5mm}\langle \Ww^{\vep_1} \Ww^{\vep_2} \rangle = \frac{n_5^2}{4|z_{12}|^4} \langle V_1 V_2\rangle
    \langle V_1' V_2'\rangle \times 
    \left\{\begin{array}{cc}
       j_1 (1-2j_1) (j_1^2 - m_1^2) \times \cc & \quad \vep_1 = \vep_2 = 1 \\[1ex]
        \dfrac{(j_1-1)(1-2j_1)}{(j_1-1)^2 - m_1^2}\times \cc  & \quad \vep_1 = \vep_2 = -1 ~~\\
        0 & \vep_1 = -\vep_2\;.
    \end{array}
    \right. 
\end{equation}
From Eq.~\eqref{2ptMbasisSL2} we find that the coefficients are exactly those needed to produce the shift $j\to j+\vep$ in the two-point function. Hence, Eq.~\eqref{MellinSL2} shows that the weight of the corresponding holographic dual is $h=j+\vep$ \cite{Gaberdiel:2007vu}. In particular, the  
operator $\Ww^{-}$ with maximal SU(2) charge has $h = j-1 = m'$, and thus corresponds to a chiral primary operator of the holographic CFT. 

The computation of the $\langle \Xx \Xx \rangle$ correlators is analogous; we obtain 
 \begin{equation}
  \hspace{-1mm}
     \langle \Xx^{\vep_1} \Xx^{\vep_2} \rangle = \frac{n_5^2}{4|z_{12}|^4} \langle V_1 V_2\rangle
     \langle V_1' V_2'\rangle \times 
     \left\{\begin{array}{cc}
 \dfrac{(j_1'+1)(1+2j_1')}{(j_1'+1)^2 - {m_1'}^{\!2} } \times \cc & \vep_1 = \vep_2 = 1 \\
      j_1' (1+2j_1') ({j_1'}^{\!2} - {m_1'}^{\!2}) \times \cc~  & \vep_1 = \vep_2 = -1 \\
         0 & \vep_1 = -\vep_2 = \pm 1 
     \end{array}
     \right.
     \label{XX2pt}
 \end{equation}
For $\Xx^+$ at highest SU(2) weight we have $h=j=j'+1=m'$, leading to a second family of spacetime chiral states. We will discuss the corresponding operators in the holographic CFT theory and fix their normalization below.

So far, we have constructed chiral operators whose boundary weights $h=j-1$ and $h=j$ are bounded from above by $h<\frac{n_5+1}{2}$, see Eq.~\eqref{Djrange}. However, in the D1D5 CFT one can have chiral primaries in $n$-twisted sectors with $n$ up to $n_1 n_5$, and where $h$ grows linearly with $n$, as discussed around Eq.~\eqref{eq:cp-bound}. Thus, it seems that so far we are missing most of the heavier chiral operators. However, as discussed in \cite{Giribet:2007wp}, such states lie in the sectors of the worldsheet theory with non-trivial spectral flow charges, as we now review.

In the supersymmetric theory, spectrally flowed primary operators are built by combining the bosonic flowed primaries introduced in Eqs.~\eqref{flowedOPESL2} and \eqref{flowedOPEsSU2} with fermionic excitations. The bosons $H_I$ allow us to express the spectral flow operation in the fermionic sectors of SL(2,$\R$) and SU(2), Eq.\;\eqref{flowfermionmodes}, in the following form,
\begin{equation}
    \psisl^\pm_\w \,=\, \psisl^\pm e^{- i \w H_1} \qqquad 
    \psisu^\pm_{\w'} \,=\, \psisu^\pm e^{ i \w' H_2},
    \label{flowH12}
\end{equation}
while the other fermions remain unchanged. Indeed, the OPEs between the operators in \eqref{flowH12} and the fermionic currents are analogous to those in \eqref{flowedOPESL2} and \eqref{flowedOPEsSU2}. 
Once factors of $e^{- i \w H_1}$ and $e^{ i \w' H_2}$ are included, the corresponding weights  
take the form in \eqref{defDeltaw} and \eqref{defDeltapw}, with $k-2=n_5=k'+2$.

In principle, one could simply ignore the possibility of including spectral flow in SU(2) since it does not give any new representations. However, for discrete series states it is useful to do so in order to solve the modified Virasoro condition, as discussed in~\cite{Martinec:2018nco,Bufalini:2021ndn}.
We use the spectral flow operator with equal amount of spectral flow in SL(2,$\R$) and SU(2),
\begin{equation}
    \exp \left(-i \w H_1 + \w \sqrt{\frac{n_5+2}{2}} \phi
    + i \w H_2 + i \w \sqrt{\frac{n_5-2}{2}} \phi'\right), \label{susyflowoperator2}
\end{equation}
which is mutually local with the supercharges, thus producing flowed singly-excited states which will also survive the GSO projection. The corresponding Virasoro condition for the flowed vertex operators is
\begin{equation}
\label{eq:vir-flowed-ns}
 \frac{1}{2} + \frac{1}{2} - \frac{j(j-1)}{n_5} - m \w - \frac{n_5}{4}\w^2 
 + \frac{j'(j'+1)}{n_5} + m'\w +\frac{n_5}{4}\w^2  \;=\; 1 \, .
\end{equation}
We seek to solve this for general $n_5$. We thus impose $j=j'+1$ as well as $m=m'$. The latter constraint is quite restrictive, since by definition we have
\begin{eqnarray}
    {\cal{V}}\text{-type operators:}~~& \ \ 
    |m|\geq j
    \ , \ \quad~~ |m'|\leq j' = j-1\, , \nn\\
    {\cal{W}}\text{-type operators:}~& \ \ 
    |m|\geq j-1
    \ , \ ~~ |m'|\leq j' = j-1\, , \nn\\
    {\cal{X}}\text{-type operators:}~~& \ \ 
    |m|\geq j
    \ , \ \quad~~ |m'| \leq j'+1 = j\, . \nn
\end{eqnarray}
Consequently, our only candidates are highest/lowest-weight ${\cal{W}}^-$-type operators with $m=m'=j'=j-1$ and ${\cal{X}}^+$-type operators with $m=m'=j'+1=j$.  Their explicit expressions are given by
\begin{subequations}
\label{WXvertexW}
\begin{eqnarray}
    {\cal{W}}^\w_j &=&  
    e^{-\varphi} \psi^- e^{-i \w H_1 }e^{i \w H_2} V^\w_{j,j} V'^\w_{j-1,j-1}\, ,
    \\
    {\cal{X}}^\w_j &=& 
    e^{-\varphi} e^{-i \w H_1 } \chi^+ e^{i \w
    H_2} V^\w_{j,j} V'^\w_{j-1,j-1}\, .
\end{eqnarray}
\end{subequations}
These flowed states are also BRST-invariant \cite{Giribet:2007wp} since the supercurrent $G$ can be written in the flowed frame as 
\begin{equation}
\label{BRSTflowG}
G(z) = \tilde{G}(z) + \frac{\w}{z}\left(\chi^3-\psi^3\right), 
\end{equation}
such that the extra terms on the RHS of this equation act trivially on highest/lowest weight states. 

The two-point functions of these spectrally flowed operators can be determined straightforwardly from the corresponding bosonic ones. This is because the latter impose $\w_2 = - \w_1$, such that the charge conservation rules for the $H_{I}$ exponentials are automatically satisfied.
We define the conjugate operators $\hat{\Ww}^{\w}$, $\hat{\Xx}^{\w}$, and obtain
\begin{equation}
    \langle \Ww^{\w_1}_1 \hat{\Ww}^{\w_2}_2\rangle \,=\, 
    \langle \Xx^{\w_1}_1 \hat{\Xx}^{\w_2}_2\rangle \,=\, 
    \langle V_{j_1,j_1}^{\w_1} V_{j_1,-j_1}^{\w_2}\rangle
    \langle 
    V_{j_1-1,j_1-1}^{'\w_1} V_{j_1-1,1-j_1}^{'\w_2}
    \rangle
    \frac{n_5^2}{|z_{12}|^{4(1+
    \w_1+\w_1^2)}} \, . 
\end{equation}
Spectral flowed primaries are always annihilated by $J_0^-$, and are thus lowest-weight with respect to the SL(2,$\R$) zero mode algebra. 
After the supersymmetric spectral flow \eqref{susyflowoperator2}, similarly to the bosonic transformations~\eqref{h-sf-2}, \eqref{flowedOPEsSU2-b}, the spectral flowed primaries have 
quantum numbers $h$ and $m'$ that have  increased by 
$\,\frac{n_5}{2}\omega\,$ from their values for the unflowed vertex operators.
We thus conclude that these vertex operators correspond exactly to the additional chiral operators we were looking for, with
\begin{equation}
    {\cal{W}}^\w_j ~:~~ h  = m' = j-1+\frac{n_5}{2}\w \qqquad 
   {\cal{X}}^\w_j ~:~~ h  = m' = j+\frac{n_5}{2}\w \, .
    \label{eq:h-sf}
\end{equation}
These quantum numbers extend to large values, by raising $\w$. 
In the holographic CFT there are states with conformal weight of order $n_1n_5$, however in our worldsheet models $n_1$ is of order $g_s^{-2}$ (c.f.~\eqref{eq:q1-qp}) and we work in perturbation theory in $g_s$, so finite $n_1$ physics is not accessible.
Moreover, when considering holographic CFT operators with conformal weight of order $n_1n_5$, the dual bulk configuration is not a light probe on the original background, but rather a different background.
The rest of the modes associated to such boundary operators  are obtained by acting with the global current $J_0^+$ as in Eq.~\eqref{flowedxbasis}, and do not have simple expressions in the $m$-basis since they are not flowed primaries.

\subsubsection{Ramond sector}
\label{sec:RsectorAdS3}

We now review the Ramond sector physical operators of the worldsheet theory, in the $m$-basis. To our knowledge, this construction has only been carried out explicitly in the literature for the case of highest/lowest-weight states~\cite{Dabholkar:2007ey,Giribet:2007wp}; we shall present explicit expressions for more general Ramond sector operators. 

We will make use of the spin fields introduced in \eqref{supercharges}, and distinguish the slightly more involved AdS$_3 \times \SSS^3$ sector, for which we write the relevant factors as
\begin{equation}
    S_{\vep_1  \vep_2 \vep_3} \,=\, e^{\frac{i}{2}(\vep_1 H_1 + \vep_2 H_2 + \vep_3 H_3 )} \, .
\end{equation}
We denote the AdS$_3 \times \SSS^3$ chirality by $\vep \equiv \vep_1 \vep_2 \vep_ 3 $. We shall implement this by considering $\vep_ 3$ to be fixed to be $\vep_ 3=\vep \vep_1 \vep_2$.
We impose the chiral GSO projection via the mutual locality condition $\prod_{I=1}^5
\vep_I=1$, and we implement this by fixing $\vep_5 = \vep \:\! \vep_4$. 
We introduce a generic linear combination of bosonic primaries and spin fields of AdS$_3 \times \SSS^3$ of fixed chirality,
\begin{equation}
\label{eq:svv}
    \left( S V V' \right)^\vep_{J,m,J',m'} = \sum_{\vep_1, \vep_2 = \pm 1} f^{\vep}_{\vep_1\, \vep_2} S_{\vep_1 \vep_2 \vep_3} V_{j,m-\frac{\vep_1}{2}} V'_{j',m'-\frac{\vep_2}{2}} \qqquad
    \vep_3 = \vep\:\! \vep_1\vep_2 \,,
\end{equation}
where the total spins $(J,J')$ will be related to $(j,j')$ in various ways momentarily. Note that for highest/lowest weight states, there may be only one allowed choice of $\vep_1$ and/or $\vep_2$, as we shall see in an example below. In the canonical ``$-\frac{1}{2}$'' picture, the Ramond sector vertex operators then take the form
\begin{equation}
\label{YyAdS3}
    \Yy^{\vep,\vep_4}_{J,m,J',m'} = e^{-\frac{\varphi}{2}} \left( S V V' \right)^\vep_{J,m,J',m'} e^{\frac{i \vep_4}{2} \left( H_4 + \vep H_5 \right)} \, .
\end{equation}
The Clebsch--Gordan coefficients
$   f^{\vep}_{\vep_1\, \vep_2}$ are computed by requiring that the $\Yy$ operators transform appropriately under the action of the currents $J^{\pm}$, $K^{\pm}$. In our conventions, this gives four linear combinations. In the equations below, the first bracket specifies how $(J,J')$ are related to $j$ and $j'$; for instance, for case $A$, $(J=j-1/2, J'=j'+1/2)$. For each case we write the coefficients as a list, $   f^{\vep}_{\vep_1\, \vep_2}  = \left(f^{\vep}_{+ +},f^{\vep}_{+ -},f^{\vep}_{- +},f^{\vep}_{- -} \right) $. We obtain
\begin{subequations}
\label{eq:RRcoeff}
\begin{align}
    A\!: \; ( j-\tfrac12,\;\!  j' + \tfrac12)\;\!,  \quad~ f^{\vep, A}_{\vep_1\, \vep_2} & \,=\, \big(1, \, i , \, \vep , \, \vep  \;\! i  \big)\;\!, \\[1.5mm]
    B\!: \; (j+\tfrac12, \;\!j' + \tfrac12)\;\!,  \quad~
    f^{\vep, B}_{\vep_1 \, \vep_2} & \,=\, \left(f^{B_1}  ,\, i\;\! f^{B_1}  ,\, \vep \;\! f^{B_2}  , \, \vep \;\! i \;\! f^{B_2} \right), \\[1.5mm]
    C\!: \;  (j-\tfrac12,\;\! j' - \tfrac12)\;\!,  \quad~
    f^{\vep, C}_{\vep_1 \, \vep_2} & \,=\, \left(f^{C_1}, \, -i \;\! f^{C_2}, \, \vep \;\! f^{C_1} , \, \vep (-i) f^{C_2} \right), \\[1.5mm]
    D\!: \;  (j+\tfrac12,\;\! j' - \tfrac12)\;\!,  \quad~
    f^{\vep, D}_{\vep_1 \, \vep_2 } & \,=\, \left(f^{B_1} f^{C_1}, (-i) f^{B_1} f^{C_2} ,  \, \vep \;\! f^{B_2} f^{C_1}, \, \vep(-i) f^{B_2} f^{C_2} \right),
\end{align}
\end{subequations}
where
\begin{equation}
    f^{B_1} = m + j - \half \, , \quad f^{B_2} = m - j + \half\, , 
    \quad
    f^{C_1} = j' - m' + \half \, ,  \quad  f^{C_2} = j' + m' + \half \,.
\end{equation}
We note that in all cases we have $f^{+}_{\vep_1 \vep_2} = \vep_1 \,f^{-}_{\vep_1 \vep_2} $.
In addition, and using $j = j'+1$, BRST-invariance gives four equations for each chirality, out of which only two are linearly independent, namely 
\begin{subequations}
\label{BRSTAdS3}
\begin{align}
    f_{-+}^{\vep} & \,=\, \frac{1}{\left(j+m-\half\right)} \left[ f_{++}^\vep\left(\vep\, m + m'\right) - i \,  f_{+-}^\vep \left(j' - m' + \half\right)\right]  ,\\
    f_{--}^{\vep} & \,=\, \frac{1}{\left(j+m-\half\right)} \left[ i \,f_{++}^\vep \left(j' + m' + \half\right) + f_{+-}^\vep \left(\vep \, m - m '\right)      \right] .
\end{align}
\end{subequations}
These are satisfied by only half of the states in Eq.~\eqref{eq:RRcoeff}. The physical states in the ``$-\frac{1}{2}$'' picture are given by the $A$ and $D$ states with $\varepsilon=1$ (and either choice of $\varepsilon_4=\pm 1$), plus the $B$ and $C$ states with $\varepsilon=-1$ (and again either sign of $\varepsilon_4$), making the correct eight physical polarizations.

The full list of expressions in \eqref{eq:RRcoeff} is useful in order to construct the representatives of such operators in the ``$-\frac{3}{2}$'' ghost picture, necessary for computing two-point functions. 
To obtain the ``$-\frac{3}{2}$'' picture operators, we make an educated guess for their expressions, and then apply the picture \textit{raising} operator, i.e. 
\begin{equation}
    \Phi^{(-\frac{1}{2})}(w) \,=\, \lim_{z\to w} \left(e^{\varphi}G\right)(z) \;  \Phi^{(-\frac{3}{2})}(w) \, .
\end{equation}
In order to get a non-trivial propagator, and up to an overall constant, the appropriate guess is that they are given by the states with the same spins but opposite chirality. Explicitly, we have 
\begin{equation}
    \Yy^{\vep,\vep_4 \,  (-\frac{3}{2})}_{J,m,J',m'} \,=\, \pm \frac{\sqrt{n_5}}{2j-1} e^{-\frac{3\varphi}{2}} \left( S V V' \right)^{-\vep}_{J,m,J',m'} e^{\frac{i \vep_4}{2} \left( H_4 + \vep H_5 \right)} \, ,
    \label{Ypicture3half} 
\end{equation}
where the negative (positive) sign holds for the cases $A$ and $B$ ($C$ and $D$).

We can now compute the two-point functions in the unflowed Ramond sector. Only diagonal pairings are non-zero, by construction. Denoting the antiholomorphic sector contributions by ``c.c.'', we obtain
\begin{subequations}
\label{AdSRam2pt}
\begin{align}
\langle \Yy_{[A]}^{\vep_4,(-\half)} \Yy_{[A]}^{-\vep_4,(-\frac{3}{2})} \rangle &\,=\, \frac{n_5}{|z_{12}|^4} 
\braket{V_1 V_2}\braket{V'_1 V'_2} \left(
\dfrac{(2j-1)}{(j-m-\half)(j'+m'+\half)} \times \text{c.c.} \right)  
 , \label{2ptYyA}\\
\langle \Yy_{[B]}^{\vep_4,(-\half)} \Yy_{[B]}^{-\vep_4,(-\frac{3}{2})} \rangle &\,=\, \frac{n_5}{|z_{12}|^4}
\braket{V_1 V_2}\braket{V'_1 V'_2} \left(
\dfrac{(2j-1)(j+m-\half)}{(j'+m'+\half)} \times \text{c.c.} \right)
,\\
\langle \Yy_{[C]}^{\vep_4,(-\half)} \Yy_{[C]}^{-\vep_4,(-\frac{3}{2})} \rangle &\,=\, \frac{n_5}{|z_{12}|^4}
\braket{V_1 V_2}\braket{V'_1 V'_2}
\left(\dfrac{(2j-1)(j'-m'+\half) }{(m-j+\half)} \times \text{c.c.}\right)
 ,\\
\langle \Yy_{[D]}^{\vep_4,(-\half)} \Yy_{[D]}^{-\vep_4,(-\frac{3}{2})} \rangle &\,=\, \frac{n_5}{|z_{12}|^4}
\braket{V_1 V_2}\braket{V'_1 V'_2} \left(
(2j-1)(m+j-\half)(j'-m'+\half)  \times \text{c.c.} \right)
 ,
\end{align}
\end{subequations}
where here $\braket{V_1 V_2} = \braket{V_{m_1-1/2} V_{-m_1+1/2}}$
and $\braket{V'_1 V'_2} = 
\braket{V'_{m_1'-1/2} V'_{-m_1'+1/2}}$.
As expected, the coefficients resulting from the linear combinations effectively shift the spins $j \to J$ and $j' \to J'$ in the gamma functions coming from the bosonic correlators. 

Among the states described above, the only chiral one corresponds to the SU(2) highest-weight operator of type $A$. To simplify notation and for later convenience, we suppress the SU(2) labels and use the label $j$ rather than $J$ (here $J=j-1/2$). This operator has quantum numbers
\begin{equation}
\label{eq:q-nos-rr}
    \Yy_{j,m[A]}^{+,\vep_4} ~:~~~~ h \:\!=\:\! J \:\!=\, j-\half \,=\, j'+\half \,=\:\! J' \:\!=\:\! m' \, . 
\end{equation}
The explicit form of this operator is simpler than the generic Ramond sector operator, and is given by 
\begin{equation}
\label{eq:rr-A-cpo}
    \Yy_{j,m[A]}^{+,\vep_4} \,=\, 
    e^{-\frac{\varphi}{2}}
    \left(S_{+++} V_{j,m-\half} + 
    S_{-+-} V_{j,m+\half}
    \right)_{j-\half,m}V'_{j-1,j-1} e^{\frac{i \vep_4}{2} \left( H_4 + H_5 \right)} \, . 
\end{equation}
As in the NS sector, the rest of the chiral operators belong to the  spectrally flowed sectors. These are obtained by acting with the spectral flow operator \eqref{susyflowoperator2}.
From the flowed Virasoro condition, similar to Eq.\;\eqref{eq:vir-flowed-ns}, the resulting operators must have $m = m'$ (together with the relations in \eqref{eq:q-nos-rr}), and so only the second term in \eqref{eq:rr-A-cpo} is non-vanishing, giving rise to the flowed Ramond operators (we now suppress also the label $m=J=j-1/2$)
\begin{equation}
    \Yy^{+,\vep_4,\w}_{j\:\! [A]} \,=\, 
    e^{-\frac{\varphi}{2}}
    S_{-+-}^\w \:\! V_{j,j}^\w \;\!
    V_{j-1,j-1}^{'\w}\:\! e^{\frac{i \vep_4}{2} \left( H_4 + H_5 \right)} \, , 
\end{equation}
where
\begin{equation}
    S_{- + -}^\w \,\equiv\, e^{\frac{i}{2}\left[(1+2\w)(-H_1 +H_2) - H_3 \right]} \, . 
    \label{spinwmbasis}
\end{equation}
The corresponding two-point function is equivalent to the highest-weight case of \eqref{2ptYyA}, up to the usual additional $\delta_{\w_1,-\w_2}$ factor.

\subsubsection{Holographic dictionary for light chiral primaries} \label{sec:dict}

We have reviewed three sets of $m$-basis vertex operators corresponding to chiral primaries of the holographic CFT. Two sets are in the NS sector: $\Ww^-_{j,m}$ and $\Xx^+_{j,m}$, together with the corresponding spectral flowed operators
$\Ww^{\w}_{j}$ and $\Xx^{\w}_{j}$. The third set is in the Ramond sector, $\Yy_{j,m[A]}^{+,\vep_4}$
and its spectral flow,
$\Yy^{+,\vep_4,\w}_{j\:\! [A]}$. From now on we shall omit the label $A$ and the AdS$_3\times \SSS^3$ chirality $\epsilon=+$, denoting this operator by $\Yy^{\w,\vep_4}_{j}$. Recall that in the spectral flowed sectors, the remaining states in the zero-mode algebra are obtained by acting with $J_0^+$, as discussed around Eqs.~\eqref{flowedxbasis} and \eqref{eq:h-sf}.

In order to reconstruct the corresponding local operators of the spacetime CFT, we need to combine such modes by going to the $x$-basis, as done in Eqs.~\eqref{ExpSL2} and \eqref{flowedxbasis} in the bosonic SL(2,\R) model.\footnote{In this paper we are interested in operators of fixed R-charge. Hence, and in contrast to \cite{Dabholkar:2007ey,Giribet:2007wp}, we do not introduce isospin variables in the SU(2) sector.} For the operators at hand, the sum over $m$ in the analog of Eqs.~\eqref{ExpSL2} and \eqref{flowedxbasis} factorizes between fermionic and bosonic contributions, leading to expressions of the following form: 
\begin{subequations}
\label{chiralxbasisAdS3}
\begin{eqnarray}
    \Ww_j^{\w}(x) &=& e^{-\varphi} \psi^{\w}(x) e^{i \w H_2} V_j^{\w}(x)
    V^{\w}_{j-1,j-1} \, ,
    \\ [1ex]
    \Xx_j^{\w}(x) &=& e^{-\varphi} \psi^{\w-1} (x) e^{i (\w+1) H_2} V_j^{\w}(x)
    V^{\w}_{j-1,j-1} \, ,
    \\ [1ex]
    \Yy_j^{\w,\vep_4}(x) &=& e^{-\frac{\varphi}{2}}
    S^{\w}(x) V_j^{\w}(x)
    V^{\w}_{j-1,j-1} e^{\frac{i \vep_4}{2} \left( H_4 + H_5 \right)} \, .
\end{eqnarray}
\end{subequations}
Here $\psi^{\w}(x)$ and $S^{\w}(x)$ are defined as follows. First, note that the fermions $\psi^a$ introduced in \eqref{fermionsSL2SU2def}, which generate an affine $\widehat{\sl(2,\R)}_{-2}$ algebra with level $k_\psi = -2$, constitute affine primaries with spin $J_\psi=-1$, on which, however, the zero-mode currents act as in \eqref{zeromodesSL2} but with\footnote{This convention is perhaps more natural from the SL(2,$\R$) point of view, but we have decided to employ the conventions used in the most relevant literature for us, i.e.~\cite{Maldacena:2000hw,Maldacena:2001km,Dabholkar:2007ey,Giribet:2007wp}.} $J \to 1-J$, i.e. $J^{\pm}_0|J,m\pm 1\rangle =  (m\pm J) |J,m\pm 1\rangle$. As a consequence, and in contrast with what happens with bosonic primaries, the action $J_0^+$ on $\psi^-$, the lowest-weight state, is truncated. Identifying $\psi^{\w=0}(0) = \psi^-$, the resulting $x$-basis operator has only three terms:
\begin{equation}
\label{eq:psi-w}
    \psi^{\w=0}(x) \,\equiv\, \psi^-(x) \,=\, e^{x J_0^+} \psi^- e^{-x J_0^+} 
    \,=\, \psi^- - 2x \:\! \psi^3 + x^2 \psi^+ \, .
\end{equation}
Of course, we already knew the action of the currents on $\psi^a$ from \eqref{fermionsSL2SU2def}, but the advantage of the above discussion is that it extends to the spectrally flowed sectors. Indeed, $\psi^{\w} (0) = \sqrt{n_5} e^{-i(1+\w) H_1}$ is the lowest-weight component of a spin $J_\psi^\w = -1-\w$ field. The corresponding $x$-basis operator is of the form 
\begin{equation}
    \psi^{\w} (x) \,\equiv\, \sqrt{n_5} \; e^{x J_0^+} e^{-i(1+\w) H_1} e^{-x J_0^+} \, ,
    \label{psiwxbasis}
\end{equation}
and contains $1-2J_\psi^\w = 2\w +3$ terms. 

Similarly, the spin field \eqref{spinwmbasis} is the lowest-weight component of a representation with SL(2,$\R$) and SU(2) spins $(J^{\w}_S,J'^{\w}_S)=(-\frac{1}{2}-\w,\frac{1}{2}+\w)$, such that the $x$-basis operator is
\begin{equation}
    S^{\w}(x) \,\equiv\, e^{x J_0^+} S_{- + -}^\w e^{-x J_0^+} \, ,
    \label{spinwxbasis}
\end{equation}
and contains $2(1+\w)$ terms. 

We thus have three types of vertices, $\Ww_j^\w(x,z)$, $\Xx_j^\w(x,z)$ and $\Yy_j^{\w,\pm}(x,z)$, which correspond to local chiral primary operators of the boundary theory. As mentioned before, these should be completed with analogous antiholomorphic excitations, which have been omitted in the presentation. As discussed around Eqs.\;\eqref{eq:h-sf} and \eqref{eq:q-nos-rr}, their boundary weights are given by
\begin{equation}
\label{jwandhAdS3}
    h \left[\Ww_j^\w\right] = j_\w-1 \qqquad
    h \left[\Yy_j^{\w,\pm}\right] = j_\w-\frac{1}{2} \qqquad 
    h \left[\Xx_j^\w\right] = j_\w \, ,
\end{equation}
where $j = j'+1$ and so
\begin{equation}
 j_\w = j + \frac{n_5}{2}\w , \qquad 
    j = 1,\frac{3}{2},\dots,
\frac{n_5}{2}
,\qquad
\w = 0,1,\dots \, .
\label{jwrangecc}
\end{equation}

Up to normalization, which will be fixed shortly, these operators are identified with the chiral primaries of the holographic CFT listed in Eq.~\eqref{D1D5CFTweights3}.
In the $\Yy_j^{\w,\vep_4}$ tower, $\vep_4=\pm$ is identified with the boundary quantum number $\dot{A}$ in \eqref{ccops}.
Note also that the $\Ww_j^\w$ tower starts with the identity operator of the boundary theory.\footnote{As discussed in \cite{Kutasov:1999xu}, this is subtle, since the operator is actually a spectral-flow-sector dependent constant. This subtlety is related to the fact that spectral flow charge is not conserved in $n$-point functions with $n\geq 3$, and was resolved in \cite{Kim:2015gak} by performing a Legendre transform to the microcanonical ensemble, in which the total number of fundamental strings is fixed.}
The dictionary is summarized in Table \ref{ccdictionary}. 
\begin{table}[th]
\centering
\begin{tabular}{|c|c|c|c|}
\hline
Worldsheet & Weight $h$ & Twist $n$ & Dual Operator  \\ \hline
$\Ww^\w_{j}$&$j_{\w}-1 $& $2j_{\w}-1$ &$O^{-}_{n}$ \\ \hline
$\Yy^{\w,\vep_4}_{j}$&$j_{\w}-\half $&$2j_{\w}-1$&$O^{\dot{A}}_{n}$ \\ \hline
$\Xx^\w_{j}$&$j_{\w}$&$2j_{\w}-1$&$O^{+}_{n}$ \\ \hline
\end{tabular}
\caption{Dictionary between worldsheet vertex operators and 
chiral primaries of the holographically dual CFT. Here $j_\w = j + \frac{n_5}{2} \w$, $j=1,\frac{3}{2},\dots,\frac{n_5}{2}$, and $\w = 0,1,\dots$}
\label{ccdictionary}
\end{table}

Although most of the chiral primaries of the holographic CFT are accounted for by considering the ranges given in~\eqref{jwrangecc}, it is known that those belonging to the the $n$-twisted sectors with $n= p n_5$ with $p \in  \mathbb{N}$ are still missing~\cite{Dabholkar:2007ey,Giribet:2007wp}. 
These would correspond to operators sitting at the boundary of the allowed range of $j$ in Eq.\;\eqref{Djrange}, at which the spectrum becomes degenerate and the continuous representations appear~\cite{Teschner:1997ft,Giveon:2001up}. 
The absence of these states in the worldsheet spectrum has been related to the fact that the NS5-F1 model sits at a singular point in the moduli space where all RR modes are turned off~\cite{Seiberg:1999xz}. 

The twist $n$ of the holographic CFT operators is identified as~\cite{Dabholkar:2007ey,Giribet:2007wp} 
\begin{equation}
 n \,=\, 2 j - 1 + n_5 \w \,. \label{njw}
\end{equation}
Let us make a side comment regarding the limit in which there is only a single NS5 brane sourcing the background, $n_5=1$. This model is special in that it corresponds to the tensionless limit of the theory. It has to be treated with care since the usual RNS formalism outlined above breaks down due to the fact that the bosonic SU(2) level would become negative. It was shown in~\cite{Eberhardt:2018ouy,Eberhardt:2020bgq} that for $n_5=1$ the worldsheet theory is exactly dual to the supersymmetric symmetric orbifold $(\TT^4)^{n_1}/S_{n_1}$. In this model, the discrete series is absent, the spectrum truncates to $j=1/2$, physical states have $\w>0$, and the spectral flow charge is identified with $n$, i.e.~$n=\w$. Eq.~\eqref{njw} is the known generalization of this relation for $n_5>1$.

In order to fix the normalization of the operators, we compute their two-point functions. Making use of 
\begin{align}
    &\langle \psi^{\w_1} (x_1) \psi^{\w_2} (x_2)  \rangle \times {\rm c.c.} \,=\, x_{12}^{2(\w_1+1)}
    \langle e^{-i(1+\w_1)H_1} \:\!
    e^{i(1-\w_2) H_1} \rangle \times {\rm c.c.} \,=\, \delta_{\w_1,-\w_2}
    \frac{|x_{12}|^{4(\w_1+1)}}{|z_{12}|^{
    2(\w_1+1)^2}}\,  , \nn
    \\
    &\langle S^{\w_1,+} (x_1) S^{\w_2,-} (x_2)  \rangle \times {\rm c.c.}  \,=\, \frac{x_{12}^{2\w_1+2}}{z_{12}}
    \langle S_{-+-}^{\w_1}
    S_{+-+}^{\w_2} \rangle \times {\rm c.c.} \,=\, 
    \delta_{\w_1,-\w_2} \frac{|x_{12}|^{4\w_1+2}}{|z_{12}|^{
    4\w_1(\w_1+1)+\frac{5}{2}}}\, , \nn
\end{align}
we obtain 
\begin{eqnarray}
\langle \Ww_{j_1}^{\w_1} (x_1,z_1) \Ww_{j_2}^{\w_2} (x_2,z_2)\rangle &\,=\,&  \frac{n_5^2\:\! B(j_1)}{16}\frac{\delta(j_1-j_2)\:\! \delta_{\w_1,-\w_2} }{|x_{12}|^{4(j_1-1)+2n_5\w_1} |z_{12}|^4} \, ,  \\  \langle \Xx_{j_1}^{\w_1} (x_1,z_1) \Xx_{j_2}^{\w_2} (x_2,z_2)\rangle &\,=\,&  \frac{n_5^2 \:\! B(j_1)}{16} \frac{\delta(j_1-j_2)\:\! \delta_{\w_1,-\w_2}}{|x_{12}|^{4j_1+2n_5\w_1}|z_{12}|^4} \, ,  \\
\langle \Yy_{j_1}^{\w_1,\pm} (x_1,z_1) \Yy_{j_2}^{\w_2,\mp} (x_2,z_2)\rangle &\,=\,&  \frac{n_5
\:\!B(j_1)}{(2j_1-1+n_5 \w_1)^2}\frac{\delta(j_1-j_2)\:\! \delta_{\w_1,-\w_2}}{|x_{12}|^{4j_1+2(n_5\w_1-1)}|z_{12}|^4}\, . \label{eq:2pt-ads-Y} 
\end{eqnarray}
Here we have used that in the spectrally flowed R sectors the denominator in the extra factor of the corresponding vertex operator in the ``$-\frac{3}{2}$'' picture is shifted as $2j-1 \to 2j-1+ n_5 \w$ as compared to \eqref{Ypicture3half}. Moreover, the $V_{\mathrm{conf}}$ factor in Eq.~\eqref{2ptwSL2} is cancelled by the pole appearing in \eqref{2ptMbasisSL2} upon setting $m_1=j_1$. 

The string two-point function is then obtained by including an extra factor $g_s^{-2}\sim n_1/n_5$ as usual in string perturbation theory, fixing $z_1=0$ and $z_2=1$, and  dividing by a volume of the conformal group that leaves such worldsheet insertions fixed. As discussed in \cite{Kutasov:1999xu,Maldacena:2001km}, this cancels the divergence coming from $\delta(j_1-j_2)$, leaving a constant $j$-dependent factor of the form $(2j-1+n_5 \w)$. As a consequence, the holographic dictionary reads 
\begin{align}
    O_n^{--}(x,\bar{x}) &\leftrightarrow 
    A_{\mathrm{NS}}(j,\w)\Ww_j^\w(x,\bar{x}), \qquad\qquad
    O_n^{++}(x) \leftrightarrow 
    A_{\mathrm{NS}}(j,\w)\Xx_j^\w(x,\bar{x}), 
    \\[1mm]
& \qquad \qquad \qquad  O_n^{\dot{A}\dot{B}} \leftrightarrow 
    A_{\mathrm{R}}(j,\w) \Yy_j^{\w,\dot{A}\dot{B}}(x,\bar{x}),
\end{align}
with $n$ related to the worldsheet quantum numbers as in \eqref{njw} and where~\cite{Gaberdiel:2007vu,Dabholkar:2007ey}
\begin{equation}
    A_{\mathrm{NS}}(j,\w) = 
    \frac{4g_s}{\sqrt{n_5^2\, B(j)(2j-1+n_5\w)}} \qqquad
    A_{\mathrm{R}}(j,\w) = 
    g_s\sqrt{\frac{(2j-1+n_5\w)}{n_5 \, B(j)}} \, .
\end{equation}
Of course, this identification is only expected to hold at small string coupling, i.e.~for $n_1 \gg n_5$. 
Analysis and comparison of boundary and worldsheet three-point functions was carried out in~\cite{Pakman:2007hn,Dabholkar:2007ey, Gaberdiel:2007vu}.

\section{Null-gauged description and worldsheet spectrum} 
\label{sec:nullgaugedmodel}

We now proceed to describe massless excitations in the worldsheet theories associated with the heavy backgrounds we study. We describe in detail how to obtain the low-lying physical states via BRST quantization in these null-gauged models, both in the NSNS and RR sectors. 
In the subset of the backgrounds that preserve some supersymmetry,
we discuss the BPS light excitations.

In the full null-gauged models, these massless vertex operators describe linearized fluctuations around the full asymptotically linear dilaton solutions describing the heavy states, and so can be thought of as worldsheet representatives of light states belonging to the Little String Theory living on the NS5 branes.

Our main interest in this work will be computing correlators in the IR AdS$_3$ limit, in which we have reviewed the fact that the backgrounds are related to orbifolded AdS$_3\times \SSS^3\times \TT^4$ via a spacetime spectral flow large coordinate transformation.  We describe how the AdS$_3$ limit can be taken on the worldsheet vertex operators. This leads to states that can be understood holographically, in the spacetime (fractionally) spectrally flowed frame defining the heavy background, as discussed around Eq.\;\eqref{SFchargesHCFT}.

\subsection{BRST quantization}
\label{sec:BRSTquantization}

We start by reviewing the quantization of the class of worldsheet coset models introduced in Section~\ref{Section2}, which describe the propagation of superstrings in the JMaRT backgrounds and their (BPS and/or two-charge) limits~\cite{Martinec:2017ztd,Martinec:2018nco,Martinec:2020gkv,Bufalini:2021ndn}.

Before gauging, we have the WZW model associated to the (10+2)-dimensional group manifold SL(2,$\R$)$\,\times\,$SU(2)$\, \times \,  \R_t \, \times \,$U(1)$_y \, \times \, $U(1)$^4$ as introduced in~\eqref{eq: coset}. This is described simply by adding the extra time direction $t$ and spatial circle $y$ to the matter content employed in the previous section, together with the corresponding fermionic partners $\lambda^t$ and $\lambda^y$. The latter are bosonized using a canonically normalised scalar $H_6$ as
\begin{align}
      \lambda_t & = \half \left( e^{i H_6} - e^{-i H_6}\right) \qqquad \lambda_y = \half \left( e^{i H_6} + e^{-i H_6}\right) \,,\\
  & \hspace{1cm}  i \:\! \der H_{6}  = 2\,  \lambda^t\lambda^y \qqquad  H_6^{\dagger} = - H_6 \,.
\end{align} 
The holomorphic parts of their OPEs are 
\begin{equation}
    -t(z)t(w) \sim  
    y(z)y(w)\sim - \frac{1}{2}\log(z-w) \, , \quad~
    -\lambda^t(z)\lambda^t(w)\sim  
    \lambda^y(z)\lambda^y(w)\sim \frac12\frac{1}{(z-w)} \,,  
\end{equation}
and they give additional free field contributions to the matter $T$ and $G$ in \eqref{TAdS3S3T4def} and \eqref{GAdS3S3T4def}, namely
\begin{equation}
    T_{(ty)} \,=\, 
    \der t \der t - \der y \der y
     \qqquad 
    G_{(ty)} \,=\, 2 i \left(-\lambda^t \der t + \lambda^y \der y\right). 
\end{equation}
Introducing $P_{t,L} = i \:\! \der t\,$, $P_{t,R} = i \:\! \bar\der t\,$, $P_{y,L} = i\:\! \der y\,$, and $P_{y,R} = i\:\! \bar\der y$, we gauge the chiral null currents\footnote{The symbol $J$ for the current operators should not be confused with the total SL(2,$\R$) spin that has appeared in Section~\ref{sec:vertexops}. The meaning should be clear from the context.}
\begin{equation}
   J \, =\, i{\cal{J}} \,=\, J^3 + l_2 K^3 + l_3 P_{t,L} + l_4 P_{y,L} \ , \ \quad
   \bar{J} \, =\, i\bar{{\cal{J}}} \,=\, \bar{J}^3 + r_2 \bar{K}^3 + r_3 P_{t,R} + r_4 P_{y,R} \, , 
    \label{Jcurrents}
\end{equation}
which are the quantum operator versions of the classical currents in Eq.~\eqref{eq:jjbar-def}. 
The supersymmetric partners of the currents $J$ and $\bar J$ are
\begin{equation}
\label{fermioniccurrents}
    \boldsymbol{\lambda} = \psi^3 + l_2 \chi^3 + l_3 \lambda^t + l_4 \lambda^y 
    \qqquad 
    \bar{\boldsymbol{\lambda}} = \bar{\psi}^3 + r_2 \bar{\chi}^3 + r_3 \bar{\lambda}^t + r_4 \bar{\lambda}^y \,.
\end{equation}

To perform the null gauging, one introduces additional fermionic and bosonic first-order ghosts, denoted by $(\tilde{b},\tilde{c})$ and $(\tilde{\beta},\tilde{\gamma})$, with conformal weights $\Delta[\tilde{c}] = 0$ and $\Delta[\tilde{\gamma}] = 1/2$~\cite{Martinec:2020gkv}. The central charges $c_{\tilde{b}\tilde{c}} = -2$ and $c_{\tilde{\beta}\tilde{\gamma}} = -1$ cancel the additional matter contribution $c_{ty} = 3$. The $(\tilde{\beta},\tilde{\gamma})$ system has no background charge and is bosonized via
\begin{equation}
	\tilde{\beta} = e^{-\tilde{\varphi}} \del \tilde{\xi} \qqquad   \tilde{\gamma} = \tilde{\eta} \, e^{\tilde{\varphi}} \,.
\end{equation}

We shall momentarily introduce a modified BRST charge that imposes invariance under the action of the null currents \eqref{Jcurrents} and their supersymmetric partners \eqref{fermioniccurrents}. Physical operators in 9+1 dimensions will be given by states of the ungauged (10+2)-dimensional WZW model that survive the gauging procedure~\cite{Martinec:2020gkv}. Of course, and as we shall see shortly, the Virasoro conditions and the expressions of the BRST-exact states will be modified accordingly.
We consider a set of mutually local operators before the gaugings, i.e.~we perform the analog of the GSO projection in the (10+2)-dimensional model. We thereby obtain get a tachyon-free spectrum in the gauged models.

Underlying this procedure is the fact that for the case of chiral null gaugings, the Polyakov-Wiegmann identity allows one to rewrite the gauged action of the \textit{downstairs} model into a form identical to that of the \textit{upstairs} model in terms of a new gauge-invariant variables~\cite{Chung:1992mj, Israel:2004ir}, see also~\cite{Bufalini:2021ndn}.
This is achieved at the level of the path integral by means of a field redefinition with a Jacobian that is almost trivial except for a factor which, when exponentiated, gives rise to the additional ghost fields described above. 

Explicitly, physical operators in the coset model are defined by the cohomology classes of the BRST charge~\cite{Martinec:2020gkv}  
\begin{equation}
\label{BRSTnull}
    \Qq = \oint dz \left[ c \left( T + T_{\beta \gamma \bar{\beta} \bar{\gamma} } \right) + \gamma G + \tilde{c} J + \tilde{\gamma} \boldsymbol{\lambda} + \text{ghosts} \right] \:\! ,
\end{equation}
where the last two terms implement the null-gauging procedure. Whether the resulting spectrum is supersymmetric or not depends on whether some linear combination(s) of the following supercharges are BRST-invariant~\cite{Martinec:2020gkv},
\begin{equation}
    Q_\vep = \oint dz \, e^{-\left(\varphi - \tvphi\right)/2} S_\vep  \qqquad S_\vep = \exp \left(\frac{i}{2} 
    \sum_{I=1}^{6}\vep_I H_I\right).
    \label{supercharges2}
\end{equation}
For these to be mutually local, we impose the analog of the GSO projection in (10+2) dimensions,
\begin{equation}
    \prod_{I=1}^6 \vep_I = 1 \, . 
    \label{locality12d}
\end{equation}
We shall discuss the conditions for spacetime supersymmetry after we have analyzed more general Ramond sector vertex operators, around Eq.\;\eqref{susyrestricparam-0}. For now, we emphasize that only a subset of the backgrounds we consider preserve some spacetime supersymmetry.

\subsection{The unflowed NS sector}

We now analyze physical NS sector states of the gauged models, focusing on states with no spectral flow charges in SL(2,$\R$) or SU(2), and no winding charge around the $y$-circle.
As usual, the lightest physical operators come with a single fermionic excitation on top of the tachyon state
\begin{equation}
    \Tt_{\jsl,\msl,\jsu,\msu} \,=\, 
    e^{-\varphi} V_{j,m} V'_{j',m'} e^{i (-E\,t + P_y y)} \, . 
    \label{tachyon}
\end{equation} 
Note that since $t$ is a non-compact direction and $\w_y = 0$, both $E$ and $P_y$ are identical on the left and on the right sectors. For massless states, the $L_0$ and $\bar{L}_0 $ Virasoro constraints both read
\begin{equation}
\label{nullgaugedVirasoro1}
   0 \,=\, - \frac{j(j-1)}{n_5} + \frac{j'(j'+1)}{n_5} 
    - \frac{1}{4} E^2 + \frac{1}{4} P_{y}^2 \,.
\end{equation}
Moreover, operators are uncharged with respect to the null-currents $J,\bar{J}$ in \eqref{Jcurrents} if and only if their quantum numbers are related by 
\begin{equation}
\label{nullgaugeconstr}
0 = m + l_2 \, m' + \frac{l_3}{2} E + \frac{l_4}{2} P_{y} \qqquad
0= \bar{m} + r_2\, \bar{m}' + \frac{r_3}{2} E +\frac{r_4}{2} P_{y} \;.
\end{equation}
We will work in the canonical ``$-1$'' picture for the $\varphi$ ghost. On the other hand, the fact that $\tilde{\varphi}$ has background charge $Q_{\tilde{\varphi}}=0$ allows us to build NS states directly at $\tilde{\varphi}$-picture zero. BRST-closed operators must then have a vanishing second-order pole in their OPE with the supercurrent $G$, and vanishing first-order pole in their OPE with the fermionic current $\boldsymbol{\lambda}$ given in \eqref{fermioniccurrents}.

As can be expected from the fact that the $\TT^4$ is untouched by the gaugings, the simplest solutions are the 6D scalars 
\begin{equation}
    \Vv^{i}_{\jsl,\msl,\jsu,\msu} =  e^{-\varphi} \lambda^i \Vsl_{\jsl,\msl} \Vsu_{\jsu,\msu} \, e^{i (-Et + P_y y)} \qqquad i=6,\dots,9.
    \label{6DscalarsNSNS}
\end{equation}
These are direct analogs of the global AdS$_3$ states defined in Eq.~\eqref{T4statesAdS3}. They were considered in detail in \cite{Martinec:2018nco}, and their energies were matched with those of the minimally coupled scalar perturbations on top of the JMaRT background as computed in supergravity.

The remaining massless vertex operators will constitute the beginning of the main new results of this work.
They are slightly more involved to construct, due to the fact that their polarization lies in a direction in which the null currents act non-trivially.  An important consequence is that the raising/lowering operators $J^\pm_0$ and $K^\pm_0$ do not commute with the BRST charge $\Qq$ anymore. So, unlike in global AdS$_3\times \SSS^3\times \TT^4$ as reviewed in Section \ref{sec:AdS3S3T4}, physical states need not have definite SL(2,$\R$) and SU(2) spins. 
They will, however, have definite projections $m,\bar{m},m'$ and $\bar{m}'$, and also well-defined  energy $E$ and momentum $P_y$.

This situation is
a consequence of the fact that the AdS$_3\times \SSS^3$ isometries are absent in the  asymptotically linear dilaton geometry. Nevertheless, these isometries are restored in the IR, by taking $R_y$ large while keeping $ER_y$ and $P_y R_y$ fixed, see  Eqs.~\eqref{FixedAdSlimit}--\eqref{eq:ads-JMaRT-dil}. In this regime, the vertex operators of the gauged models will reduce to the AdS$_3\times \SSS^3$ expressions in Eqs.~\eqref{WstatesAdS3} and \eqref{XstatesAdS3}.

Let us consider a generic linear combination of NS sector vertex operators,
\begin{equation}
    e^{-\varphi} \left[\left(c^r \psi^r V_{j,m-r} V'_{j',m'} +
    d^r \psi^r V_{j,m} V'_{j',m'-r}\right) + 
    \left(c^t \lambda^t + c^y \lambda^y\right) V_{j,m} V'_{j',m'} \right] e^{i (-Et + P_y y)},
\end{equation}
where the notation mirrors that of the AdS$_3\times \SSS^3$ expressions in  Eq.~\eqref{linearcombsKLL}, in particular summation over $r=+1,-1,0$ is implicit, with ``$0$'' corresponding to the ``$3$'' direction of the respective algebras. Of these eight degrees of freedom, two are removed by the conditions arising from the $G$ and $\boldsymbol{\lambda}$ terms in the BRST charge, which respectively read
\begin{equation}
0 \,=\, m c^3+ (m-j)c^+ + (m+j)c^- +
        m' d^3 + (j'+m')d^+ + (j'-m')d^-
        +c^t\frac{E}{2} + c^y \frac{P_y}{2} ,   
\end{equation}         
and
\begin{equation}
    0 \,=\, n_5 \left(-c^3+l_2 d^3\right) - l_3 c^t + l_4 c^y.
\end{equation}        
This leaves six states, out of which two turn out to be BRST-exact. The first exact state comes, as usual, from the action of $G$ on the tachyon operator \eqref{tachyon}, while the second one has no global AdS$_3$ counterpart and appears due to the action of $\boldsymbol{\lambda}$ on the same state. Their explicit expressions are 
\begin{eqnarray}
    \Phi_{G} \,=\, e^{-\varphi} && \left[ 
    \frac{2}{n_5}\half V'_{j',m'} \left( 
    (m-j+1)\psi^-V_{j,m+1} + 
    (m+j-1)\psi^+V_{j,m-1}  - 2 m \psi^3  V_{j,m}\right) \right. \nn \\[1ex]
    && \,\,\, + \frac{2}{n_5}\half V_{j,m} \left( (j'+m'+1) \chi^-V'_{j',m'+1} +
    (j'-m'+1) \chi^+V'_{j',m'-1} 
    + 2 m' \chi^3  V'_{j',m'} \right)  \nn \\[1ex]
    && \,\,\, + \left.
    \half \left(- E \lambda^t + P_y \lambda^y \right) V_{j,m}V'_{j',m'} 
    \right]e^{i (-E\,t + P_y y)},
\end{eqnarray}
and 
\begin{equation}
    \Phi_{\boldsymbol{\lambda}} \,=\, 
    e^{-\varphi} \left[
    \psi^3 + l_2 \chi^3 + l_3 \lambda^t + l_4 \lambda^y
    \right] V_{j,m}V'_{j',m'} \: e^{i (-E\,t + P_y y)},
\end{equation}
respectively.
Such states are trivially BRST invariant since $G$ and $\boldsymbol{\lambda}$ square to the Virasoro constraint \eqref{nullgaugedVirasoro1} and the null condition \eqref{eq: null gauge constraints}, and the relevant term in their product is $G(z) \, \boldsymbol{\lambda}(0) \sim \boldsymbol{\lambda}(z) \, G(0) \sim J(0)/z$, whose action vanishes by means of the condition~\eqref{nullgaugeconstr}. In the end, we are left with four physical vertex operators to add to the four from the $\TT^4$ directions to give the correct eight polarizations in the holomorphic sector in 9+1 dimensions. 

We choose a basis for these four physical vertex operators such that, in the AdS$_3$ limit, they reduce to the basis of global AdS$_3$ vertex operators described around Eq.~\eqref{linearcombsKLL}.
We thus obtain 
\begin{subequations}
\label{vectorNSgauged}
\begin{eqnarray}
    \Ww^\vep &\,=\,& e^{-\varphi}\left[(\psi V_{j})_{j+\vep,m} V'_{j',m'} + \left(c^t_{\vep}\, \lambda^t + c^y_{\vep}\, \lambda^y\right) V_{j,m}V'_{j',m'}\right] e^{i \left(-E\,t + P_y  y\right)}, \\
    \Xx^\vep &\,=\,& e^{-\varphi} \left[V_{j,m} (\chi V'_{j'})_{j'+\vep,m'}  + \left(d^t_{\vep} \,  \lambda^t + d^y_{\vep} \, \lambda^y\right) V_{j,m}V'_{j',m'}\right] e^{i \left(-E \,t + P_y  y\right)},
\end{eqnarray}
\end{subequations}
where the SL(2,$\R$) and SU(2) coefficients are those given in \eqref{linearcombsKLL}--\eqref{cdcoefs}, while the novel ones are\footnote{The coefficients $c^{t,y}$ and $d^{t,y}$ were reported in the Letter~\cite{Bufalini:2022wyp} with a slightly different notation, related by $c^{t,y}_{\text{there}} = c^{t,y}_{\varepsilon, \text{here}}/c^{3}_\varepsilon$, and likewise for $d^{t,y}$.}
\begin{equation}
    c^t_{\vep} \,=\, -c^3_{\vep} \frac{ n_5 P_y}{l_4 E + l_3 P_y} \qqquad 
    c^y_{\vep} \,=\, c^3_{\vep} \frac{ n_5 E}{l_4 E + l_3 P_y} \,,
\end{equation}
\begin{equation}
    d^t_{\vep} \,=\, d^3_{\vep} \frac{ n_5 l_2 P_y}{l_4 E + l_3 P_y} \qqquad ~~
    d^y_{\vep} \,=\, -d^3_{\vep} \frac{ n_5 l_2  E}{l_4 E + l_3 P_y} \,.
\end{equation}
By construction, the resulting states are polarized transverse to the gauge directions.
As anticipated, they are built out of a linear combination of terms of spin $j$ and $j+\vep$ ($j'$ and $j'+\vep$). Moreover, at leading order in the large $R_y$ expansion, the coefficients in the $t,y$ directions go to zero, since $E,P_y \sim {\cal{O}}(1/R_y)$, $l_{3,4} \sim \Oo(R_y)$, and $l_2 \sim \Oo(1)$.

\subsection{The unflowed R sector}
\label{sec:nullRRsector}

We now describe the physical states in the R sector of the null-gauged model. The computation turns out to be more involved than in the NS sector, since the spin fields necessarily involve all $\vep$-chiralities.
As a consequence, we will not find a situation akin to \eqref{vectorNSgauged} in which a subset of coefficients are exactly those of the global AdS$_3\times\SSS^3$ operators. However, 
we will again show that in the AdS$_3$ limit the vertex operators will reduce to their global AdS$_3\times\SSS^3$ counterparts.

We introduce AdS$_3 \times \SSS^3 $ and $\R_t\times \SSS^1_y\times \TT^4$ spin fields,\footnote{The order of the spin fields in $\mathcal{S}_{\vep_6 \vep_4 \vep_5}$ has been chosen for convenience in order to reduce clutter in computations involving cocycle factors.}
\begin{equation}
S_{\vep_1 \vep_2 \vep_3} =
e^{\frac{i}{2}\left(
    \vep_1 H_1 +
    \vep_2 H_2 + \vep_3 H_3
    \right)}
\qqquad 
    \mathcal{S}_{\vep_6 \vep_4 \vep_5} = 
    e^{\frac{i}{2}\left(
    \vep_6 H_6 +
    \vep_4 H_4 + \vep_5 H_5\right)}\,.
\end{equation}
Recalling the definition of the AdS$_3$ chirality $\vep$ and the mutual locality / chiral GSO projection in (10+2) dimensions, \eqref{locality12d}, we substitute away $\vep_3$ and $\vep_6$ via
\begin{equation}
    \vep_3 = \vep \vep_1 \vep_2  \qqquad  \vep_6 = \vep \vep_4 \vep_5 \,.
    \label{nullchirality}
\end{equation}
The $H_{4,5}$ exponentials are spectators under the action of $\Qq$, so the parameters $\vep_4, \vep_5$ will label the vertex operators. For fixed $\vep_4, \vep_5$, we consider $\vep_6$ to be controlled by $\varepsilon$ through the second equation in \eqref{nullchirality}, and we will form linear combinations of different values over $\vep_1$, $\vep_2$, $\vep$.

We work with vertex operators in ghost pictures  $(q_\varphi,q_{\tilde{\varphi}}) = (-\half, +\half)$, for which the $\boldsymbol{\lambda}$-constraint is non-trival, while there is no need to worry about BRST-exact states. 
We thus make an ansatz for R sector vertices of the following form:
\begin{equation}
 \label{RRnullAnsatz}
	\Yy^{\vep_4,\vep_5} \,=\, e^{-(\varphi- \tilde{\varphi})/2}
	\sum_{\vep_1, \vep_2,\vep} F^{\vep}_{\vep_1 \vep_2 \vep_4 \vep_5} 
	S_{\vep_1 \vep_2 \vep_3} 
	\mathcal{S}_{\vep_6 \vep_4 \vep_5} V_{j,m-\frac{\vep_1}{2}}V'_{j',m'-\frac{\vep_2}{2}}
	e^{i(-E \:\! t + P_y \:\! y)} \,,
\end{equation}
Note that the coefficients $F^{\vep}_{\vep_1 \vep_2 \vep_4 \vep_5} $ are not determined by the representation theory of SL(2,$\,\R$)$\times$SU(2), since the states will not in general have definite spin.

The $c\:\!T$  and $\tilde{c}\:\!J$ terms of the BRST operator $\Qq$ \eqref{BRSTnull}  act as in the NS sector. Hence the unflowed, non-winding states of the R sector also satisfy both  the Virasoro condition \eqref{nullgaugedVirasoro1} and the bosonic null-gauge constraint  \eqref{nullgaugeconstr}. 

Next, the $e^{\tilde{\varphi}}\boldsymbol{\lambda}$ term in $\Qq$ leaves $\vep_{1,2,4,5}$ unchanged, so for this term we can treat  $\vep_{1,2,4,5}$ as fixed, and focus on the sum over $\varepsilon=\pm$. 
The resulting constraints on $F^{\pm}_{\vep_1 \vep_2 \vep_4 \vep_5}$ form a two-dimensional homogeneous linear system, which is degenerate due to the null condition on the gauge parameters, Eq.~\eqref{eq: null gauge constraints}. For each choice of $\vep_{1,2,4,5}$, we have
\begin{align}
\begin{aligned}
\label{LambdaFrelation}
	 (l_3 + \vep_4 \vep_5  l_4) F^-_{\vep_1 \vep_2 \vep_4 \vep_5} \,&-\, 
	i \sqrt{n_5} \,  (1 -  \vep_1 \vep_2 l_2) F^+_{\vep_1 \vep_2 \vep_4 \vep_5} \,=\,0 \,, \\
		 i \sqrt{n_5}(1 +  \vep_1 \vep_2 l_2) F^-_{\vep_1 \vep_2 \vep_4 \vep_5} \,&-\, 
	(l_3 - \vep_4 \vep_5  l_4) F^+_{\vep_1 \vep_2 \vep_4 \vep_5} \,=\,0 \,.
\end{aligned}
\end{align}
These constraints halve the degrees of freedom. When $|l_2|=1$ (and so $|l_3|=|l_4|$), some of the $F^\varepsilon_{\vep_1 \vep_2 \vep_4 \vep_5}$ get set to zero. For a given $\vep_{1,2,4,5}$, when neither of $F^\pm_{\vep_1 \vep_2 \vep_4 \vep_5}$ get set to zero, their ratio ${F^-_{\vep_1 \vep_2 \vep_4 \vep_5}}/{F^+_{\vep_1 \vep_2 \vep_4 \vep_5}}$ becomes determined.
So the 32 d.o.f.~remaining after imposing GSO in (10+2) dimensions have now become 16, corresponding to $\vep_{1,2,4,5}$ in our parameterization.

Let us pause to discuss how Eq.~\eqref{LambdaFrelation} behaves in the large $R_y$ limit. We have $l_2\sim \Oo(1)$ while $l_3 + l_4 \sim \Oo(R_y)$ and $l_3 - l_4 \sim \Oo(1/R_y)$, from \eqref{integersCFT1odd}--\eqref{L32def}.
When $\vep_4  \vep_5 = +1$, we obtain $F^+ \sim \Oo(1)$ and $F^- \sim \Oo(1/R_y)$, so at leading order in large $R_y$ we obtain a purely positive chirality operator. Similarly, when $\vep_4  \vep_5 = -1$, at leading order in large $R_y$ we obtain a purely negative chirality operator.
So we obtain operators of definite AdS$_3\times \SSS^3$ chirality $\vep$, with $\vep_4  \vep_5=\vep$, exactly as in Section \ref{sec:AdS3S3T4}, see Eq.~\eqref{YyAdS3}. As before, one of $\vep_4$ or $\vep_5$ remains unfixed, say $\vep_4$.

We now examine the action of $e^{\varphi}G$ on the R vertex operator ansatz \eqref{RRnullAnsatz}. This will reduce the remaining 16 degrees of freedom to the correct 8 physical polarizations in the holomorphic sector. It leads to the following set of equations (we suppress the $\vep_4,\vep_5$ subscripts on the RHS for ease of notation):
\begin{align}
   \Bb^\vep_{\vep_1 \vep_2 \vep_4 \vep_5} & \,\equiv\, \left( m + \vep_1 j - \frac{\vep_1}{2}\right) F^\vep_{(-\vep_1)\vep_2}  + i \, \vep_1 \vep_2 \,  \left( j' - \vep_2m' + \half \right) F^\vep_{\vep_1(-\vep_2)}  \nn \\
     & \qquad\qquad\quad~  - (\vep m + \vep_1 \vep_2 m') F^\vep_{\vep_1\vep_2}  + \frac{i \sqrt{n_5}}{2}
    \left( \vep_4\vep_5 P - \vep \, E \right)
    F^{(-\vep)}_{\vep_1\vep_2} \;=\; 0 \,.
    \label{BRSTFcompact}
\end{align}
Comparing to the AdS$_3 \times \SSS^3$ BRST condition, the only new term is the fourth and final one, proportional to $F^{(-\vep)}_{\vep_1\vep_2}$, which has the effect of mixing the $\varepsilon$ chiralities.
The first three terms are unchanged from the AdS$_3 \times \SSS^3$ BRST condition, so the AdS$_3 \times \SSS^3$ limit of this condition is simply to drop the fourth term.
Note that Eq.~\eqref{LambdaFrelation} implies that half of these equations are redundant, and allows us to decouple the $F^+$ from the $F^-$ coefficients.  Moreover, by using the Virasoro constraint \eqref{nullgaugedVirasoro1}, the bosonic null-gauge condition \eqref{nullgaugeconstr}, and the null constraint on the gauge parameters \eqref{eq: null gauge constraints}, one can show that for fixed $\vep_4$ and $\vep_5$, actually only two equations are linearly independent.
For generic values of quantum numbers, such that all denominators appearing below are nonzero, the linearly independent equations can be taken to be 
\begin{align}
\begin{aligned}
	\label{finalFRRgauged}
		F^+_{-+} & = -i \, \frac{j' - m ' + \half}{j + m - \half} \, F^{+}_{+-} + \frac{-j(j-1) + j' (j'+1) + m^2 - m'^2}{(j+m-\half)\left[ m-m' + \frac{l_4 - \vep_4 \vep_5 l_3}{2(l_2-1)} \left( \vep_4 \vep_5 E - P_y \right) \right]} F^{+}_{++} \, , \\
		F^+_{--} & =  \frac{m-m' + \frac{l_4- \vep_4 \vep_5 l_3}{2(l_2-1)} \left( \vep_4 \vep_5 E - P_y \right) }{j+m-\half} F^{+}_{+-}+ i \, \frac{j' + m ' + \half}{j + m - \half} \, F^{+}_{++} \, .
\end{aligned}
\end{align}
Alternatively, the two linearly independent equations can generically be taken to be
\begin{align}
F^-_{-+} & = -i \, \frac{j' - m ' + \half}{j + m - \half} \, F^{-}_{+-} + \frac{-j(j-1) + j' (j'+1) + m^2 - m'^2}{(j+m-\half)\left[ -m-m' + \frac{n_5(l_2-1)}{2(l_4 - \vep_4 \vep_5 l_3)} \left( \vep_4 \vep_5 E + P_y \right) \right]} F^{-}_{++} \, , \nonumber\\
F^-_{--} & =  \frac{-m-m' + \frac{n_5(l_2-1)}{2(l_4- \vep_4 \vep_5 l_3)} \left( \vep_4 \vep_5 E + P_y \right) }{j+m-\half} F^{-}_{+-}+ i \, \frac{j' + m ' + \half}{j + m - \half} \, F^{-}_{++} \, .
\label{finalFRRgaugedNeg}
\end{align}

Let us pause again to check consistency with the AdS$_3\times \SSS^3$ limit. Setting $\vep_4 \vep_5 = \vep$ and taking the large $R_y$ limit of Eqs.~\eqref{finalFRRgauged}, we indeed find that a solution is given by setting (the AdS$_3\times \SSS^3$ limit of) $F^{\vep}_{\vep_1 \vep_2}$ to be equal to the values $f^{\vep}_{\vep_1 \vep_2}$ specified in Eqs.~\eqref{eq:RRcoeff}--\eqref{BRSTAdS3}.

Similarly to AdS$_3\times \SSS^3$, Eqs.~\eqref{finalFRRgauged} are two equations for four unknowns, so for each $\vep_4$, $\vep_5$ there is a two-parameter family of solutions, which we take to be parameterized by the values of $F^{+}_{+\pm}$. If working with Eqs.~\eqref{finalFRRgaugedNeg}, we take the two-parameter family of solutions to be parameterized by the values of $F^{-}_{+\pm}$. Together with $\vep_4$, $\vep_5$, this gives 8 physical polarizations.

In Section \ref{sec:AdS3S3T4}, for AdS$_3\times \SSS^3$ these unfixed coefficients were chosen such that the vertex operators transform appropriately under the action of the currents $J^{\pm}$ and $K^\pm$. However, in the null-gauged worldsheet theory associated to the full asymptotically linear dilaton geometry this need not necessarily be the case.

In the cosets we fix these coefficients by requiring a reasonable IR limit. We treat the different $\vep$ chiralities separately. For $\vep=1$ we set the particular components $F^{+}_{+\pm}$ equal to their values in the AdS limit, $F^{+}_{+\pm} = f^{+}_{+\pm}$. The rest of the coefficients are then obtained using Eqs.~\eqref{finalFRRgauged} and \eqref{LambdaFrelation}. 
Alternatively, for $\vep=-1$ we set $F^{-}_{+\pm} = f^{-}_{+\pm}$ and again solve for the remaining coefficients using Eqs.~\eqref{finalFRRgaugedNeg} and \eqref{LambdaFrelation}. 

We now turn to the analysis of the spacetime supercharges preserved by the null gauging, following on from the initial discussion around Eq.\;\eqref{supercharges2} (see also~\cite{Martinec:2020gkv}). 
The supercharge analysis corresponds to the limit of the Ramond vertex operator analysis in which we take $j = j' = E = P_y = 0, \;\! m = \frac{\vep_1}{2}, \;\! m' = \frac{\vep_2}{2}$, as can be seen by comparing Eqs.\;\eqref{supercharges2} and \eqref{RRnullAnsatz}. 
In this limit, the center-of-mass wavefunction trivializes and we are left with integrated vertex operators involving only the spin fields.
As before, we parameterize the $\vep_i$ according to \eqref{nullchirality}; $\vep_4$ and $\vep_5$ are spectators that will label the supercharges; and we will sum over $\vep,\vep_1,\vep_2$ as in \eqref{RRnullAnsatz}. The $J$ constraint \eqref{nullgaugeconstr} reduces to 
\begin{equation}
\label{susyrestricparam-0}
    \vep_1 + \vep_2 \, l_2 = 0  \qquad \Rightarrow \qquad \vep_1\vep_2 = - l_2 \,, 
\end{equation}
so supersymmetry is preserved in the holomorphic sector if and only if $|l_2|=1$, and thus $|l_3|=|l_4|$. The $\gamma G$ constraint \eqref{BRSTFcompact} reduces directly to \begin{equation}
\label{susyrestricparam-1}
    \vep = -1 \,,
\end{equation}
so all supercharges have negative AdS$_3\times \SSS^3$ chirality. Then the $\lambda$ constraint in the first line of \eqref{LambdaFrelation}, with $F^+=0$, reduces to
\begin{equation}
\label{susyrestricparam-2}
l_3 + \vep_4 \vep_5 \, l_4 = 0  \,. 
\end{equation}
So when $|l_2|=1$ and $|l_3|=|l_4|\neq 0$, there are four holomorphic supercharges, labelled by say $\vep_2$ and $\vep_4$. 

Combining this analysis with the corresponding one in the antiholomorphic sector, we observe consistency with the passage below Eq.\;\eqref{LGT} describing which subset of the backgrounds are supersymmetric. In terms of the spacetime spectral flow parameters $s$, $\bar{s}$ introduced in Eq.\;\eqref{SFchargesHCFT}, we have $l_2=2s+1$, $r_2=2\bar{s}+1$.
The circular supertube backgrounds of~\cite{Maldacena:2000dr,Balasubramanian:2000rt} have $s=\bar{s}=0$ and preserve supersymmetry in both holomorphic and antiholomorphic sectors; 
the backgrounds of~\cite{Lunin:2004uu,Giusto:2004id,Giusto:2004ip,Giusto:2012yz} have $\bar{s}=0$, $s\neq 0$ and so preserve supersymmetry only in the antiholomorphic sector; the general JMaRT backgrounds~\cite{Jejjala:2005yu} have $s$ and $\bar{s}$ both nonzero, and preserve no supersymmetry.

\subsubsection*{Picture changing in the R sector}

In order to compute two-point functions of operators $\Yy$ in the Ramond sector of the gauged model, we need to define their picture-changed versions. Propagators will be non-vanishing only if the total ghost charges add up to  $\;\!-Q_\varphi=-2\;\!$ and $\;\!-Q_{\tilde{\varphi}}=0\:\!$. The  picture-changing operators are given by $P_{+1} \sim e^{\varphi} G$ and $\tilde{P}_{+1}\sim e^{\tilde{\varphi}} \boldsymbol{\lambda}$. One possible natural choice would be to compute the two-point function (superscripts denote $(q_\varphi,q_{\tilde{\varphi}})$ charges)
\begin{equation}
    \braket{\Yy^{\left( - \frac{3}{2}, - \half\right)}(z) \;\! \Yy^{\left( - \frac{1}{2}, + \half\right)}(w) } \, . 
\end{equation}
However, it turns out that looking for an explicit expression for the state $\Yy^{\left( - \frac{3}{2}, - \half\right)}$ is not the simplest way to go. This is due to the fact that such a state is automatically BRST closed, so that it must be determined by the somewhat cumbersome procedure of removing all the BRST-exact contributions. To avoid this issue, one can distribute the ghost charges in a different way inside the correlator, and consider instead the equivalent two-point function
\begin{equation}
    \braket{\Yy^{\left( - \frac{3}{2}, + \half\right)}(z) \;\! \Yy^{\left( - \frac{1}{2}, - \half\right)}(w) } \, . 
\end{equation}
Here $\Yy^{\left( - \frac{3}{2}, + \half\right)}$ is in the canonical $\tilde{\varphi}$-picture, while $\Yy^{\left( - \frac{1}{2}, - \half\right)}$ is in the canonical $\varphi$-picture. Thus, although this forces us to compute two additional R-sector operators instead of only one, these are constrained by the  $\:\!\tilde{\gamma}\boldsymbol{\lambda}\:\!$ and $\:\!\gamma G\:\!$ BRST constraints, respectively. The procedure is then similar to that employed above to construct $\Yy^{\left( - \frac{1}{2}, + \half\right)}$. We thus make the Ans\"{a}tze 
\begin{align}
\begin{aligned}
    & \Yy^{\left( - \frac{3}{2}, + \half\right), \vep_4 \vep_5}  \,=\, e^{-\left( \frac{3}{2} \varphi - \half \tilde{\varphi} \right)} \sum_{\vep, \vep_1, \vep_2} \Ll^{\vep}_{\vep_1 \vep_2 \vep_4 \vep_5} S_{\vep_1 \vep_2 \vep_3}  \mathcal{S}_{\vep_6 \vep_4 \vep_5} V_{j,m-\frac{\vep_1}{2}}V'_{j',m'-\frac{\vep_2}{2}} 
	e^{i(-E \:\! t + P_y \:\! y)} \,, \nn \\
    & \Yy^{\left( - \frac{1}{2}, - \half\right), \vep_4 \vep_5}  \,=\, e^{-\left( \frac{1}{2} \varphi + \half \tilde{\varphi} \right)} \sum_{\vep, \vep_1, \vep_2} \Gg^{\vep}_{\vep_1 \vep_2 \vep_4 \vep_5} S_{\vep_1 \vep_2 \vep_3}  \mathcal{S}_{\vep_6 \vep_4 \vep_5} V_{j,m-\frac{\vep_1}{2}}V'_{j',m'-\frac{\vep_2}{2}}  
	e^{i(-E \:\! t + P_y \:\! y)} \,. \nn
\end{aligned}
\end{align}
where again $\vep_3$, $\vep_6$ are substituted away using \eqref{nullchirality}. 
These must satisfy 
\begin{align}
\label{BRSTforYy}
    \begin{aligned}
 \oint dz :e^{\tilde{\varphi}}  \boldsymbol{\lambda} : \! (z) \: \Yy^{\left( - \frac{3}{2}, + \half\right), \vep_4 \vep_5} (w) & \,=\, 0 \, , \\       
 \oint dz :e^{\varphi}G_{\text{tot}} : \! (z) \: \Yy^{\left( - \frac{1}{2}, - \half\right), \vep_4 \vep_5} (w) & \,=\, 0 \, ,
    \end{aligned}
\end{align}
and 
\begin{align}
\label{PictureforYy}
\begin{aligned}
:e^{\varphi}G_{\text{tot}}: \! (z) \: \Yy^{\left( - \frac{3}{2}, + \half\right), \vep_4 \vep_5} (w) & \,=\, \Yy^{\left( - \frac{1}{2}, + \half\right), \vep_4 \vep_5} (w) \, , \\[1.5ex]
:e^{\tilde{\varphi}}  \boldsymbol{\lambda}: \! (z) \: \Yy^{\left( - \frac{1}{2}, - \half\right), \vep_4 \vep_5} (w) & \,=\, \Yy^{\left( - \frac{1}{2}, + \half\right), \vep_4 \vep_5} (w) \, .
\end{aligned}
\end{align}
By solving the above constraints, all the coefficients $\Ll^\vep$ and $\Gg^\vep$ can be expressed explicitly in terms of the $F^\vep $ coefficients in Eq.~\eqref{finalFRRgauged}.

In addition, one can explicitly check that in the AdS limit they correctly reproduce the expected behaviour. From the definition of the corresponding coset states $\Yy^{\left( - \frac{3}{2}, + \half \right)} $ and $\Yy^{\left( - \frac{1}{2}, - \half\right)} $, one might reasonably expect that they would reduce to the states $\Yy^{\left( - \frac{3}{2}\right)} $ and $\Yy^{\left( - \frac{1}{2}\right)} $ of Section \ref{sec:RsectorAdS3} respectively. However, care is needed when comparing both chiralities and normalisations in the UV and IR. To explain this, let us consider for instance $\Yy^{\left( - \frac{1}{2}\right)}_A $, whose $\vep$-chirality is $\vep = +1$. (Analogous comments hold for other operators.) First of all, recall that already in the case of AdS$_3 \times \SSS^3 \times \TT^4$, the picture-changing operator induces a change in chirality of the state. Indeed, in case ``A'' of the analysis in \eqref{eq:RRcoeff}, the physical states in the ``$-\frac{3}{2}$'' picture have \textit{negative} $\vep$-chirality. 

In the full coset models, an analogous pattern occurs with the two picture-changing operators $P_{+1}, \tilde{P}_{+1}$. The coset state $\Yy^{\left( - \frac{3}{2}, + \half \right)}$ that correctly reduces to $\Yy^{\left( - \frac{3}{2}\right), \vep = -1}_A $ in the AdS limit indeed has \textit{positive} $\vep$-chirality. This means that, in our case under study, the coefficients $\Ll^{+}_{\vep_1 \vep_2}$ reduce to $- \frac{\sqrt{n_5}}{j + j'} f^{-}_{\vep_1 \vep_2}$. Similarly, the coset state $\Yy^{\left( - \frac{1}{2}, - \half \right)}$ with negative $\vep$-chirality reduces to the $\Yy^{\left( - \frac{1}{2}\right)}_A$ state. However, in the latter case there is a normalisation factor $(i \, \k R_y)\inv$ to account for. This is removed by taking into account the normalisation of the picture-changing operator, which indeed contains a term which is dominant in the AdS limit, $l_3 \lambda_t + l_4 \lambda_y \sim (\k R_y)(\lambda_t + \lambda_y) $.

We illustrate the example of the $\Gg^{-}_{++}$ coefficient, appearing in the coset state $\Yy^{\left( - \frac{1}{2}, - \half \right)}$. The argument holds analogously for all the other coefficients. The explicit expression of $\Gg^{-}_{++}$ in terms of positive coefficients $F^+_{\vep_1 \vep_2}$ is obtained by solving the constraints Eq.~\eqref{BRSTforYy} and Eq.~\eqref{PictureforYy}, without using Eq.~\eqref{nullgaugedVirasoro1} and Eq.~\eqref{nullgaugeconstr}. For generic quantum numbers such that the denominators below are non-zero, one finds
\begin{align}
    & \frac{(l_3+l_4)^2}{i (E - P_y) n_5 (1+l_2)} \; \Gg^{-}_{++}   \\
    &= \frac{ 
    (m+j - \half)F^{+}_{-+} + i (j'-m'+\half)F^{+}_{+-} 
   - \left( \frac{2(j-j'+1)(j+j')(l_3 + l_4)}{(E-P_y) n_5 (1+l_2)} + \frac{(-1+l_2)}{(1+l_2)}(m-m') \right) F^+_{++} }{ \left( -j(j-1) + j'(j'+1) + n_5 \frac{(m + l_2 m')(E-P_y)}{l_3+l_4}  + n_5 \frac{n_5(-1+l_2^2)(E-P_y)^2}{4(l_3+l_4)^2}  \right)} \nn \, , 
\end{align}
and thus for $R_y \gg 1$ one has $\Gg^{-}_{++} \simeq (i \k R_y)\inv f^+_{++} + \Oo(R_y^{-2}) $, as claimed above.

The expressions for the coefficients $\Gg, \Ll$ are quite lengthy, and we leave the computation of correlators in the full coset model for future work. 
Nevertheless, it is easy to check that in all cases the coefficients reduce to the expected expressions when going into the IR regime. 
Consequently, we find that, to leading order in $R_y$, the coset two-point functions in the RR sector reproduce the $m$-basis expressions in Eq.~\eqref{AdSRam2pt}, as they should. As will become clear in Section \ref{Section 5} below, this does \textit{not} mean that the physics in the IR regime of the coset model is that of global AdS$_3\times \SSS^3\times\TT^4$; the bosonic null gauge condition \eqref{nullgaugeconstr} will lead to substantially different correlators in the appropriately defined $x$-basis.

\subsection{Flowed/winding sectors}
\label{sec:flowedwindingsectors}

We now briefly discuss the states with non-trivial spectral flow we will be interested in, that is, those corresponding to the description of the higher-weight chiral primaries described in Section \ref{sec:AdS3S3T4}. 

A generic state with excitation numbers ($\frac{1}{2}$,$\frac{1}{2}$) in the null-gauged worldsheet theory must satisfy the 
the $L_0$ and $\bar{L}_0$ Virasoro constraints
\begin{subequations}
\label{VirasoroNullWinding}
\begin{align}
    0 &= 
    \frac{j'(j'+1)-j(j-1)}{n_5} - m \w
    + m' \w' + \frac{n_5}{4}\left(\w^{'2}-\w^2\right) - \frac{1}{4} \left(E^2 - P_{y,L}^2\right)
    \, , \\
    0 &= 
    \frac{j'(j'+1)-j(j-1)}{n_5} - \bar{m} \w
    + \bar{m}' \bar{\w}' + \frac{n_5}{4}\left(\bar{\w}^{'2}-\w^2\right) - \frac{1}{4} \left(E^2 - P_{y,R}^2\right) 
    ,  
\end{align}
where (we reuse the notation $P_{y,L}
$, $P_{y,R}$ for the eigenvalues of the corresponding operators)
\end{subequations}
\begin{equation}
P_{y,L/R} =  
\frac{n_y}{R_y}\pm \w_y R_y 
\, , 
\end{equation}
with $\w_y\in \mathbb{Z}$ the winding on the $y$-circle. The level-matching $L_0-\bar{L}_0$ constraint thus reads 
\begin{equation}
\label{LevelMatching}
 0 =    \w (\bar{m} - m) + m' \w' - \bar{m}' \bar{\w}' + 
 \frac{n_5}{4}(\w^{'2}-\bar{\w}^{'2}) +
  n_y \w_y \ .  
\end{equation}
Let us try to follow the global AdS$_3$ procedure as close as possible, and consider states with $\w = \w' = \bar{\w}'$ and $m=m'=\bar{m} = \bar{m}'$. Then, \eqref{LevelMatching} forces us to set $\w_y = 0$. Moreover, for states constructed by spectrally flowing highest/lowest weight primaries, the discussion around Eq.~\eqref{BRSTflowG} shows that the $e^{\varphi}G$ part of the BRST charge acts as in the unflowed case. Moreover, the action of $e^{\tilde{\varphi}}\boldsymbol{\lambda}$ is left unchanged. The derivation of the coefficients involved in the definition of the vertex operators constructed above then goes through without changes, and we only need to restrict to the highest (lowest) possible values of $m$ ($m'$) in each case.

Regarding the gauge constraints, recall that, both in SL(2,$\R$) and in SU(2), the different modes of the spectrally flowed operators are obtained by acting with the raising/lowering operators $J_0^\pm$ and $K_0^\pm$ on the flowed primary.
Although this does not lead to operators that can be expressed in a simple way in the $m$-basis, the presence of these different modes is crucial in order to obtain the set of physical modes that satisfy the gauge constraints. Focusing on (lowest-weight) discrete states corresponding to operators of spacetime weight $h$ in the chiral multiplets, this gives modes described by worldsheet operators with projections
$\:\!m_\w \:\!=\:\! J + \frac{n_5}{2}\w + n \:\!=\:\! h + n\:\!$ and $\:\!m_\w' \:\!=\:\! J'+ \frac{n_5}{2}\w - n'$, with $n,n'\in \mathbb{N}_0$, and similarly in the antiholomorphic sector. The bosonic gauge constraints now read 
\begin{equation}
\label{nullgaugeconstrW}
0 = m_\w + l_2 \, m_\w' + \frac{l_3}{2} E + \frac{l_4}{2} P_{y} \qqquad
0= \bar{m}_\w + r_2\, \bar{m}_\w' + \frac{r_3}{2} E +\frac{r_4}{2} P_{y} \;.
\end{equation}
As discussed below Eq.\;\eqref{6DscalarsNSNS}, this implies that $J_0^\pm$ and $K_0^\pm$ do not commute with the BRST charge. Consequently and importantly, only the subset of modes satisfying Eqs.~\eqref{nullgaugeconstrW} will be physical. The implications will be discussed at length in the following section.

Before concluding this section, let us review the fact that there is a residual discrete gauge symmetry in these models, which implies that operators related by shifts of the following form describe the same physical state~\cite{Martinec:2018nco}, see also~\cite{Bufalini:2021ndn}. Parameterizing $(l_4,r_4)$ through $\pp$ and $\k$ as in \eqref{integersCFT2-b}, the symmetry is \begin{equation}
    \delta \left(\w,\w',\bar{\w}',E,n_y,\w_y\right) = \left(1,
    -l_2,-r_2,l_3,-\pp,\k
    \right) .
    \label{discretegauge}
\end{equation}
In particular, one can trade a unit of SL(2,$\R$) spectral flow for $\:\! -\k$ units of winding $\w_y$, together with corresponding shifts in $\w'$ and $\bar{\w}'$. The energy also acquires a term linear in $R_y$, namely $\delta E = - \k R_y + {\cal{O}}(1/R_y)$. The interpretation of the factor $\k$ relating $\w$ and $\w_y$ can be traced back, for instance, to the $\mathbb{Z}_k$ orbifold appearing in the IR, see Eq.~\eqref{eq:ads-JMaRT-met}. It reflects the fact that the CFT state associated with the background lives in the $\k$-twisted sector of the D1D5 CFT.

The operators discussed in this section do not exhaust the spectrum of the worldsheet model; for instance we have not discussed operators that do not not satisfy $\w_y \equiv 0$ mod $\k$, which were analyzed in~\cite{Martinec:2018nco}.
However, the operators described above 
comprise a large set of light operators in spectral flowed sectors, in parallel to the analysis of global AdS$_3\times \SSS^3$, which will be general enough for our purposes in the present work.

\section{Novel heavy-light correlators from the worldsheet}
\label{Section 5}

In this section we describe the computation of two-point correlators in the null gauged models, corresponding to HLLH correlators of the holographic CFT. To do so, we take a set of physical coset operators derived in the previous section and flow them to the IR, in which the geometry is locally an orbifold of AdS$_3\times \SSS^3$. 
We develop a proposal to define coset operators in an appropriate $x$-basis corresponding to local operators of the holographic CFT.
We then use this definition to compute a large set of HLLH correlators. We observe precise agreement between a subset of these and known results computed in supergravity and holographic CFT, and significantly extend these results.

\subsection{Light states in the AdS$_3$ regime}

We begin by describing in more detail the vertex operators of the null-gauged model in the AdS$_3$ limit. As discussed around Eq.~\eqref{FixedAdSlimit}, we send $R_y \to \infty$, keeping $\tilde{t} = t/R_y$ and $\tilde{y}=y/R_y$ fixed. After choosing the gauge $\tau=\sigma=0$, this leads to a geometry described by the six-dimensional  metric \eqref{eq:ads-JMaRT-met}, which is related to $\mathbb{Z}_\k$-orbifolded AdS$_3\times \SSS^3 \times \TT^4$ by the large coordinate transformation Eq.~\eqref{LGT}.

We focus initially on light states with no winding or worldsheet spectral flow. As we have argued in the previous section,  the different polarizations and the associated coefficients simply reduce to those described in Section \ref{sec:AdS3S3T4} in the AdS$_3$ limit. Here we further describe what happens to their quantum numbers in the regime of interest. In general, the Virasoro condition \eqref{nullgaugedVirasoro1} determines $j$ via the solution 
\begin{equation}
    j = \frac{1}{2} + \sqrt{ 
    \left(j'+\frac{1}{2}\right)^2 + 
    \frac{n_5}{4}\left(
    P_y^2 - E^2
    \right)} \, ,
    \label{jVir}
\end{equation}
where we have fixed the sign in order to have $j$ in the range \eqref{Djrange}. 
As $R_y \to \infty$, we hold fixed the rescaled energy and momentum
\begin{equation}
\label{defeeny}
    {\cal{E}} = E R_y \qqquad n_y = P_y R_y \, . 
\end{equation}
Hence, the second term inside the square root in \eqref{jVir} is ${\cal{O}}\left(1/R_y^2\right)$, and at large $R_y$ the solution becomes $j = j'+1+{\cal O}\left(1/R_y^2\right)$, which to leading order is the usual AdS$_3\times \SSS^3$ relation.
The ${\cal{O}}\left(1/R_y^2\right)$ corrections to $j$ are non-zero when $|{\cal{E}}| \neq |n_y|$, which is generically the case.
To see this, note that at large $R_y$
the gauging parameters associated to the $t$ and $y$ directions become
\begin{equation}
    l_3 = r_3 = l_4 = -r_4 = - \k R_y +  {\cal O}\left(1/R_y\right).
\end{equation}
On the other hand, those associated to the $\SSS^3$ angular directions do not scale with $R_y$, and remain $l_2 = 2s+1$ and $r_2 = 2\bar{s}+1$. Hence, at leading order at large $R_y$, Eqs.~\eqref{nullgaugeconstr} take the form 
\begin{equation}
\label{nullgaugeconstrAdS}
0 = m + (2s+1)\, m' - \frac{\k}{2}\left({\cal{E}} + n_y \right) \qqquad
0= \bar{m} + (2\bar{s}+1)\, \bar{m}' - \frac{\k}{2}\left({\cal{E}} - n_y \right) \, , 
\end{equation}
which fix ${\cal{E}}$ and $n_y$ in terms of $m$, $m'$, $\bar{m}$, $\bar{m}'$,  such that indeed generically ${\cal{E}} \neq \pm n_y$. 

Although for simplicity we restricted to light states with no winding or worldsheet spectral flow in Eqs.~\eqref{jVir} and \eqref{nullgaugeconstrAdS}, the present discussion and the computations in the rest of this section are analogous for winding states after replacing the projections $m \to m_\w$, etc, where $m_\w$ was defined above Eq.~\eqref{nullgaugeconstrW}. Our results will be valid for the full set of chiral primaries that can be described within the usual AdS$_3\times \SSS^3\times \TT^4$ worldsheet theory, as well as their descendants under the global part of the chiral algebra, that fill out the short multiplet. 
We shall comment further on states with non-trivial winding and/or worldsheet spectral flow in due course.

\subsection{Identifying the spacetime modes}

Let us discuss the identification of the spacetime modes. We shall work in a gauge in which the upstairs SL(2,$\R$) time $\tau$ and angular direction $\sigma$ are fixed. Then, importantly, the asymptotic boundary of the physical downstairs AdS$_3$ is parameterized by $t/R_y$ and $y/R_y$, at a fixed point on the $\SSS^3$. We therefore define
\begin{equation}
\label{mymodes}
    m_y = \frac{1}{2} \left({\cal{E}} + n_y \right) \qqquad
    \bar{m}_y = \frac{1}{2} \left({\cal{E}} - n_y \right) \, ,
\end{equation}
and we interpret these as the asymptotic mode labels.
We will see that this gives rise to a rich set of correlators that agree with and extend previous results.

Let us first consider $\k=1$, and continue to focus mainly on the holomorphic sector. Given a holographic CFT chiral primary with spacetime weight $h$ and definite (left) R-charge $m'$, we wish to construct the dual worldsheet operator by summing over the corresponding modes. As reviewed in Section \eqref{sec:AdS3S3T4}, in global AdS$_3\times \SSS^3$ this leads to $x$-basis operators of the form ${\cal{O}}_{h,m'}(x) = \sum_{m} x^{m-h} {\cal{O}}_{h,m,m'}$, where the modes  ${\cal{O}}_{h,m,m'}$ are identified as either  $\Ww_{j,m,j',m'}$, $\Xx_{j,m,j',m'}$ or $\Yy_{j,m,j',m'}$, where $j$ and $h$ are related by Eq.~\eqref{jwandhAdS3}. 
For simplicity, we collect all of these modes under the notation $\Vv_{j,m,\bar{m}}\Vv'_{j',m',\bar{m}'}$.  
In the null-gauged models, we replace $m\to m_y$ in the exponent of $x$ in the sum, defining the worldsheet operator
\begin{eqnarray}
\label{modesAdSk1}
    \Oo_{h,m'}(x,\bar{x}) &\,\equiv\,& \frac{1}{\k^{h + \bar{h}}} \sum_{m_y,\bar{m}_y} x^{m_y-h}\bar{x}^{\bar{m}_y - \bar{h}} \Vv_{j,m,\bar{m}}\Vv'_{j',m',\bar{m}'} e^{-i m_y (\tilde{t}-\tilde{y})}e^{-i \bar{m}_y (\tilde{t}+\tilde{y})} \,,
\end{eqnarray}
where we have temporarily included the antiholomorphic dependence to emphasize the coupling between the left- and right-moving sectors due to the null gauge constraints, Eq. \eqref{nullgaugeconstr}.
The normalization factors of $\k$ have been introduced for later convenience and will be discussed below Eq.~\eq{finalHLLH}.
We emphasize that, from the point of view of the worldsheet theory, $x$ is an auxiliary complex variable, while $\tilde{t}$ and $\tilde{y}$ are scalar fields.

Note that combining the bosonic null constraint \eqref{nullgaugeconstrAdS} and the definition of the modes \eqref{mymodes}, we obtain $m_y = m + (2s+1) m'$. This relation parallels the supergravity spectral flow large gauge transformation \eqref{LGT}, and we will make use of this observation when discussing the relation to the holographic CFT in due course.

For $\k>1$ we follow the same logic, and make the same definition \eq{modesAdSk1}. This time however, combining Eqs.~\eqref{nullgaugeconstrAdS} and \eqref{mymodes} we obtain 
\begin{equation}
\label{eq:my}
m_y = \frac{1}{\k}\left(m + (2s+1) m'\right) \,, \qquad \bar{m}_y = \frac{1}{\k}\left(\bar{m} + (2\bar{s}+1)\bar{m}'\right) \,.
\end{equation}
We shall shortly see that this gives rise to an important technical complication relating the holomorphic and antiholomorphic sectors.

Before computing our first example of a HLLH correlator, let us briefly return to operators with non-zero spectral flow and/or winding charge. 
In Section \ref{sec:flowedwindingsectors} we analyzed a set of coset vertex operators in sectors with non-zero worldsheet spectral flow, corresponding to chiral primaries of higher twists, similar to those in global AdS$_3\times \SSS^3$.
In that section we worked with $\w_y=0$, and reviewed that the large gauge spectral flow transformation~\eqref{discretegauge} relates these to operators with $\w_y\in \k \mathbb{Z}$.

When $\w_y\neq 0$, in general one should be careful both when examining whether the states survive the AdS$_3$ limit, and also when defining the AdS$_3$ energy $\mathcal{E}$ and angular momentum $n_y$, since fundamental string $y$-winding charge can be exchanged for background flux~\cite{Martinec:2018nco}.
However, for operators which have $\w_y\in \k \mathbb{Z}$, the situation is straightforward: we shall simply use gauge spectral flow to always work in a frame in which $\w_y=0$, and then use the definitions of $\mathcal{E}$ and $n_y$ in \eqref{defeeny} and the modes $m_y$, $\bar{m}_y$ in \eqref{mymodes}.

The discussion becomes more complicated when considering global SL(2,$\R$)$\times$SU(2) descendants of the spectrally flowed primaries.  The isomorphism between the affine modules gives a simple identification for the highest/lowest weight states, but this structure becomes more complicated for rest of the of the multiplet. Indeed, global descendants of the spectrally flowed affine primary state are identified with affine/Virasoro descendants of the corresponding unflowed equivalent state with non-trivial $y$-winding, such that one needs to include string oscillator excitations. The situation is similar to what happens for the usual series identification $V_{j,j}^{\w=0} \sim V_{\frac{k}{2}-j,j-\frac{k}{2}}^{\w=1}$ in bosonic SL(2,$\R$), where, for instance, one has $V_{j,m}^{\w=0} \sim (j^+_{0})^{m-j}V_{j,j}^{\w=0} \sim  (j^+_{-1})^{m-j}V_{\frac{k}{2}-j,j-\frac{k}{2}}^{\w=1}$, see \cite{Martinec:2020gkv}. 
We leave a more detailed exploration of these features in the coset models for future work. In the reminder of this paper we will  work with operators that have $\w_y=0$.

\subsection{HLLH correlators: first example}
\label{HLLHh1/2}

We now compute a first explicit example of a worldsheet two-point function in the cosets corresponding to the heavy backgrounds under consideration. For this purpose, we focus on a particular light operator probing the backgrounds with $\bar{s}=0$ (hence $r_2 = 1$), but general $s$. These describe supersymmetric spectral flowed supertubes~\cite{Lunin:2004uu,Giusto:2004id,Giusto:2004ip,Jejjala:2005yu,Giusto:2012yz}. We shall demonstrate that this worldsheet correlator agrees with both the supergravity and symmetric product orbifold CFT HLLH four-point functions computed in \cite{Galliani:2016cai}. We will also significantly extend beyond the set of correlators computed in \cite{Galliani:2016cai}.

The light operator in this first example is a massless RR operator of $\Yy_{[A]}$ type with $h=m'=1/2$ and hence $j=1$, see Eq.\;\eqref{eq:q-nos-rr}. We shall denote this operator by $\Yy_{\frac{1}{2}}$. Together with the analogous antiholomorphic part, this vertex operator is 
dual to a particular $(h,\bar h)=(\half,\half)$ chiral primary of the HCFT denoted by $O^{++}$, which we introduced in Eq.~\eqref{O++ operator}. 
In the six-dimensional supergravity arising from reduction on $\TT^4$, this corresponds to a 
particular combination of fluctuations of a scalar and anti-self-dual two-form potential, in a tensor multiplet that is not turned on in the backgrounds we consider. 
In type IIB supergravity, in the present NS5-F1-P duality frame, these fields correspond to a particular combination of supergravity fluctuations of the RR axion and certain components of the RR two-form and four-form potentials. 
(In the D1-D5-P frame, the fluctuations are of the RR axion and certain components of the RR four-form and NS-NS two-form potentials). 
The light operator corresponds to a particular scalar spherical harmonic of these fields~\cite{Galliani:2016cai,Giusto:2019qig,Rawash:2021pik}.

In the large $R_y$ limit, we have shown that the $\Yy$ operators simply reduce to their AdS$_3\times \SSS^3$ cousins of Section \ref{sec:AdS3S3T4}, times the additional exponentials in $t$ and $y$. These exponentials give trivial contributions to the two-point functions $\langle \Yy_{\frac{1}{2}} \Yy_{\frac{1}{2}} \rangle$ once the charge conservation conditions $m_{y,1} = - m_{y,2}$ and $\bar{m}_{y,1} = - \bar{m}_{y,2}$ are imposed. So do the SU(2) parts of the vertex operators for $m'_1 = - m'_2$, upon using the appropriate normalization. Hence, at the level of the two-point function of $m$-basis operators, i.e.~the mode correlators from the spacetime point of view, the only non-trivial contribution comes from the SL(2,\,$\R$) part of the upstairs theory. 

To illustrate this more explicitly and also more generally, we introduce the following notation: $\Oo_{h,m'}$ will denote a generic massless worldsheet vertex operator in the AdS limit of the null-gauged model with spacetime weight $h$ and spacetime R-charge $m'$. When $m'=h$, we shall suppress the label $m'$ and write $\Oo_h$. The corresponding holographic CFT chiral primaries will be denoted by $O_h$. 
Let us then consider the following worldsheet correlator,
\begin{align}
    \bra{s,\bar{s},\k} &\Oo_{h_1,m'_1} (x_1,z_1) \Oo_{h_2, m'_2} (x_2,z_2) \ket{s,\bar{s},\k} \nn \\ &\,\equiv\, \frac{1}{\k^{2h_1 + 2h_2}} \sum_{m_{y,i},\bar{m}_{y,i}} \prod_{i=1,2}x_i^{m_{y,i}-h_i}
    \bar{x}_i^{\bar{m}_{y,i}-h_i} \lim_{R_y \to \infty}\langle \Vv_1 (z_1) \Vv_2 (z_2) \rangle \, ,
        \label{511}
\end{align}
where $\Vv$ denotes a generic $m$-basis massless vertex operator of the full coset model.
The spacetime correlator corresponds to the worldsheet-integrated version of \eqref{511}. 

As discussed below \eqref{eq:2pt-ads-Y}, the normalization of the vertex operators is chosen so that the overall factors coming from the worldsheet integration procedure cancel out. 
After setting $h_1 = h_2 = h$, $x_1=1$ and $x_2 = x$, computing the free-field correlators, 
and imposing the various charge conservations, the integrated correlator becomes 
\begin{equation}
    \bra{s,\bar{s},\k}  \Oo_{h,m'} (1) \Oo_{h,m'}^{\dagger} (x) \ket{s,\bar{s},\k}  \,=\, \frac{1}{\k^{4h}} \sum_{m_{y},\bar{m}_{y}} x^{m_{y}-h}
    \bar{x}^{\bar{m}_{y}-h}
    \lim_{R_y \to \infty}\langle \Vv_1 \:\! \Vv_2 \rangle \, ,
\end{equation}
where $\langle \Vv_1 \;\! \Vv_2 \rangle$ stands for the  $m$-basis two-point function with the $z$-dependence stripped out, c.f.~Eqs.~\eqref{eq: W2pt},~\eqref{AdSRam2pt}. Note that on the right-hand side, the sum over $m_y$ involves the R-charge and other quantum numbers of the operator located at $x$.

In turn, the remaining correlator is  particularly simple in our context. 
As argued around Eq.~\eqref{2ptYyA}, the $m$-basis two-point functions of worldsheet chiral primary operators reduce to the Gamma functions expression in the bulk term of \eqref{2ptMbasisSL2} with the replacement $j \mapsto J = h$. These are simply the coefficients obtained by Mellin-transforming the usual propagator $|1 - x|^{-4h}$, which further all become equal to one when $h=\half$.
Thus, for $\Yy_\half$ 
in the $\bar{s}=0$ backgrounds,
we obtain
\begin{equation}
   \bra{s,\k} \Yy_{\half} (1) \Yy_{\half}^\dagger(x) \ket{s,\k} \,=\, \frac{1}{\k^2} \sum_{m_{y},\bar{m}_{y}} x^{m_{y}-\frac12}
    \bar{x}^{\bar{m}_{y}-\frac12} \, .
\end{equation}
To be fully explicit, we take the operator at $x_2 = x$ to be a discrete series state $\mathcal{D}^+_j$ corresponding to an anti-chiral primary which, having set $J = h = \bar{h}$, has $m=h+n$, $\bar{m}=h+\bar{n}$ and $m'=\bar{m}'=-h$, where $n,\bar{n}$ are non-negative integers. For the supersymmetric backgrounds with
$\bar{s} = 0$ and setting $h=\half$, the relation \eq{eq:my} becomes
\begin{equation}
\label{eq:mmy}
m_y = \frac{n-s}{\k} \,, \qquad \bar{m}_y = \frac{\bar{n}}{\k}\,.
\end{equation}
Hence, the correlator takes the form
\begin{equation}
   \label{HLLHws}
   \bra{s,\k} \Yy_{\half} (1) \Yy_{\half}^\dagger (x) \ket{s,\k} \,=\,  \frac{1}{\k^2} \sum_{n,\bar{n}}\phantom{}^{'} x^{\frac{n - s}{\k} -\frac{1}{2}} \, \xb^{\frac{\bar{n}}{\k}-\frac{1}{2}} \,,
\end{equation}
where we have denoted the sum with a prime because the range of summation over $n,\bar{n}$ is constrained. We now determine this constraint. Subtracting the two equations in \eqref{eq:mmy} and comparing with \eqref{mymodes}, we obtain
\begin{align}
\begin{aligned}
\label{eq:my-3}
m_y-\bar{m_y}  \,=\, \frac{1}{\k}(n-\bar{n}-s) &\,=\, n_y \cr
\Rightarrow n-\bar{n} &\,=\, \k n_y + s\,.
\end{aligned}
\end{align}
Thus the allowed values of $n-\bar{n}$ are constrained by $n_y \in \mathbb{Z}$. To see this in detail, let us first write $s=\k p + \hat{s}$ with $0\le \hat{s} < \k $ and $p \in \mathbb{N}$. For convenience here and later, we define $a\equiv\k-\hat{s}$, so that $s=\k p - a$, and $1 \le a \le \k$. Then the sum in \eqref{HLLHws} is restricted to be over non-negative integers $n,\bar{n}$ satisfying
\begin{equation}
    \bar{n}-n  \;\equiv\;a \quad \mathrm{mod}~ \k .
    \label{nnbaralpha-0}
\end{equation}

We now demonstrate that the correlator \eqref{HLLHws}, \eqref{nnbaralpha-0} agrees precisely with the supergravity and orbifold CFT expressions derived in \cite{Galliani:2016cai}. In the holographic CFT, we have a HLLH four-point function $\langle O_H(x_3) O_L(x_1) O_L^\dagger(x_2) O_H^\dagger (x_4) \rangle$. Using M\"obius symmetry we set $x_3=0$ and $x_4\to \infty$, in which case the heavy operators are interpreted as in/out states which we similarly denote as 
$|s,\k\rangle$, such that the four-point function becomes a two-point function in the heavy background. We further set $x_1=1$,  while $x_2 = x$ parametrizes the usual cross-ratio. Then the supergravity result of \cite[Eq.~(4.25)]{Galliani:2016cai} is 
\begin{equation}
\label{HLLHGalliani}
    \langle s,\k|O_\half(1)O_\half^\dagger(x)|s,\k\rangle \;=\;
    \frac{x^{(\hat{s}-s)/\k}}{|x||1-x|^2} \frac{1-|x|^{2(1-\hat{s}/\k)}+\bar{x}\left(|x|^{-2\hat{s}/\k}-1\right)}{1-|x|^{2/\k}} \;,
\end{equation}
where the overall normalization of the supergravity amplitude was not fixed in \cite{Galliani:2016cai}.
Eq~\eqref{HLLHGalliani} was further shown to coincide with the corresponding symmetric orbifold CFT calculation for the particular cases $\hat{s}=0$ and $\hat{s} = \k-1$. 

To demonstrate agreement between \eqref{HLLHws} and \eqref{HLLHGalliani}, it is easier to work from \eqref{HLLHGalliani} towards our expression \eqref{HLLHws}. 
We start by rewriting \eqref{HLLHGalliani} in terms of sums akin to those involved in the definition of the $x$-basis, i.e.~the mode expansion of local operators in spacetime as seen from the worldsheet theory. Recalling that $\hat{s} = \k-a$, the correlator becomes
\begin{eqnarray}
    \langle s,\k|O_\half(1)O_\half^\dagger(x)|s,\k\rangle
   &=&
     \frac{x^{1-p}}{|x||1-x|^2} \frac{1-|x|^{2a/\k}+\bar{x}\left(|x|^{2a/\k-2}-1\right)}{1-|x|^{2/\k}} \nn \\
    &=&
    \frac{x^{-p}}{|x|} \left[\frac{1}{|1-x|^2}\frac{1-|x|^2}{1-|x|^{2/\k}} - \frac{1}{1-\bar{x}}\frac{1-|x|^{2a/\k}}{1-|x|^{2/\k}}\right].
    \label{HLLH2terms}
\end{eqnarray}
Assuming $|x|<1$, the RHS can then be expressed as
\begin{equation}
      \sum_{\hat{\bar{n}}=0}^{\infty}
   \left[ \, 
   \sum_{\hat{n}=0}^{\infty} \, \sum_{\delta = 0}^{\k-1} \, - \,   \delta_{\hat{n},0} \, \sum_{\delta = 0}^{a-1} \, 
   \right] x^{\hat{n} + \frac{\delta}{\k}-p-\frac{1}{2} }
   \, \xb^{\hat{\bar{n}} + \frac{\delta}{\k}-\frac{1}{2} }
\; . 
    \label{HLLHstotalsum}
\end{equation}
The second term in \eqref{HLLHstotalsum} is understood as subtracting the $\hat{n}=0$ and $\delta=0,\dots,a-1$ coming from the first term. As a consequence, we can further rewrite the sum over $\hat{n},\hat{\bar{n}}$ and $\delta$ in Eq.~\eqref{HLLHstotalsum} as a restricted double sum,
\begin{equation}
     \sum_{n,\bar{n}}\phantom{}^{'} x^{\frac{n - s}{\k} -\frac{1}{2}} \, \xb^{\frac{\bar{n}}{\k}-\frac{1}{2}} \ ,
    \label{HLLHssumN}
\end{equation}
over pairs of non-negative integers $(n, \bar{n})$ satisfying the following restrictions
\begin{equation}
    \bar{n}-n  \;\equiv\;a \, \mathrm{mod}\, \k \qqquad n,\bar{n} \in \mathbb{N}_0  \, .
    \label{nnbaralpha}
\end{equation}
Indeed, for $a=\k$ we simply parametrize $n = \k \hat{n} + \delta$ and $\bar{n} = \k \hat{\bar{n}} + \delta$, which enforces the mod$~\k$  condition \eqref{nnbaralpha}, so that the sum gives the first term in \eqref{HLLHstotalsum}. On the other hand, for $a<\k$ we can take $n = \k \hat{n} + \delta-a$ and $\bar{n} = \k \hat{\bar{n}} + \delta$, so that the factors of $a$ coming from $s$ and $n$ cancel each other out in the exponent. However, in this case we need to explicitly subtract the contribution of all pairs  ($\hat{n},\delta$) which lead to $n<0$, thus again giving \eqref{HLLHstotalsum}. 
Therefore we observe agreement between \eqref{HLLHws}, \eqref{nnbaralpha-0} and \eqref{HLLHGalliani}, up to the overall normalization that was not fixed in the supergravity calculation of~\cite{Galliani:2016cai}. Here and throughout the paper, we shall not keep track of overall normalization factors.

For $\hat{s}=0$ and $\hat{s} = \k-1$, since Eq.\;\eqref{HLLHGalliani}  matches the corresponding symmetric orbifold CFT correlator, we have demonstrated an explicit match between the worldsheet and symmetric orbifold CFT correlators. This is striking, as it is an agreement across moduli space for a correlator that a priori is not covered by an existing non-renormalization theorem. 

This agreement is almost certainly due to the special nature of the heavy states we consider. Indeed, let us compare the worldsheet $x$-basis operator Eq.~\eqref{modesAdSk1} with the discussion of holographic CFT spectral flow in \cite[App.~A]{Avery:2009tu}.
Spectral flow in the holographic CFT is an automorphism of the small $(4,4)$ superconformal algebra, that is a useful tool to relate different states and operators. 
For instance, the heavy backgrounds we consider are related by fractional spectral flow to the $\k$-orbifolded NSNS vacuum, as discussed around Eq.\;\eqref{SFchargesHCFT}.

Given a symmetric orbifold CFT correlator, one can perform spectral flow on both the operators and the background states. The value of the correlator is invariant under this operation. One can use this to map the correlator in one of our heavy backgrounds to a correlator in the $\k$-orbifolded NSNS vacuum. Of course, for $\k=1$, after undoing the spectral flows one obtains a vacuum correlator.

Taking for simplicity $\k=1$ and $\bar{s}=0$, the transformation of a chiral primary operator under this operation is~\cite[App.~A]{Avery:2009tu}
\begin{equation}
\label{Averyshift}
    \tilde{{{O}}}_{h}(x) 
    \,=\, x^{ (2s+1) m'} {O}_{h}(x) \, . 
\end{equation}
The exponent of $x$ directly parallels the $x$ factors appearing in Eq.~\eqref{modesAdSk1}. 
This observation generalizes straightforwardly to $\bar{s}\neq 0$ and to $\k>1$, whereupon operators have fractional modes taking values in $\mathbb{Z}/\k$. We will comment further on the relation between worldsheet and symmetric product orbifold CFT correlators in due course.

\subsection{Non-BPS HLLH correlators for $h = \half$} 

The correlator presented in the previous subsection can be readily generalized to compute a set of novel HLLH correlators involving the same light operators, but probing the more general class of non-supersymmetric backgrounds given by the JMaRT solutions, in which both spacetime (fractional) spectral flow parameters $s$ and $\bar{s}$ are non-trivial. 
Note that the parameters $s, \bar{s}, \k$ defining the background must satisfy $s(s+1)-\bar{s}(\bar{s}+1) \in \k \, \mathbb{Z}$, from combining Eqs.\;\eqref{mnk-quant} and \eqref{SFchargesHCFT}.

The same steps as described in the previous subsection lead directly to the following generalization of Eq.~\eqref{HLLHws}:
\begin{equation}
    \bra{s,\bar{s},\k} \Yy_{\half} (1) \Yy_{\half}^\dagger (x) \ket{s,\bar{s},\k} = \frac{1}{\k^2}\sum_{n,\bar{n}}\phantom{}^{'} x^{\frac{n - s}{\k} -\frac{1}{2}} \, \xb^{\frac{\bar{n}-\bar{s}}{\k}-\frac{1}{2}} \ .
    \label{HLLHssumN2}
\end{equation}
To make precise the restricted summation, analogously to $s=\k p-a$ we write $\bar{s} = \k \bar{p} - \bar{a}$, with
$1\leq \bar{a} \le \k$.
We parametrize $n = \k \hat{n} + \delta - a$ and $\bar{n} = \k \hat{\bar{n}} + \delta - \bar{a}$ in order to satisfy the condition coming from the subtracted gauge constraint generalizing Eq.~\eqref{nnbaralpha-0}, namely
\begin{equation}
    \bar{n}-n \;\equiv\; ( a - \bar{a} ) \, \mathrm{mod}\, \k \, .
    \label{nnbaralpha2}
\end{equation}
Then, by summing over all possible values of $\hat{n}$, $\hat{\bar{n}}$ and $\delta$ such that $n$ and $\bar{n}$ are non-negative and satisfy \eqref{nnbaralpha2}, defining $b \equiv \mathrm{min}(a,\bar{a})$ we obtain 
\begin{align}
    \langle s,\bar{s},k| & \Yy_{\half}(1) \Yy_{\half}^\dagger(x,\xb) |s,\bar{s},k\rangle \,=\, \frac{1}{\k^2}\frac{x^{-p}\xb^{-\bar{p}}}{|x|} \, \times \label{HLLH2terms2} \\ 
    & \left[\frac{1}{|1-x|^2} \frac{1-|x|^2}{1-|x|^{2/\k}} - \frac{1}{1-\bar{x}}\frac{1-|x|^{2a/\k}}{1-|x|^{2/\k}} 
    - \frac{1}{1-x}\frac{1-|x|^{2\bar{a} /\k}}{1-|x|^{2/\k}} \, + \, \frac{1-|x|^{2 b/\k}}{1-|x|^{2/\k}}
    \right] \nn.
\end{align}
As before, the second and third terms remove contributions for which either $n$ or $\bar{n}$ become negative, while the fourth one compensates for the over-counting of cases in which both $n$ and $\bar{n}$ are negative.

At first sight, Eq.~\eqref{HLLH2terms2} may seem to depend on the values of $a$ and $\bar{a}$ separately, in apparent contradiction with the fact that, as is implied by \eqref{nnbaralpha2}, only their difference matters. However, the RHS of Eq.~\eqref{HLLH2terms2} can be rewritten as 
\begin{eqnarray}
    &&
    \frac{1}{\k^2} \frac{x^{-s/\k}\xb^{-\bar{s}/\k}}{|x||1-x|^2} \frac{1}{
    \left(1-|x|^{2/\k}\right)} 
    \left(\frac{x}{\bar{x}}\right)^{(\bar{a}-a)/2\k} \times \label{HLLH2terms3}  \\
    && \hspace{25mm}
    \left[ (1-x)|x|^{(a-\bar{a})/\k}
    + (1-\bar{x})|x|^{-(a-\bar{a})/\k}
    -|1-x|^2 |x|^{-|a-\bar{a}|/\k}
    \right], \nn
\end{eqnarray}
which explicitly depends only the orbifold parameter $\k$, the spectral flow parameters $s$ and $\bar{s}$, and the difference $a - \bar{a} \equiv \bar{s}-s \, \mathrm{mod}\, \k$, as expected. Note that \eqref{HLLH2terms3} is symmetric under the simultaneous replacements $x \leftrightarrow \bar{x}$ and $s \leftrightarrow \bar{s}$.

\medskip

The worldsheet correlators\;\eqref{HLLH2terms2}--\eqref{HLLH2terms3} are one of the main results of this paper. 
Unlike the $\bar{s}=0$ supersymmetric example of the previous subsection, generically the corresponding supergravity or holographic CFT correlators have not been computed in the literature. 

Since the backgrounds are non-supersymmetric when $s$ and $\bar{s}$ are both non-zero, again a priori there is no obvious reason to expect the correlators to be protected across moduli space.

However, better-than-expected agreement between supergravity and holographic CFT has already been observed for a closely related observable describing the analog of the Hawking radiation process~\cite{Chowdhury:2007jx,Avery:2009xr,Chakrabarty:2015foa}; we shall comment further on this in due course.

Moreover, for the particular cases where $s$ and $\bar{s}$ are congruent to either $0$ or $\k-1$ mod $\k$, the holographic CFT correlator follows straightforwardly from the techniques in App. A of \cite{Galliani:2016cai}, 
providing another exact match even for the non-supersymmetric backgrounds.
We shall generalize this further in the next section.

\section{More general heavy-light correlators}
\label{sec:Sec6}

In this section we present HLLH correlators for generic chiral primaries of conformal weights $h>1/2$. We then progress to describe higher-point heavy-light correlators. 
As an application, we compute the analogue of the Hawking radiation process from the backgrounds under consideration. We also compute a five-point HLLLH correlator of the symmetric product orbifold CFT, and demonstrate precise agreement with the corresponding worldsheet correlator.

\subsection{HLLH correlators for general $h$}
\label{sec:kthroots}

We now consider HLLH correlators where the light operators are chiral primaries with $h \geq 1$. For these correlators the $m$-basis SL$(2,\R)$ two-point functions do not trivialize, and the resulting sums become more complicated.

By following the method outlined in the previous sections, the worldsheet computation leads to restricted sums of the form
\begin{equation}
   \frac{1}{\k^{4h}} \sum_{n,\bar{n}}\phantom{}^{'} x^{\frac{n - 2 h s}{\k}- h} \xb^{ \frac{\bar{n} - 2 h \bar{s}}{\k}- h} \frac{\Gamma(2h+n)\Gamma(2h+\bar{n})}{\Gamma(2h)^2\Gamma(n+1) \Gamma(\bar{n}+1)} \ .
    \label{HLLHssumNh}
\end{equation}
For $\k=1$, there are no restrictions on the allowed values of the mode numbers $n$ and $\bar{n}$, and so the sum can be performed straightforwardly to obtain
\begin{equation}
      \langle s,\bar{s}, \k = 1|\Oo_h(1)\Oo^\dagger_h(x,\xb)|s,\bar{s},\k = 1\rangle 
     \,=\,   
     \frac{x^{-2 h s}\xb^{-2 h \bar{s}}}{|1-x|^{4h}} \, .
\end{equation}
This expression agrees with the corresponding HLLH correlator of the symmetric product orbifold CFT, as a direct consequence of the discussion around Eq.~\eqref{Averyshift}.

On the other hand, for $\k>1$ the sum becomes more difficult to carry out explicitly since $n$ and $\bar{n}$, which appear in the arguments of the Gamma functions, must satisfy the constraint 
\begin{equation}
\label{nnbaralphah}
\bar{n} - n \,\equiv\, 2 h \left(\bar{s}-s\right) \  \mathrm{mod} \  \k \ ,    
\end{equation} 
which is the direct generalization of Eq.\;\eqref{nnbaralpha2}.
We shall first briefly describe a method to construct these correlators iteratively, starting from the $h=\frac{1}{2}$ case obtained above, then present an improved method.

Our iterative construction operates by expressing the additional coefficients in the sum in Eq.\;\eqref{HLLHssumNh} in terms of differential operators acting on  results for lower values of $h$. 
Let us illustrate how this works for the simplest non-trivial case $h=1$. From the general expression \eqref{HLLHssumNh}, we have 
\begin{eqnarray}
      \langle s,\bar{s}, \k |\Oo_{h=1}(1)\Oo^\dagger_{h=1}(x,\xb)|s,\bar{s},\k\rangle 
     &\,=\,&  \frac{1}{\k^{4}}\sum_{n,\bar{n}}\phantom{}^{'} x^{\frac{n - 2 s}{\k}- 1} \xb^{ \frac{\bar{n} - 2 \bar{s}}{\k}-1} (n+1)(\bar{n}+1)  \\
     &\,=\,& \frac{1}{\k^{4}}\frac{x^{-\frac{2s}{\k}} \bar{x}^{-\frac{2\bar{s}}{\k}}}{|x|^2}
     (\k x \der_x+1)(\k \xb \der_{\xb}+1) 
     \sum_{n,\bar{n}}\phantom{}^{'} x^{\frac{n}{\k}} \xb^{ \frac{\bar{n}}{\k}} \nn \ .
\end{eqnarray}
Hence, the differential operators act on a sum similar to the one analyzed in the previous section. For the general case, the procedure iterates. We redefine $a$ and $\bar{a}$ to be  generalizations of the $a$ and $\bar{a}$ used in the $h=1/2$ correlators (see above Eqs.\;\eqref{nnbaralpha-0} and~\eqref{nnbaralpha2}), where we replace $s\mapsto 2hs$, $\bar{s}\mapsto 2h\bar{s}$, such that $a - \bar{a} \equiv 2h(\bar{s}-s) \, \mathrm{mod}\, \k$. Then we obtain
\begin{equation}
     \langle s,\bar{s}, \k |\Oo_{h}(1)\Oo_h^{\dagger}(x,\xb)|s,\bar{s},\k\rangle = \frac{1}{\k^{4h}}\frac{x^{-\frac{2 h s}{\k}} \bar{x}^{-\frac{2h \bar{s}}{\k}}}{|x|^{2h} } \frac{D_{h,\k}\bar{D}_{h,\k}}{\Gamma(2h)^2} \sum_{n,\bar{n}}\phantom{}^{'} x^{\frac{n}{\k}} \xb^{ \frac{\bar{n}}{\k}} \, , 
     \label{generalHLLH1a}
\end{equation}
where we have introduced differential operators of order $2h-1$ defined as
\begin{equation}
    D_{h,\k} \equiv (\k x \der_x+2h-1) \cdots (\k x \der_x+1) \, ,
    \label{generalHLLH1b}
\end{equation}
and where    
\begin{equation}
    \sum_{n,\bar{n}}\phantom{}^{'} x^{\frac{n}{\k}} \xb^{ \frac{\bar{n}}{\k}} \,=\, \frac{(1-x)|x|^{(a-\bar{a})/\k}
    + (1-\bar{x})|x|^{-(a-\bar{a})/\k}
    -|1-x|^2 |x|^{-|a-\bar{a}|/\k}}{|1-x|^2
    \left(1-|x|^{2/\k}\right)} 
    \left(\frac{x}{\bar{x}}\right)^{(\bar{a}-a)/2\k} .
    \label{generalHLLH1c}
\end{equation}

Although it leads to correct results, the procedure outlined above quickly becomes cumbersome, and leads to seemingly complicated expressions for higher values of $h$. 
In addition, it does not appear to give any insight into whether the results are likely to match with computations in the symmetric product orbifold CFT. 
However, we can improve on both these points with a different method. We now describe this method by first rederiving the $h=1/2$ correlators of the previous section, and then generalizing the improved method
to arbitrary values of $h$, and also to higher-point heavy-light correlators.

Let us thus re-examine the general expression Eq.~\eqref{HLLHssumNh}, and consider the case in which $a-\bar{a} = 0$ for simplicity. We can take into account the restriction on the allowed values of $n$ and $\bar{n}$ by considering an \textit{unrestricted} sum over arbitrary positive integers by making use of a ``Kronecker comb''. In other words, we impose that $n-\bar{n} = 0 \ \mathrm{mod} \ \k$ by including extra coefficients of the form 
\begin{equation}
    \sum_{q \in \mathbb{Z}} \delta_{n-\bar{n},\k q} \,=\,  \frac{1}{\k} \sum_{r = 0}^{\k-1} e^{2 \pi i r \, \frac{n - \bar{n}}{\k}}, 
    \label{Kroneckercomb}
\end{equation}
where the final equality is obtained by Fourier transformation, and represents a simple form of the discrete Poisson summation formula. The RHS in Eq.~\eqref{Kroneckercomb} is interesting, because the exponentials can be absorbed into terms involving powers of $x$ and $\bar{x}$. Explicitly, 
we can rewrite the expression \eqref{HLLHssumNh} with $a-\bar{a} = 0$ as
\begin{equation}
\label{eq:6-9}
    \frac{1}{\k^{4h+1}} \frac{x^{-\frac{2 h s}{\k}} \bar{x}^{-\frac{2h \bar{s}}{\k}}}{|x|^{2h} } \, \
    \sum_{n,\bar{n}\geq 0} \sum_{r =0}^{\k-1} u_r^n \bar{u}_r^{\bar{n}} \frac{\Gamma(2h+n)\Gamma(2h+\bar{n})}{\Gamma(2h)^2\Gamma(n+1) \Gamma(\bar{n}+1)} \ ,
\end{equation}
where $u_r$, $\bar{u}_r$ are the $\k^{\mathrm{th}}$ roots of $x$ and $\bar{x}$, respectively; writing $x^{\frac{1}{\k}} \equiv |x|^{\frac{1}{\k}}e^{2\pi i \frac{{\rm Arg} (x)}{\k}}$,
\begin{equation}
    u_r \,\equiv\, x^{\frac{1}{\k}} e^{2 \pi i \, \frac{r}{\k}} \qqquad 
    \bar{u}_r \,\equiv\, \bar{x}^{\frac{1}{\k}} e^{-2 \pi i \, \frac{r}{\k}} \;. 
    \label{yrdef}
\end{equation}
Thus, inside the convergence region $|x|<1$ we can exchange the order of the sums, such that the \textit{unrestricted} sum over integers $n$ and $\bar{n}$ leads to the usual expression for the CFT two-point function. However, it is evaluated at the different values of $u_r$, 
instead of the insertion point itself. Thus, the expression \eqref{eq:6-9} becomes
\begin{equation}
    \frac{1}{\k^{4h+1}} \frac{x^{-\frac{2 h s}{\k}} \bar{x}^{-\frac{2h \bar{s}}{\k}}}{|x|^{2h} } \sum_{r=0}^{\k-1} \frac{1}{|1-u_r|^{4h}} \, .
\end{equation}

In fact, we can rewrite this in a slightly more general form. Indeed, it is easy to ``unfix'' the first insertion point and write the full expression of the HLLH correlator in terms of $x_1$ and $x_2$. To do so, we introduce $\k^{\mathrm{th}}$ roots of $x_1$, $x_2$, and $x_2/x_1$ via
$u_{1,r_1}^\k=x_1$, $u_{1,r_2}^\k=x_2$, and $u_{21,r}^\k=x_2/x_1$, and then
make use of the identity 
\begin{equation}
    \frac{1}{|x_1|^{\frac{4h}{\k}}}\sum_{r=0}^{\k-1} \frac{1}{|1-u_{21,r}|^{4h}} \,=\ \frac{1}{\k} \sum_{r_1,r_2=0}^{\k-1} 
    \frac{1}{|u_{1,r_1}-u_{2,r_2}|^{4h}}
    \; .
    \label{idx1x2a}
\end{equation}
This gives
\begin{align}
    \begin{aligned}
    & \hspace{-1mm} \langle s,\bar{s}, \k |\Oo_h(x_1,\bar{x}_1)\Oo^\dagger_h(x_2,\bar{x}_2)|s,\bar{s},\k\rangle\Big|_{\,
    2 h (s-\bar{s})\, = \, 0 \, \mathrm{mod} \, \k } \,  \, \\[2ex]
    &\hspace{9mm}=\, \frac{1}{\k^{4h+2}}\left(\frac{x_2}{x_1}\right)^{-h \frac{(2s+1)}{\k}}
    \left(\frac{\bar{x}_2}{\bar{x}_1}\right)^{-h \frac{(2\bar{s}+1)}{\k}}
    |x_1 x_2|^{2h \left(\frac{1}{\k}-1\right)} 
 \sum_{r_1,r_2=0}^{\k-1} 
    \frac{1}{|u_{1,r_1}-u_{2,r_2}|^{4h}} \ .
\end{aligned}
\end{align}

The general case is computed entirely analogously. We must simply replace $n - \bar{n} \mapsto n - \bar{n} + (a-\bar{a})$ in Eq.~\eqref{Kroneckercomb}, which induces some extra phases. The appropriate generalization of Eq.~\eqref{idx1x2a} is given by 
\begin{equation}
    \frac{1}{|x_1|^{\frac{4h}{\k}}}\sum_{r=0}^{\k-1} \frac{e^{2\pi i r 
    (a-\bar{a})/\k 
    } }{|1-u_{21,r}|^{4h}} \,=\, \frac{1}{\k} \sum_{r_1,r_2=0}^{\k-1} 
    \frac{e^{2\pi i (r_2-r_1) 
    (a-\bar{a})/\k 
    }}{|u_{1,r_1}-u_{2,r_2}|^{4h}} \;.
    \label{idx1x2b}
\end{equation}
Then the HLLH correlator with generic values of the orbifold parameter $\k$, the spectral flow parameters $s$ and $\bar{s}$, and the weight of the light chiral primary operator $h$, takes the form 
\begin{align}
\label{HLLHCPprefinal}
    \begin{aligned}
    & \hspace{-3mm} \langle s,\bar{s}, \k |\Oo_h(x_1,\bar{x}_1)\Oo^\dagger_h(x_2,\bar{x}_2)|s,\bar{s},\k\rangle \,  \, \\[2ex]
    & \hspace{4.5mm} =\, \frac{1}{\k^{4h+2}}
    \left(\frac{x_2}{x_1}\right)^{-h \frac{(2s+1)}{\k}}
    \left(\frac{\bar{x}_2}{\bar{x}_1}\right)^{-h \frac{(2\bar{s}+1)}{\k}}
    |x_1 x_2|^{2h \left(\frac{1}{\k}-1\right)} 
    \sum_{r_1,r_2=0}^{\k-1} 
    \frac{e^{2\pi i (r_2-r_1) 
    (a-\bar{a})/\k 
    }}{|u_{1,r_1}-u_{2,r_2}|^{4h}} \;,
\end{aligned}
\end{align}
where the sum is over the $\k$-th roots of the insertion points $x_1$ and $x_2$, as defined above \eqref{idx1x2a}, and where $2 h (\bar{s}-s) \equiv a - \bar{a}$ mod $\k$.

Note that we can relax the chiral primary condition and consider operators in which $m'\neq \pm h$.
We shall continue to focus on massless vertex operators, however this could be generalized further.
In addition, by making use of the phases and the $x_{1,2}$ powers on the RHS of \eqref{HLLHCPprefinal}, we can rewrite the result in a cleaner form, 
\begin{align}
\begin{aligned}
     &\hspace{-4mm}\langle s,\bar{s}, \k |\Oo_{h,m'}(x_1,\bar{x}_1)\Oo_{h,m'}^\dagger(x_2,\bar{x}_2)|s,\bar{s},\k\rangle  \\  
      & \hspace{15mm}=\, 
      \frac{1}{\k^{2}} 
    \sum_{r_1,r_2=0}^{\k-1} 
    \left(\frac{u_{2,r_2}}{u_{1,r_1}}\right)^{-m'(2s+1)}
    \left(\frac{\bar{u}_{2,r_2}}{\bar{u}_{1,r_1}}\right)^{-\bar{m}'(2\bar{s}+1)}
    \frac{|u_{1,r_1} u_{2,r_2}|^{2h(1-\k)}}{\k^{4h}|u_{1,r_1}-u_{2,r_2}|^{4h}} \; .
    \label{finalHLLH}
\end{aligned}
\end{align}

\subsection{Matching between worldsheet and symmetric product orbifold}
\label{sec: matchingwoldsheetorbifold}

The appearance of the $\k^\mathrm{th}$ roots of the physical insertions $x_{1,2}$ in Eq.~\eqref{finalHLLH} is related to the fact that the 
holographic description of the heavy backgrounds involves heavy states in 
$\k$-twisted sectors of the boundary CFT. The same feature appears in certain computations performed using the Lunin-Mathur covering space technique~\cite{Lunin:2001pw}, specifically when there are operators of twist $\k$ inserted at the origin and infinity of the CFT plane, and when there are other untwisted operators in the correlator. Then the coordinate transformation to the $\k$-fold covering space is precisely $x=u^\k$. 

Thus, when the light worldsheet operators correspond to untwisted operators of the symmetric product orbifold CFT, it is natural to identify $u$ with the coordinate on the $\k$-fold covering space that trivializes the twist operators involved in the definition of the heavy states.

The sum over the different roots generates the usual phases included in the definition of fractional modes by summing over the different copies of the theory~\cite{Lunin:2001pw},
\begin{equation}
\label{eq:o-frac}
    O_{\frac{m}{\k}} = \oint \frac{dx}{2\pi i} \sum_{r=1}^\k 
    O_{(r)}(x)
    e^{\frac{2 \pi i m}{\k}(r-1)} x^{h+\frac{m}{\k}-1} \,.
\end{equation}
Moreover, the \textit{fractional} spectral flow defining the background, when mapped to a $\k$-fold covering space, becomes  \textit{integer} spectral flow with parameters $2s+1$ and $2\bar{s}+1$~\cite{Giusto:2012yz,Chakrabarty:2015foa}.
Hence, one can generalize the discussion around Eq.~\eqref{Averyshift} and simply consider the appropriate powers of $u_{i,r_i}$ to arise from performing spacetime spectral flow on the operators, in the $\k$-fold covering space. Finally, the last factor on the RHS of \eqref{finalHLLH} corresponds to the usual two-point function evaluated at the roots, including the necessary Jacobian factors arising from mapping to the $\k$-fold covering space, $|\partial u / \partial x|^{2h}$.
Obtaining precisely this Jacobian is the justification for the factors of $\k$ introduced in the definition of the $x$-basis operators in Eq.~\eq{modesAdSk1}.
Thus, we see that symmetric orbifold CFT HLLH correlators for which the covering space is $x=u^\k$ agree in both structure and value with the worldsheet correlator~\eqref{finalHLLH}.

By contrast, for twisted operators, the interpretation of our worldsheet result \eqref{finalHLLH} is more involved: the Lunin-Mathur covering map for such correlators is \textit{not} $x=u^\k$.
To understand the precise relation, we focus on light operators of twist two, and show that Eq.~\eqref{finalHLLH} nevertheless matches with the symmetric orbifold CFT also in this case.
The relevant four-point function was studied recently in 
\cite{Lima:2020nnx,Lima:2021wrz,AlvesLima:2022elo} 
in the Sym$^{N}(\TT^4)$ CFT. At leading order in large $N$, the correlator is 
dominated by a contribution from a covering space with genus zero, where the copy indices of the light twist-two operator act on different $\k$-strands corresponding to the heavy state. One of the light insertions effectively joins together two $\k$-strands into a $2\k$-strand, and the other light insertion effectively cuts the $2\k$-strand back to two strands of length $\k$. 
For this process, and setting for simplicity $s=\bar{s}=0$ as done in \cite{Lima:2021wrz}, the relation between the physical-space cross-ratio $x$ and the covering-space cross ratio $v$ is\footnote{Note that in \cite{Lima:2021wrz,AlvesLima:2022elo}, the base (physical) space coordinates are denoted by $z$ or $u$ rather than our $x$, while the covering space coordinates are denoted by $t$ or $x$ rather than our $v$.}
\begin{equation}\label{eq: covering Lima}
    x(v) = \left(\frac{v+1}{v-1}\right)^{2\k} \, . 
\end{equation}
The correlator of interest involves the function 
\begin{equation}
    \frac{(v+1)^{2+2\k}(v-1)^{2-2\k}}{v^{2}} \, ,
\end{equation}
where $v(x)$ is defined through Eq.~\eqref{eq: covering Lima}. 
The correlator itself is obtained 
by summing over the $2\k$ pre-images of $x$ and including an $N$- and $\k$-dependent overall factor. 
However, due to the $v \to 1/v$ symmetry of the map \eqref{eq: covering Lima}, there are actually only $\k$ distinct contributions \cite{AlvesLima:2022elo}, corresponding to distinct ramified coverings of the base space. 
When normalisation factors are taken into account, this corresponds to what in \cite{AlvesLima:2022elo} is called a `Hurwitz block function'. Upon inserting the explicit solutions
\begin{equation}
    v_r(x) \,=\, \frac{x^{\frac{1}{2\k}}e^{\frac{i \pi r }{\k}}+1}{x^{\frac{1}{2\k}}e^{\frac{i \pi r }{\k}}-1}  \qqquad r = 0, \dots, \k - 1 \,,
\end{equation}
where as before $x^{\frac{1}{2\k}}$ stands for a particular $(2\k)^{\mathrm{th}}$ root of $x$, the final expression remarkably coincides with the $s=\bar{s}=0$ case of Eq.~\eqref{finalHLLH}.
The analysis for the JMaRT states and for more general light insertions can be carried out analogously.

Recall that, as reviewed in Section~\ref{secCPD1D5CFT}, at a generic spacetime dimension $h$ there is a degeneracy in the twist $n$ of light states in the symmetric product orbifold CFT. 
An interesting feature of the worldsheet correlator \eqref{finalHLLH} is that it is independent of this twist $n$. 
Recall also that, for untwisted light operators, the worldsheet correlator has the same structure as the covering space method of the symmetric product orbifold CFT.
The fact that the agreement of HLLH correlators extends to (at least some) twisted light operators is thus remarkable from the point of view of the holographic CFT. Despite the more complicated covering map,
the above discussion demonstrates how, for these correlators, the end result agrees with an expression whose structure is that of the simple map $x=u^\k$.

\subsection{Higher-point heavy-light correlators}
\label{sec:higher-point}

Our general expression for HLLH correlators, Eq.~\eqref{finalHLLH}, together with the matching to the symmetric product orbifold CFT that we have observed so far, motivate a deeper exploration. Thus, we now describe how local $x$-basis operators are seen from the spectrally flowed frame as indicated by our null-gauged worldsheet models. This will allow us to extract consequences for worldsheet three-point and higher-point functions, corresponding to holographic CFT correlators with two heavy states and three or more light operators.

The AdS$_3$ limit of the holomorphic gauge condition, \eqref{nullgaugeconstrAdS}, upon using the definition of $m_y$ in Eq.~\eqref{mymodes}, reads
\begin{equation}
    0 = m + (2s+1) m' - \k m_y \, .
    \label{gaugexbasis1}
\end{equation}
We wish to re-interpret this constraint in the local coordinate basis of the holographic CFT. A priori, it is perhaps not obvious that this is a useful thing to do, since the usual $x$-basis operators are constructed by resumming the action of $J_0^\pm$, which does not commute with the BRST charge in the coset theory. However we shall see that it will be very useful.

Let us observe that there are two notions of $x$-type local coordinates in the worldsheet model. The one used so far in Section \ref{Section 5} and the present section is the physical $x$ coordinate of the gauged models.  However, before gauging, there is an analogous coordinate for the  \textit{upstairs} SL(2,$\R$) algebra. We will denote the associated coordinate by the complex variable $u$; we will see momentarily that $u^\k = x$, so that there will be no clash with the $u$ used above.

The differential operator $x \der_x + h$ corresponds to the quantity $m_y$, as can be seen by comparing~Eqs.~\eqref{diffSL2}, \eqref{ExpSL2}, \eqref{modesAdSk1}. On the other hand, the upstairs SL(2,$\R$) projection $m$ corresponds to an analogous operator in the $u$ variable: we write this as $u \der_u + h - \beta$, where we have allowed for a shift $\beta$, whose precise form will become clear shortly, as will the reason for its existence. Then Eq.~\eqref{gaugexbasis1} can be expressed in terms of these differential operators as
\begin{equation}
\label{eq:x-u}
    \k x \der_x \,=\, u \der_u + (2s+1)m' + h(1-\k) - \beta  \, . 
\end{equation}
In order that this condition is solved by $u^\k = x$, we choose 
\begin{equation}
    \label{beta}
    \beta = h(1-\k) + (2s+1)m'  \, .
\end{equation}
Thus the role of $\beta$ is two-fold. On the one hand, the first term in \eqref{beta} effectively replaces the weight by $h \to h_u \equiv \k h$, which further supports the discussion above about $u$ corresponding to a covering space coordinate in the holographic CFT. It also generates the Jacobian factor obtained in Eq.~\eqref{finalHLLH}. On the other hand, the second term in \eqref{beta} takes into account the shift arising from spacetime spectral flow.

We now use this to obtain an improved construction of gauge-invariant operators directly in the $x$-basis, built upon $u$-basis operators of the upstairs SL(2,$\R$), i.e.~without relying on their spacetime Virasoro mode expansion as in~\eqref{modesAdSk1}. Although such a construction gives equivalent results at the level of worldsheet two-point functions~\eqref{finalHLLH}, its importance for higher-point functions was highlighted recently in~\cite{Dei:2021xgh}. The construction proceeds as follows:
\begin{enumerate}
    \item We consider an operator whose upstairs SL(2,$\R$) part is expressed in the usual local SL(2,$\R$) basis, $\mathcal{V}_h(u,\bar{u})$, where for simplicity we set $h=\bar{h}$. We multiply this by an SU(2) vertex operator $\mathcal{V}'_{j',m',\bar{m}'}$. We suppress the exponentials of $t$ and $y$, since they have weight zero in the AdS$_3$ limit, and their only effect is taken into account through \eqref{gaugexbasis1} and its antiholomorphic counterpart. We introduce the notation
    \be
    \hat{\Oo}_{h,m',\bar{m}'}(u,\bar{u})\,\equiv\, \cV_h(u,\bar{u})\mathcal{V}'_{j',m',\bar{m}'}\,.
    \ee
    \item We introduce the above $\beta$-shift by multiplying by an extra factor $u^\beta \bar{u}^{\bar{\beta}}$.
    \item We sum the resulting operator over all insertion points $u$ such that $u^\k=x$.
\end{enumerate}
Explicitly, we define
\begin{equation}
\label{shiftwithk}
    \Oo_{h,m',\bar{m}'}(x,\bar{x}) 
    \,\equiv\,
    \frac{1}{\k^{2h+1}} \sum_{u^\k = x} u^{\beta} \bar{u}^{\bar{\beta}} \hat{\Oo}_{h,m',\bar{m}'}(u,\bar{u}) \, .
\end{equation}
Comparing with Eq.~\eqref{modesAdSk1}, using the Kronecker comb \eqref{Kroneckercomb} to impose the constraints as above, we indeed have
\begin{align}
    {{\cal{O}}}_{h,m',\bar{m}'}(x,\bar{x}) & \;\!\equiv\;\! \frac{1}{\k^{h + \bar{h}}}  \sum_{m_y,\bar{m}_y}
    x^{m_y-h}\bar{x}^{\bar{m}_y-h} \;\!\Vv_{j,m,\bar{m}}\Vv'_{j',m',\bar{m}'} \nn \\
    &\;\!=\;\! \frac{1}{\k^{2h}} \sum_{m,\bar{m}}\phantom{}^{'} 
    x^{\frac{1}{\k}\left[m + (2s+1)m'\right]-h}\bar{x}^{\frac{1}{\k}\left[\bar{m} + (2\bar{s}+1)\bar{m}'\right]-h} \;\! \Vv_{j,m,\bar{m}}\Vv'_{j',m',\bar{m}'} \nn \\
    &\;\!=\;\! \frac{1}{\k^{2h+1}} \!\!\: \sum_{u^\k\,=\,x} \sum_{m,\bar{m}}
    u^{m-h+(2s+1)m'+h(1-\k)}\bar{u}^{\bar{m} -h+(2\bar{s}+1)\bar{m}'+h(1-\k)} \;\! \Vv_{j,m,\bar{m}}\Vv'_{j',m',\bar{m}'} \nn \\
    &\;\!=\;\! \frac{1}{\k^{2h+1}} \!\!\: \sum_{u^{\k}\,=\,x} u^{\beta}\bar{u}^{\bar{\beta}} \; \! \hat{\Oo}_{h,m',\bar{m}'}(u,\bar{u})   \,.
    \label{eq:op-beta}
\end{align}

We note that in the symmetric product orbifold CFT, when mapping the $S_{\k}$-invariant untwisted operators $O(x) = \sum_{r=1}^{\k}O_{(r)}(x)$ to the $\k$-fold covering space, using the 
inverse relation to \eqref{eq:o-frac},
\begin{equation}
\label{eq:or-inv}
    O_{(r)}(x) \,=\, \frac{1}{\k} \sum_m O_{\frac{m}{\k}}  x^{-\frac{m}{\k}-h} e^{-\frac{2 \pi i m}{\k}(r-1)}, 
\end{equation}
one obtains an expression closely analogous to Eq.\;\eqref{shiftwithk}.

We now exploit the expression \eqref{eq:op-beta} to study higher-point functions. We first rewrite the HLLH correlator \eqref{finalHLLH} in the simple form 
\begin{equation}
\label{finalHLLHbeta}
     \langle s,\bar{s}, \k |\Oo_{h,m'}(x_1,\bar{x}_1)\Oo_{h,m'}^{\dagger}(x_2,\bar{x}_2)|s,\bar{s},\k\rangle  \,=\,  \frac{1}{\k^{4h+2}} 
    \sum_{u_i^\k=x_i}
    \dfrac{u_1^{ \beta_1}\bar{u}_1^{ \bar{\beta}_1} u_2^{ \beta_2}\bar{u}_2^{ \bar{\beta}_2}   }{|u_{1}-u_{2}|^{4h}}  \, ,
\end{equation}
with $\beta_i = h_i(1-\k) + (2s+1)m_i' , \, \bar{\beta}_i = h_i(1-\k) + (2\bar{s}+1)\bar{m}_i' $, and where the charge conservation $m_1' + m_2' = 0$ is understood. 
We then observe that the worldsheet correlator with $n$ light insertions with weights $h_i$ and charges $m_i', \bar{m}_i'$ is given by the following straightforward generalization of \eqref{finalHLLHbeta} (in which we partially suppress antiholomorphic quantities):
\begin{align}
\label{finalHLLLLLH}
    \langle s,\bar{s}, \k |  \Oo_{h_1,m'_1}(x_1) &\dots \Oo_{h_n,m'_n}(x_n)  |s,\bar{s},\k\rangle  \nn\\
    & =\frac{1}{\k^{H+\bar{H}+n}} 
    \sum_{u_i^\k=x_i} 
    \left(\prod_{\ell=1}^n u_\ell^{\beta_\ell}
    \bar{u}_\ell^{\bar{\beta}_\ell} 
    \right)
    \langle \hat{\Oo}_{h_1,m'_1}(u_1) \dots \hat{\Oo}_{h_n,m'_n}(u_n)\rangle
    \, , 
\end{align}
where $H=h_1 + \dots + h_n$, and $\langle \hat{\Oo}_{h_1,m'_1}(u_1) \dots \hat{\Oo}_{h_n,m'_n}(u_n)\rangle$ stands for the global AdS$_3 \times \SSS^3$ vacuum $n$-point function evaluated at the roots of the original insertion points. 

The expression \eqref{finalHLLLLLH}, which holds for generic values of $s,\bar{s},\k$ and generic light weights and charges $h_i,m_i',\bar{m}_i$, constitutes one of the main results of this paper. In Eq.~\eqref{finalHLLLLLH}, the $n=3$ case can be made quite explicit, as we shall do so in the next subsection.

The above result can straightforwardly seen to include spectrally-flowed vertex operators, as follows. Setting $\w'=\bar{\w}'=\w$ for simplicity, the bosonic null-gauge condition Eq.~\eqref{nullgaugeconstrW} in the AdS limit becomes
\begin{equation}
    0 = m_\w + (2s+1) m'_\w - \k m_y \qqquad 0 = \bar{m}_\w + (2\bar{s}+1) \bar{m}'_\w - \k \bar{m}_y \, , 
    \label{gaugexbasis1spectralflow}
\end{equation}
where, for discrete states in the lowest weight representation, $m_\w = h_\w + n $, $  h_\w = J + n_5\, \w/2$ and $m'_\w = h_\w' + n_5\, \w/2 - n'$. 
As a consequence, the exponent $\beta$ of the covering space coordinate $u$ gets replaced by $\beta \mapsto \beta_\w = h_\w(1-\k) + (2s+1) m'_\w$, and the power of $\k$ in the normalisation factor is modified accordingly. Thus, for the vertex operators in the coset models, the net effects of the spectral flow procedure are the replacements $h \mapsto h_\w, m \mapsto m'_\w$. This is understood by the fact that when a boundary light operator has the spacetime dimension $h = J$ that renders the SL(2,$\R$) spin above the unitary bound Eq.~\eqref{Djrange}, it corresponds holographically to a spectrally-flowed worldsheet vertex operator \cite{Maldacena:2000hw}. This implies that the structure of the correlator in Eq.~\eqref{finalHLLLLLH} is not drastically modified when $\w \ne 0$. 

It is important to note, however, that the entire computational complication due to worldsheet spectral flow remains present in the resulting \textit{vacuum} correlator. Indeed, we see that the $n$-point function on the heavy state is now written in terms of a vacuum $n$-point function of spectrally-flowed states. It is thus natural to expect that the AdS$_3$ selection rules carry over to $n$-point functions in the JMaRT microstates. We conclude that the generalisation of Eq.~\eqref{finalHLLLLLH} to the case of worldsheet spectrally-flowed states reads 
\begin{align}
\label{finalHLLLLLHspectralflow}
    \langle s, \bar{s}, \k | \Oo_{h_1,m'_1}^{\w_1}(x_1) & \dots \Oo_{h_n,m'_n}^{\w_n}(x_n) | s, \bar{s}, \k \rangle  = \\
    & \frac{1}{\k^{H_\w + \bar{H}_\w+n}} 
    \sum_{u_i^\k=x_i} 
    \left( \prod_{\ell=1}^n u_\ell^{\beta_{\w,\ell}}
    \bar{u}_\ell^{\bar{\beta}_{\w,\ell}} \right)
    \langle \hat{\Oo}_{h_1,m'_1}^{\w_1}(u_1) \dots \hat{\Oo}_{h_n,m'_n}^{\w_n}(u_n)\rangle \nn \,,
\end{align}
where $H_\w = \sum_{i} h_{\w,i}$ and the light operators $\Oo_{h_i,m'_i}^{\w_i}$ are $x$-basis spectrally-flowed worldsheet vertex operators. 

We emphasize that the construction we have outlined in this section only holds in the IR AdS$_3 \times \SSS^3$ limit. In the full asymptotically linear dilaton geometry, the identification of the modes $m_y$ and $\bar{m}_y$ as defined in \eqref{mymodes} breaks down, and the $t$ and $y$ exponentials can no longer be ignored. This is consistent with the fact that in the UV the dual holographic theory is not a CFT, but is instead a little string theory. Since little string theories are non-local, it is correct that the above definition of local operators does not apply. Note, however, that the \textit{mode} correlators computed in the $m$-basis still make perfect sense, and carry information about string perturbation theory in the full geometry.

Let us speculate on which subset of the above correlators can be expected to agree with those of the symmetric product orbifold theory.
Since our expressions for the general correlators~\eqref{finalHLLLLLH}, \eqref{finalHLLLLLHspectralflow} involve vacuum correlators, it is natural to conjecture that for these particular heavy backgrounds, the heavy-light correlator is protected whenever the global AdS$_3 \times \SSS^3$ vacuum correlator appearing in~\eqref{finalHLLLLLH}, \eqref{finalHLLLLLHspectralflow} is protected. 
Recall that, in the global AdS$_3 \times \SSS^3$ vacuum, two-point and three-point correlation functions of chiral primaries are protected~\cite{Baggio:2012rr}, while four-point and higher-point functions are generically renormalized. 
So heavy-light correlators with two or three light insertions on these backgrounds may be protected between worldsheet and symmetric product orbifold CFT.
It may even be possible to prove a non-renormalization theorem generalizing~\cite{Baggio:2012rr}; work in this direction is in progress. %
For now however, we next compute a heavy-light correlator with three light insertions in both worldsheet and holographic CFT.

\subsection{An HLLLH correlator in worldsheet and holographic CFT}
\label{sec:HLLLH}

We now investigate the general expression for our worldsheet correlator \eqref{finalHLLLLLH}, in a particular example with three light insertions, and compare it to the symmetric product orbifold CFT. We shall observe another highly non-trivial agreement.

We consider three light insertions living in the untwisted sector of the holographic CFT, with weights $(h_1,h_2,h_3)=(\frac{1}{2},\frac{1}{2},1)$. In the dual CFT notation, we are then interested in computing the correlator $\langle O_{\frac{1}{2}} (x_1) O_{\frac{1}{2}}(x_2) O^\dagger_1(x_3) \rangle_H$. We further focus on heavy backgrounds with $s=\k p$ with $p \in \mathbb{Z}$ and $\bar{s}=0$.  

We start by evaluating the general expression \eqref{finalHLLLLLH} for this particular worldsheet correlator. In the worldsheet theory associated to the global AdS$_3\times \SSS^3$, the $O_{\frac{1}{2}}$ correspond to two RR states, while $O_1$ is an NSNS state polarized on the $\SSS^3$ directions. The (integrated) vacuum three-point functions for these chiral primaries were studied in \cite{Dabholkar:2007ey}. In our notation, they take the form
\begin{equation}
    \langle \Oo_{h_1}^{\mathrm{RR}} (x_1) \Oo_{h_2}^{\mathrm{RR}}(x_2) \Oo_{h_3}^{\dagger \, \mathrm{NSNS}}(x_3) \rangle \,=\,  
    \frac{1}{N^{1/2}}\frac{\sqrt{(2j_1-1)(2j_2-1)(2j_3-1)^{-1}}}{|x_{12}|^{2(h_1+h_2-h_3)}|x_{13}|^{2(h_1+h_3-h_2)}|x_{23}|^{2(h_2+h_3-h_1)}}\,, 
\end{equation}
where $j_1 = h_1+\frac{1}{2}$, $j_2 = h_2+\frac{1}{2}$, and $j_3 = h_3$. The relevant values for us are simply $j_i=1$, and upon a global SL(2,$\mathbb{C}$) transformation to set $x_3=1$, we have  
\begin{equation}
    \langle \Oo_{\frac{1}{2}} (x_1) \Oo_{\frac{1}{2}}(x_2) \Oo^\dagger_1(x_3) \rangle\,=\, 
    \frac{1}{N^{1/2}}\frac{1}{|1-x_{1}|^{2}|1-x_{2}|^{2}}\,.
    \label{3ptAdS3}
\end{equation}
In order to compute the HLLLH correlator in the worldsheet coset models corresponding to the JMaRT backgrounds, we must sum \eqref{3ptAdS3} evaluated at all $\k^{\mathrm{th}}$-roots of the insertion points. An explicit expression can be obtained following the arguments of Sec.~\ref{HLLHh1/2}, using the following generalisation of Eq.~\eqref{idx1x2a} and Eq.~\eqref{idx1x2b} for the case of three insertions,
\begin{equation}
    \frac{\k}{|x_3|^{\frac{2 (\alpha + \beta)}{\k}}} \sum_{r_{1,2} = 0}^{\k-1} \frac{1 }{|1-u_{13,r}|^{2 \alpha}|1-u_{23,r}|^{2 \beta}} \,= \sum_{r_{1,2,3} = 0}^{\k-1} \frac{1}{|u_{3,r_3} - u_{1,r_1}|^{2 \alpha} |u_{3,r_3} - u_{2,r_2}|^{2 \beta}}\,,
\end{equation}
where $u_{i,r_i} = x_i^{1/\k} e^{2 \pi i \, r_i/\k }$ and $u_{j\ell,r} = (x_j/x_\ell)^{1/\k} e^{2 \pi i \, r/\k }$ with $\alpha = \beta = 1$, one obtains
\begin{equation}
    \langle \Oo_{\frac{1}{2}} (x_1) \Oo_{\frac{1}{2}}(x_2) \Oo^\dagger_1(x_3) \rangle_H \,=\, \frac{1}{\k^{7}} \frac{
    (x_1x_2)^p \, |x_1 x_2|^{\frac{2}{\k}-1}}{
    |1-x_{1}|^{2}|1-x_{2}|^{2}} 
    \, \frac{1-|x_1|^2}{1-|x_1|^{\frac{2}{\k}}} \, \frac{1-|x_2|^2}{1-|x_2|^{\frac{2}{\k}}}\,. 
    \label{HLLLHformWS}
\end{equation}
The result \eqref{HLLLHformWS}
constitutes the first computation of a
heavy-light worldsheet correlator with three light insertions probing a black hole microstate. 

We now show that the same result can be obtained from the HCFT at the symmetric orbifold point. We follow the method used in~\cite[App.~A]{Galliani:2016cai} for the HLLH correlator reviewed in Section~\ref{HLLHh1/2}. 
The heavy states we use indicate that we should work in the $\k$-twisted sector of the theory. 
The operators can be written in terms of the fermions introduced in Eq.~\eqref{rhofermions}. For the $h=\frac{1}{2}$ chiral primaries, and for each strand of $\k$ copies of the theory, this simply reads 
\begin{equation}
    O_{\frac{1}{2}} \,=\, \sum_{r=1}^\k O_{\frac{1}{2},(r)} \,=\, -\frac{i}{\sqrt{2}} \sum_{r=1}^{\k} \psi_{(r)}^{+ \dot{A}} \tilde{\psi}_{(r)}^{+ \dot{B}} \epsilon_{\dot{A}\dot{B}} \,=\, 
    -\frac{i}{\sqrt{2}} \sum_{\rho=0}^{\k-1} \psi_\rho^{+ \dot{A}} \tilde{\psi}_\rho^{+ \dot{B}} \epsilon_{\dot{A}\dot{B}} \; , 
\end{equation}
while for the $h=1$ operator we find 
\begin{eqnarray}
     O^\dagger_{1} \,=\, \sum_{r=1}^\k O_{1,(r)} &=& \frac{1}{4}\sum_{r=1}^{\k} \psi_{(r)}^{- \dot{A}}
     \psi_{(r)}^{- \dot{B}}
     \tilde{\psi}_{(r)}^{- \dot{C}}
     \tilde{\psi}_{(r)}^{- \dot{D}}
     \epsilon_{\dot{A}\dot{B}}
     \epsilon_{\dot{C}\dot{D}}
     \nn  \\
     &=& \frac{1}{4\k}\sum_{\rho_i=0}^{\k-1} 
     \delta_{\rho_1+\rho_2,\rho_3+\rho_4} \, \psi_{\rho_1}^{- \dot{A}}
     \psi_{\rho_2}^{- \dot{B}}
     \tilde{\psi}_{\rho_3}^{- \dot{C}}
     \tilde{\psi}_{\rho_4}^{- \dot{D}}
     \epsilon_{\dot{A}\dot{B}}
     \epsilon_{\dot{C}\dot{D}}.
\end{eqnarray}
We will work in the bosonized language, in which 
\begin{equation}
    \psi_\rho^{+\dot{1}} = i e^{iH_\rho}\, , \qquad 
    \psi_\rho^{-\dot{2}} = i e^{-iH_\rho}\, , \qquad 
    \psi_\rho^{+\dot{2}} =  e^{iK_\rho}\, , \qquad 
    \psi_\rho^{-\dot{1}} =  e^{-iK_\rho}\, , 
\end{equation}
Here $H_\rho$ and $K_\rho$ are canonically normalized bosonic fields, in terms of which the (unit normalized) heavy states take the form \cite{Galliani:2016cai}
\begin{equation}
    |H\rangle = |s=\k p,\k\rangle = \left[ \Sigma_{\k}\tilde{\Sigma}_\k
    \prod_{\rho=0}^{\k-1}
    e^{i \left(p+\frac{1}{2}-\frac{\rho}{\k}\right)(H_\rho + K_\rho)}
    e^{i \left(\frac{1}{2}-\frac{\rho}{\k}\right)(\tilde{H}_\rho + \tilde{K}_\rho)} \right]^{\frac{N}{\k}}
    |0\rangle
    \, ,
\end{equation}
where $\Sigma_{\k}$ and $\tilde{\Sigma}_\k$ are the twist operators. Note that the contribution of $\Sigma_{\k}$ and $\tilde{\Sigma}_\k$ to the correlators will simply factorize, since the $\psi_\rho$ fermions diagonalize the twisted boundary conditions. Choosing the labelling of the insertion points for later convenience, the correlator to be computed is then
\begin{eqnarray}
    &&\langle  O_{\frac{1}{2}} (x_1) O_{\frac{1}{2}}(x_2) O^\dagger_1(x_3) O_H(x_4) O^\dagger_H(x_5) \rangle = \frac{1}{\k}  \prod_{\rho,\rho'=0}^{\k-1}
    \sum_{\rho_i=0}^{\k-1} 
     \delta_{\rho_3+\rho_4,\rho_5+\rho_6} 
     \\[1ex]
    && \Big\langle 
    \left[\psi_{\rho_1}^{+ \dot{A}_1} \tilde{\psi}_{\rho_1}^{+ \dot{B}_1} \epsilon_{\dot{A}_1\dot{B}_1}\right] (x_1) 
    \left[\psi_{\rho_2}^{+ \dot{A}_2} \tilde{\psi}_{\rho_2}^{+ \dot{B}_2} \epsilon_{\dot{A}_2\dot{B}_2}\right] (x_2)
    \left[\psi_{\rho_3}^{- \dot{A}_3}
     \psi_{\rho_4}^{- \dot{B}_3}
     \tilde{\psi}_{\rho_5}^{- \dot{C}_3}
     \tilde{\psi}_{\rho_6}^{- \dot{D}_3} 
      \epsilon_{\dot{A}_3\dot{B}_3}
       \epsilon_{\dot{C}_3\dot{D}_3}\right] (x_3)
    \nn \\[2ex]
    &&  \, e^{i \left(p+\frac{1}{2}-\frac{\rho}{\k}\right)(H_\rho + K_\rho)} (x_4)
    e^{-i \left(p+\frac{1}{2}-\frac{\rho'}{\k}\right)(H_{\rho'} + K_{\rho'})} (x_5)
    e^{i \left(\frac{1}{2}-\frac{\rho}{\k}\right)(\tilde{H}_\rho + \tilde{K}_\rho)} (x_4)
    e^{-i \left(\frac{1}{2}-\frac{\rho'}{\k}\right)(\tilde{H}_{\rho'} + \tilde{K}_{\rho'})} (x_5) \nn
    \Big\rangle. 
    \end{eqnarray}
Clearly, charge conservation implies $\rho = \rho'$. For the same reason, the correlator vanishes unless $\rho_1 = \rho_3$ and $\rho_2 = \rho_4$, or
$\rho_1 = \rho_4$ and $\rho_2 = \rho_3$, or both. An analogous statement holds with $\rho_3,\rho_4$ replaced by $\rho_5,\rho_6$, hence all contributions trivially satisfy the $\rho_3 + \rho_4=\rho_5+\rho_6$ constraint. Consequently, we only really need to sum over all possible values of, say, $\rho_1$ and $\rho_2$, and also compute the product over $\rho$. 
In this way, up to an irrelevant numerical factor, the holomorphic free field contractions give
\begin{equation}
    \sum_{\rho_1,\rho_2=0}^{\k-1} 
    \frac{1}{|x_{45}^{2h_H}x_{13}x_{23}|^2} \left(\frac{x_{41}x_{35}}{x_{51}x_{34}}\right)^{p+\frac{1}{2}-\frac{\rho_1}{\k}}
    \left(\frac{x_{42}x_{35}}{x_{52}x_{34}}\right)^{p+\frac{1}{2}-\frac{\rho_2}{\k}}
    \left(\frac{\bar{x}_{41}\bar{x}_{35}}{\bar{x}_{51}\bar{x}_{34}}\right)^{\frac{1}{2}-\frac{\rho_1}{\k}}
    \left(\frac{\bar{x}_{42}\bar{x}_{35}}{\bar{x}_{52}\bar{x}_{34}}\right)^{\frac{1}{2}-\frac{\rho_2}{\k}}
    \label{HLLLHD1D5}
\end{equation}
where $h_H$ is the weight of the heavy state. 

We can now take $x_3\to 1$, $x_4\to 0$ and $x_5\to \infty$, and perform the sums over $\rho_1$ and $\rho_2$ explicitly. Upon doing so, we find that the structure of this orbifold CFT correlator Eq.~\eqref{HLLLHD1D5} precisely matches the worldsheet correlator~\eqref{HLLLHformWS}.

\subsection{Hawking radiation from the worldsheet}
\label{sec:Hawking}

As a final application of our results, we now use the HLLH correlator~\eqref{finalHLLH} to compute the amplitude that describes the analogue of the Hawking radiation process for the JMaRT backgrounds~\cite{Chowdhury:2007jx,Avery:2009xr,Avery:2009tu,Chakrabarty:2015foa}. In the bulk, this process is ergoregion radiation, which is a feature of the full asymptotically flat JMaRT solutions \cite{Cardoso:2005gj}. 
The ergoregion does not survive the fivebrane decoupling limit~\cite{Martinec:2018nco} or AdS$_3$ decoupling limit, however aspects of the process can still be studied quantitatively in those limits.
This process has been interpreted as an enhanced analogue of Hawking radiation, since both are described by the same microscopic process in the holographic CFT~\cite{Chowdhury:2007jx}. 
Indeed, acting on a thermal state, this vertex operator gives precisely the spectrum and rate of Hawking radiation of the corresponding black hole, while acting on the states dual to the JMaRT solutions yields their characteristic spectrum and rate of emission~\cite{Chowdhury:2007jx,Avery:2009xr,Avery:2009tu,Chakrabarty:2015foa}.

The emission spectrum and rate for general $\k,s,\bar{s}$ was computed in supergravity and symmetric product orbifold CFT in~\cite{Chakrabarty:2015foa}, building on the results of~\cite{Chowdhury:2007jx,Avery:2009xr,Avery:2009tu}.
We will reproduce these results from the worldsheet CFT.

We start with a specific HLLH correlator in which the light operators are given by minimally coupled scalars in six dimensions, after reducing on the $\TT^4$. The corresponding vertex operators were defined in Eq.~\eqref{6DscalarsNSNS}. These are not chiral primaries of the boundary theory, but are their superdescendants within the short multiplet, so the holographic correlator arising in the AdS$_3$  limit is easily computed by using the techniques outlined in the previous sections. 
The amplitude of interest involves an initial state consisting of a probe excitation on top of the JMaRT background, a vertex operator $\Vv$ associated to a light insertion, and a final state given by the black hole microstate. Schematically we have
\begin{equation}
    \bra{s, \bar{s}, \k}\Oo(x) \ket{s, \bar{s}, \k + \text{probe}} =
    \bra{s, \bar{s}, \k}\Oo(x) \, \Oo^\dagger(0)\ket{s, \bar{s}, \k} \, .
    \label{HawkingSchematic}
\end{equation}

To begin with, we work with $\k=1$.
Up to an overall sign, and considering the lowest energy state, the holomorphic part of the amplitude for the Hawking emission of a single quanta of dimension $h = \frac{l}{2} + 1$ and whose corresponding vertex operator has charge $m' = k-\frac{l}{2}$ reads \cite{Avery:2009tu}
\begin{equation}
\label{HawkingAvery}
	\Aa_L(x) = \frac{1}{x^{(1+\alpha) \frac{l}{2} -\alpha \;\! k +1}} = \frac{1}{x^{\frac{l}{2} + 1 - \alpha (k - \frac{l}{2})}} \, . 
\end{equation}
Here $\frac{l}{2}$ denotes the total angular momentum of the probe on the $\SSS^3$ part of the geometry, while $k$ is the number of $\Jd_0^+$ operators acting on the state with the lowest projection, appearing in the definition of the vertex operator. To compare their computation with our worldsheet result, one uses the following (notation) map: 
\begin{align}
	k - \frac{l}{2} \mapsto m' \qqquad
	\frac{l}{2} + 1 \mapsto h \qqquad
	\alpha & \mapsto l_2 = \m + \n = 2s + 1 \, .
\end{align}
Taking care of the cylinder-to-plane conversion factor $x^{-\frac{l}{2}-1}$, one obtains
\begin{equation}
\label{HawkingFinal}
	\Aa_L(x) = \frac{1}{x^{2h - m'(2s+1)}} \, .
\end{equation}

We now perform the analogous computation in the worldsheet cosets. From Eq.~\eqref{HawkingSchematic}, in the worldsheet formalism all we need to do is to insert the second operator at the boundary origin, i.e. to take the $x_2\to 0$ limit in Eq.~\eqref{finalHLLH} (with $\k=1$ for now).
This gives
\begin{align}
     \lim_{x_2\to 0} \;  x_2^{m'(2s+1)} \bar{x}_2^{\bar{m}'(2\bar{s}+1)} \; & \langle s,\bar{s}, 1 |\Oo^{(m',\bar{m}')}_h(x_1)\Oo_h^{(m',\bar{m}')\dagger}(x_2)|s,\bar{s},1 \rangle  \nn \\
     &\hspace{5mm}=\, \frac{1}{x^{2h-m'(2s+1)} }
    \frac{1}{\bar{x}^{2h - \bar{m}'(2\bar{s}+1)}}
    \, ,
    \label{HawkingHLLH}
\end{align}
in agreement with \eqref{HawkingFinal} upon including the antiholomorphic contribution.

The procedure is analogous for general $\k,s, \bar{s}$.
We again evaluate the amplitude for $x_2 \to 0$ by including the appropriate Jacobian factor for the light state, and obtain 
\begin{align}
\label{HawkingGeneral}
   \lim_{x_2\to 0} \;\!  \k^{2h} & x_2^{m_2'\frac{(2s+1)}{\k} + h(1-\frac{1}{\k})} \, \bar{x}_2^{\bar{m}_2'\frac{(2\bar{s}+1)}{\k} + h(1-\frac{1}{\k})}  \,  \langle s,\bar{s}, \k |\Oo^{(m',\bar{m}')}_h(x_1)\Oo_h^{(m',\bar{m}')\dagger}(x_2)|s,\bar{s},\k\rangle \nonumber \\
   &=\, \frac{1}{\k^{2h+2}} \sum_{r_1= 0}^{\k-1} \dfrac{e^{2 \pi i\,\frac{r_1}{\k} ( m'(2s+1) - \bar{m}'(2\bar{s}+1) )} }{x_1^{h(1+ \frac{1}{\k})- m' \frac{(2s + 1)}{\k}} \, \bar{x}_1^{h(1+ \frac{1}{\k})- \bar{m}' \frac{(2\bar{s} + 1)}{\k}}   } 
   \sum_{r_2= 0}^{\k-1} e^{2 \pi i\,\frac{r_2}{\k} ( -m'(2s+1) + \bar{m}'(2\bar{s}+1) )}
   \nonumber \\
     &=\, \frac{1}{\k^{2h}} 
   \dfrac{1}{x^{h(1+\frac{1}{\k}) - m'\frac{(2s+1)}{\k}} \; \bar{x}^{h(1+\frac{1}{\k}) - \bar{m}'\frac{(2\bar{s}+1)}{\k}}}\, \sum_{\ell \in \ZZ}\delta_{m'(2s+1) - \bar{m}'(2\bar{s}+1),\,  \k \; \! \ell} \, , 
\end{align}
where in the first equality we have exchanged the finite sum with the limit, and $x = x_1, \, m = m_1 = - m_2$. When $\k = 1$, this reduces to Eq.~\eqref{HawkingHLLH}. 
The Kronecker comb enforces the constraint $
    (2s + 1) m' - (2\bar{s} + 1) \bar{m}' \in \k \, \ZZ $,
which is a direct consequence of the difference beween left and right null-gauge constraints \eqref{nullgaugeconstrAdS} in the regime of interest. Moreover, by first multiplying the correlator in Eq.~\eqref{finalHLLH} by $x^n \, \bar{x}^{\bar{n}}$ we can also consider descendant insertions. This condition is in agreement with the results present in \cite{Avery:2009xr,Chakrabarty:2015foa} (see also~\cite{Martinec:2018nco}), where our $n_y$ has to be identified with their $\lambda$ from the supergravity analysis.

When considering the case of multi-particle emission, the above amplitude must be multiplied by a combinatorial factor, as explained in \cite{Chowdhury:2007jx,Avery:2009tu,Avery:2009xr}. To obtain the emission rate, one needs to consider the unit amplitude evaluated at $(x,\bar{x}) = (1,1)$, implying that the spatial dependence trivialises. Nevertheless, the crucial feature related to the presence of the prefactor $\k^{-2h}$, which enters the final expression of the emission rate\footnote{To compare to the final results of \cite{Avery:2009xr,Chakrabarty:2015foa} one must include the additional factor $\sqrt{\k \nu}$, where $\nu$ is related to the Bose enhancement, which is not visible for a single-particle process.}, is reproduced by~\eqref{HawkingGeneral}.

Even though the spatial dependence of the two-point function Eq.~\eqref{HawkingGeneral} plays a trivial role in the emission rate, the power of $x$ has a precise meaning in terms of the energy spectrum of the nearly unstable Hawking quanta \cite{Avery:2009tu,Chakrabarty:2015foa}. Indeed, consider the holomorphic part of the energy of these modes. In the conventions of \cite{Chakrabarty:2015foa}, the corresponding spectrum reads 
\begin{equation}
    \omega \;\! \k R_y  \,=\, \half \alpha \k (m_\phi - m_\psi) - \half \bar{\alpha} \k (m_\phi + m_\psi) - 2 \left(\frac{l}{2}+1\right) \, . 
\end{equation}
In our notation, $(m_\phi - m_\psi) = 2 m', ~\;(m_\phi + m_\psi) = - 2 \bar{m}'\,$, and $\, \frac{l}{2}+1 = h $, so this becomes
\begin{equation}
    -   \omega R_y \,=\,  \frac{2 h}{\k } - \alpha \, m' - \bar{\alpha} \, \bar{m}' \, ,
\end{equation}
where $\alpha, \bar{\alpha}$ are the same as in Eq.~\eqref{SFchargesHCFT}. Finally, taking care of the cylinder-to-plane conversion factor for a field of spacetime conformal dimension $h$, we obtain
\begin{equation}
\label{EnergyEmissionSugra}
     - \omega R_y  \,=\, 2 h \left(1 + \frac{1}{\k}\right) - \alpha \, m' - \bar{\alpha} \, \bar{m}'  \, . 
\end{equation}
The RHS is exactly the sum of the exponents of $x$ and $\bar{x}$ in Eq.~\eqref{HawkingGeneral}. Furthermore, we note that this relation is precisely the sum of the left and right bosonic null gauge constraints Eq.~\eqref{nullgaugeconstrAdS} for discrete states with $n = \bar{n} = 0$.

The emission takes place when the energy is positive, $\omega > 0$, and corresponds to quanta leaving the AdS region; in a near-decoupling limit, these quanta escape into the asymptotically flat region.
Indeed, the exponent of $x$ in Eq.~\eqref{HawkingGeneral} becomes positive and the amplitude diverges at large $x$, such that the energy indeed turns from negative to positive. This is consistent with the description of the ergoregion radiation process as pair creation~\cite{Chowdhury:2008bd}.

\section{Discussion and outlook}
\label{sec:discussion}

In this paper we have computed a large set of worldsheet correlators describing the dynamics of light modes probing a class of highly-excited supergravity backgrounds, the JMaRT solutions, in the fivebrane decoupling limit.
The results are exact in $\alpha'$ and were obtained by exploiting the solvability of the null-gauged WZW models corresponding to these backgrounds.

These coset models provide a powerful method to calculate HLLH correlators, since the heavy states are already taken into account in the worldsheet CFT itself. 
Thus spacetime HLLH correlators are two-point functions on the worldsheet, which can be computed once the vertex operators have been constructed.

We constructed physical vertex operators in both NS and R sectors, and then computed several families of correlators in the full coset models. 
We primarily focused on short strings belonging to discrete representations of the affine SL(2,$\R$) algebra, as well as a tower of modes generated by worldsheet spectral flow. 
Our main techniques can also be employed in more general sectors of the theory.

In the IR AdS$_3$ limit, due to the non-trivial gauging, the identification of the $x$ variable dual to the local coordinate of the holographic CFT requires some care.
Once we made this identification, we computed several non-trivial HLLH correlators explicitly, and analyzed them in the context of AdS$_3$/CFT$_2$.

Vertex operators that are local on the AdS$_3$ boundary are constructed by summing over all allowed values of the spacetime modes. 
An important step in our analysis consists of identifying these modes. 
We chose a gauge in which the IR AdS$_3$ boundary coordinates are $(t,y)$ of the timelike $\mathbb{R}$ and spacelike $\SSS^1$ directions of the (10+2)-dimensional model before gauging. 
We therefore identified the spacetime mode indices with the quantum numbers $m_y$ and $\bar{m}_y$ defined in \eqref{mymodes}. 
Then the gauge constraints \eqref{nullgaugeconstr} satisfied by the physical states imply that the $m_y$ mode numbers take values in $\mathbb{Z}/\k$. 
This is how the worldsheet coset models capture the fact that when $\k>1$, the heavy background states of the symmetric product orbifold CFT are in the $\k$-twisted sector
\cite{Giusto:2012yz,Chakrabarty:2015foa}.

We observed that, at large $N$, several correlators agree exactly between worldsheet and symmetric product orbifold CFT. 
The fact that our correlators are exact in $\alpha'$ significantly strengthens previous results that compared HLLH correlators between the separate supergravity and symmetric product orbifold CFT regimes.

To demonstrate our method, we presented a detailed example with an $(h,\bar{h})=(\half,\half)$ chiral primary.
The worldsheet correlator involves a non-trivial structure in terms of the boundary coordinate $x$, Eq.~\eqref{HLLHGalliani}. 
When the background is BPS, the correlator agrees precisely with the supergravity and symmetric product orbifold CFT correlators computed in~\cite{Galliani:2016cai}.
The non-BPS JMaRT backgrounds were not considered in~\cite{Galliani:2016cai}, however we demonstrated that the agreement extends also to those backgrounds.

Similarly to correlators on the background of the global AdS$_3$ vacuum, the holomorphic and antiholomorphic sectors are related through the constraint $m_y - \bar{m}_y = n_y$, where $n_y$ is the quantized momentum on the $y$ circle. 
Thus, while the spacetime modes $m_y$, $\bar{m}_y$ are fractional, their difference must be an integer. 
This mirrors the $m-\bar{m} \in \mathbb{Z}$ condition in the SL(2,$\mathbb{R}$)/U(1) cigar coset and in global AdS$_3$, which ensures that the wavefunctions are single-valued. 
In our models, the difference of the left and right gauge constraints leads to the mod $\k$ condition in Eq.~\eqref{nnbaralphah}, constraining which of the SL(2,$\R$) modes can contribute.
The HLLH correlator is then obtained by summing over a specific linear combination of $m$-basis worldsheet two-point functions.

Our analysis of these correlators involving the $h=1/2$ light operator indicated a way to obtain similar expressions for more general correlators. 
We considered general massless vertex operators, which correspond to symmetric product orbifold CFT operators in short multiplets whose top component is a chiral primary of arbitrary weight $h$, including those that live in twisted sectors.   
We computed all HLLH correlators where the light operators are massless, and where the heavy states correspond to any of the general family of orbifolded JMaRT configurations, including their BPS limits.
The result assumes a remarkably simple form, presented in Eq.~\eqref{finalHLLH}. 
It is built from three distinct factors: (1) the global AdS$_3 \times \SSS^3$ \textit{vacuum} two-point function of the light operators inserted at the $\k$-th roots of the original insertion points $x_i$, (2) the Jacobian factor associated with the corresponding change of coordinates, and (3) an additional factor coming from the way in which operators of definite R-charge transform under spectral flow. The product of these factors is then summed over all such roots. 
This structure reflects that one can formulate the computation in a $\k$-fold covering space of the target space.

We then obtained a similar expression for all higher-point functions of the schematic form $\langle H|\Oo_{1}(x_1,\bar{x}_1)\dots \Oo_{n}(x_n,\bar{x}_n)|H\rangle$, with heavy JMaRT states, and $n$ massless insertions. This is presented in Eq.~\eqref{finalHLLLLLH}.  We expect this to be valid for an arbitrary number of massless insertions of weights $h_i$ and charges $m'_i$ and $\bar{m}'_i$, and also arbitrary parameters $(\k,s,\bar{s})$ for which a consistent background exists. In this way, we have provided a recipe for computing such $(n+2)$-point heavy-light correlators in terms of $n$-point global AdS$_3\times \SSS^3$ \textit{vacuum} correlation functions of the corresponding light insertions.

It is known that vacuum two- and three-point functions of chiral primary operators are protected~\cite{Baggio:2012rr}. We therefore conjectured that heavy-light correlators in JMaRT heavy states are protected whenever the corresponding vacuum correlator in our general formula~\eqref{finalHLLLLLH} is protected.  We have investigated a particular HLLLH five-point function---the first of its kind in the literature---finding that worldsheet and symmetric product results agree. We leave a more general investigation of this proposal to future work. 

As an application, we have shown that our results describe the analog of the Hawking radiation process for the general family of non-BPS JMaRT black hole microstates, generalizing the analysis in~\cite{Avery:2009tu,Avery:2010qw,Chakrabarty:2015foa}.

In addition to these main results, our work has clarified some important technical details. For instance, the full asymptotically linear dilaton JMaRT backgrounds do not have AdS$_3\times \SSS^3$ isometries. 
Correspondingly, in the worldsheet cosets, the SL(2,$\RR$) and SU(2) raising and lowering operators $J^\pm$, $K^\pm$ of the (10+2)-dimensional ungauged model do not commute with the gauging.
Thus the NS sector vertex operators of the cosets do not have well-defined SL(2,$\R$) or SU(2) spins, see for instance Eq.\;\eqref{vectorNSgauged}. The same holds for the chirality quantum number $\vep$ in the R sector, as discussed around Eq.~\eqref{RRnullAnsatz}.

The absence of the SL(2,$\R$) spin has important implications also from the holographic point of view. It underlines the fact that the construction of $x$-basis operators is only appropriate in the AdS$_3$ limit, and breaks down otherwise. The breakdown of the $x$ coordinate is a signal of the non-locality of the non-gravitational little string theory that lives on the worldvolume of the NS5 branes, dual to the full asymptotically linear-dilaton models. Thus the states we have constructed contain valuable information about the dual LST and, more generally, about non-AdS holography \cite{Giveon:2017myj,Giveon:2017nie,Asrat:2017tzd}. 
We have nevertheless demonstrated how, in the AdS$_3$ limit, our vertex operators acquire definite spins and reduce to the appropriate expressions.

Our results suggest several directions for future investigations. First, it would be interesting to compute more general worldsheet correlators, both in the AdS$_3$ limit and in the full models. 
Our correlators are likely to generalize to a larger set of worldsheet vertex operators that correspond to operators in the symmetric product orbifold CFT that transform nicely under spacetime spectral flow~\cite{Guo:2022ifr}.
In global AdS$_3$, correlators are known to involve a highly non-trivial structure related to the non-conservation of the spectral flow number~\cite{Maldacena:2001km}. 
More generally, one would like to describe the physics of long/winding strings and their correlators in these backgrounds. 
A number of interesting techniques recently developed in \cite{Dei:2021xgh,Dei:2021yom} (for the bosonic case) are likely to have interesting implications for computations in the coset theories, for which SL(2,$\R$) constitutes a crucial building block. 

It would also be interesting to study such correlators by using conformal perturbation theory on top of a putative dual CFT explicitly associated to the NSNS singular point \cite{Seiberg:1999xz} of the moduli space, defined along the lines of \cite{Balthazar:2021xeh,Eberhardt:2021vsx}. Doing so would require an understanding of how to define the JMaRT heavy states in such a theory. Separately, it would be interesting to investigate the case $n_5=1$, which would require going beyond the RNS formalism, as done in related recent developments~\cite{Gaberdiel:2018rqv,Eberhardt:2018ouy,Gaberdiel:2022oeu}. Here one should go though the coset construction starting with the supergroup PSU(1,1|2).

In the full asymptotically linear dilaton models, more general correlators can be computed by using the vertex operators constructed in Section \ref{sec:nullgaugedmodel}. However, a shift in perspective will be needed, since the $x$ coordinate seems unlikely to be of any use in this regime. Although \textit{a priori} in our case it is more natural to work in the $m$-basis, it seems plausible to relate our results to the momentum-space correlators studied in \cite{Giveon:2017nie,Giveon:2017myj}, see also \cite{Giribet:2017imm}. In those papers, the authors work with a related null-gauged model, and further interpret their holographic LST correlators in terms of an irrelevant (single-trace) $T\bar{T}$-deformation of the IR CFT$_2$.

Separately, it will be interesting to investigate our proposal for the subset of heavy-light correlators that we expect to be protected by considering the dual computations in the symmetric product orbifold CFT.

Last, but not least, one would like to explore further how these correlation functions encode more detailed information about the physics of the microstate backgrounds we are working with. For instance, two-point functions are expected to probe the multipole ratios of the geometry~\cite{Bena:2020see,Bianchi:2020bxa}, while certain worldsheet three-point functions should be related to the Penrose process in the JMaRT backgrounds~\cite{Bianchi:2019lmi}.

Although the JMaRT backgrounds are atypical microstates, 
the HLLH correlators we have computed approach black-hole-like behaviour at large $\k$, reflecting the properties of the backgrounds in this limit. We expect that the techniques developed in this work will help further the study of more typical black hole microstates in string theory.

\acknowledgments

For discussions, we thank I.~Bena, S.~Chakraborty, N.~\v{C}eplak, A.~Dei, 
S.~Giusto, S.~Hampton, 
M.~Guica, Y.~Li, 
S.~Massai,  S.~Rawash, R.~Russo,  M.~Santagata. We  thank E.~Martinec for comments on an early draft of the manuscript.
The work of D.B.~was supported by the Royal Society Research Grant RGF\textbackslash R1\textbackslash181019. 
The work of D.T.~was supported by a Royal Society Tata University Research Fellowship.
D.B. and D.T. thank IPhT Saclay for hospitality.


\bibliographystyle{JHEP}
\bibliography{refs}

\end{document}